\documentclass[11pt,a4paper]{article}

\usepackage[utf8]{inputenc}
\usepackage[T1]{fontenc}
\usepackage{booktabs}
\usepackage{tabularx}
\usepackage{array}
\usepackage{lmodern}
\usepackage{tikz}
\usetikzlibrary{positioning,calc,arrows.meta}
\usepackage[bottom]{footmisc}
\usepackage{longtable}  
\usepackage[table]{xcolor}
\usepackage{graphicx}
\usepackage{booktabs}
\usepackage{makecell}
\usepackage{caption}
\usepackage{float}
\usepackage{placeins}
\usepackage[utf8]{inputenc}
\usepackage[T1]{fontenc}
\usepackage{lmodern}
\usepackage{ragged2e}
\usepackage{changepage}
\usepackage{fontawesome5}
\usepackage{threeparttable}
\usetikzlibrary{positioning, calc, arrows.meta}
\usepackage[most]{tcolorbox}
\usepackage{caption}
\usepackage[most]{tcolorbox}
\usepackage{xcolor}

\newtcolorbox{AIprompt}{
    enhanced,
    breakable,
    colback=white,
    colframe=gray!45,
    boxrule=0.5pt,
    arc=2pt,
    left=12pt,
    right=12pt,
    top=8pt,
    bottom=8pt,
    before skip=10pt,
    after skip=10pt,
    fontupper=\small\itshape
}

\newtcolorbox{AIresponse}{
    enhanced,
    breakable,
    colback=gray!6,
    colframe=gray!6,
    boxrule=0pt,
    borderline west={2pt}{0pt}{gray!65},
    arc=0pt,
    left=12pt,
    right=12pt,
    top=10pt,
    bottom=10pt,
    before skip=12pt,
    after skip=12pt,
    fontupper=\small
}

\usepackage{booktabs}
\usepackage{amssymb}
\usepackage{array}

\newcommand{\filled}{\ensuremath{\bullet}}
\newcommand{\emptycircle}{\ensuremath{\circ}}

\newcommand{\AIscale}[1]{%
  \ifcase#1
    \emptycircle\emptycircle\emptycircle\emptycircle\emptycircle
  \or
    \filled\emptycircle\emptycircle\emptycircle\emptycircle
  \or
    \filled\filled\emptycircle\emptycircle\emptycircle
  \or
    \filled\filled\filled\emptycircle\emptycircle
  \or
    \filled\filled\filled\filled\emptycircle
  \or
    \filled\filled\filled\filled\filled
  \fi
}

\usepackage[margin=1in]{geometry}
\usepackage{adjustbox}
\usepackage{setspace}

\usepackage[english]{babel}
\usepackage{csquotes}     

\usepackage{amsmath,amssymb,amsthm}

\usepackage{graphicx}
\usepackage{booktabs}
\usepackage{caption}
\usepackage{float}
\usepackage{booktabs}
\usepackage{pdflscape}
\usepackage{subcaption}
\usepackage[authoryear,round]{natbib}
\usepackage[backref=page]{hyperref}
\usepackage{backref}

\title{\textbf{LLMs and the Organization of Scientific Work}}
\author{%
  Stefano Bianchini\\
  \normalsize BETA, University of Strasbourg\\
  \normalsize \texttt{s.bianchini@unistra.fr}
  \and
  Georgiana-Ioana Coroama\\
  \normalsize BETA, University of Strasbourg\\
  \normalsize \texttt{gicoroama@unistra.fr}
}
\date{}

\begin{document}

\maketitle

\begingroup
\renewcommand{\thefootnote}{}
\footnotetext{%
\noindent We are particularly indebted to Henry Sauermann for suggesting, over coffee, the use of CRediT contribution data to study how AI may have affected the organization of scientific work, an idea from which this research originated. Henry also provided valuable feedback on earlier versions of this work. We are grateful to Giacomo Damioli for his thoughtful comments, which helped us refine our theoretical framework. We also thank participants at seminars and conferences held at the Sant'Anna School of Advanced Studies (Pisa, 2025), Collegio Carlo Alberto (Turin, 2025), OECD Rethinking Innovation in the Age of AI (Paris, 2025), KITZ International PhD Workshop (Karlsruhe, 2026), KID Summer School (Nice, 2026), the OIS Research Conference (Copenhagen, 2026), and the International Schumpeter Society Conference (Strasbourg, 2026) for their constructive suggestions. This research was supported by the ENACT MANTRA Chair (Mapping the Innovation Landscape: Towards Responsible AI).}

\endgroup

\begin{abstract}
\noindent The advent of large language models (LLMs) represents a technological revolution for science. Their diffusion may reshape not only how scientists conduct research, but also how they organize their work within teams. Here, we develop a conceptual framework in which LLMs expand individual capabilities across a range of research activities, reducing reliance on complementary human inputs within tasks and facilitating a more specialized division of labor among team members. We further propose that teams' history of collaboration may constrain their ability to reorganize in response to the new technological capabilities. Using CRediT contribution statements from approximately 120,000 PLOS ONE articles published between 2018 and 2025, we investigate changes in the division of labor within scientific teams following the release of ChatGPT in late 2022. We document three main findings. First, participation in scientific tasks falls below its previous trajectory, particularly in conceptualization, execution, and writing. Second, researchers become more specialized, exhibiting narrower task portfolios and less overlap in their contributions with co-authors. Third, these changes are concentrated among newly formed teams, whose members have no prior collaboration ties, whereas established teams, whose members have all collaborated previously, exhibit considerably greater organizational inertia. Our findings suggest that, in the wake of LLM diffusion, scientific work is being reorganized around greater individual autonomy and an increasingly specialized division of labor within teams.


\end{abstract}

\newpage
\doublespacing      

\section{Introduction}

The history of science and the history of technology are conjoined. To each era, its instruments. Scientists have always extended the reach of their own capacities through tools that allow them to observe, calculate, and experiment beyond what the unaided mind could achieve. Telescopes to see farther, microscopes to see smaller, and computers to calculate faster. But AI for what? Here, the boundary between tool and user begins to blur. With AI---and large language models (LLMs) in particular---we have created technologies that reach beyond narrow, well-defined operations into open-ended cognitive domains once confined to our own minds. No wonder, then, that some observers easily imagine these creations becoming autonomous ``scientists'' in their own right, endowed with minds as agile as ours, or perhaps more so \citep{kitano2021nobel, krenn2026philosophy, lu2026towards}. What happens to the scientific enterprise when intelligent machines begin to participate in activities once reserved for human minds?

A growing literature has begun to examine this question through the lens of knowledge production, asking how AI affects the productivity, novelty, and direction of scientific research \citep[see, among others,][]{furman2020automation,bianchini2022gmi,gao2024quantifying,bianchini2026ai,hao2024ai,mohnen2026ai}. However, the consequences of AI as a new ``method of invention'' may extend well beyond \textit{what} science produces to \textit{how} science is organized, a dimension of its impact that remains unexplored. Here, we turn to this organizational dimension and study how the diffusion of LLMs is reshaping the division of labor within scientific teams. Who does what within a team, how participation in different tasks expands or contracts across team members, and how labor is (re)organized across the research pipeline are the overarching questions we seek to address.

Our inquiry builds on two long-standing insights: (i) that science relies fundamentally on a division of labor, and (ii) that realizing the potential of major new technologies requires complementary organizational change. Regarding the former, we know that science has grown in scale and complexity over the past century and the mounting ``burden of knowledge'' has driven a long-term shift from the ``lone genius'' toward larger, often interdisciplinary, teams \citep{moody2004structure, wuchty2007increasing, jones2009burden, fortunato2018science}. This expansion of teamwork has been accompanied by a deeper division of labor, in which tasks and responsibilities are increasingly distributed across team members based on factors such as expertise, seniority, and gender \citep{lariviere2015team, macaluso2016, sauermann2017authorship}. Regarding the latter, the history of major technological transitions shows that new technologies rarely transform production simply by being inserted into existing workflows; rather, their full potential materializes only gradually as organizations restructure their internal processes and task boundaries around the new technological paradigm \citep{david1990dynamo,brynjolfsson2000beyond,perez2010technological,brynjolfsson2021productivity}.\footnote{Electrification provides the classic example. Its early adoption largely preserved existing factory arrangements and its full productivity gains emerged only with the subsequent redesign of factories and production processes around electric power \citep{david1990dynamo}.}

The launch of ChatGPT in late 2022 offers a particularly useful setting to observe such a process unfolding within science. From early 2023 onward, LLMs diffused rapidly across virtually all domains, at a pace rarely seen for previous general-purpose technologies \citep{bick2026rapid}. Science is no exception. The sudden arrival of a new (largely unexpected) technology into established ways of organizing scientific work, combined with its rapid (near-instantaneous) adoption, provides a unique opportunity to study how research teams have begun to reorganize around it. What makes this transition particularly consequential is the breadth of tasks that LLMs can perform. Indeed, whereas earlier AI/ML applications were typically developed for specific scientific tasks (e.g., classifying medical images, predicting protein structures, or detecting patterns in astronomical datasets), LLMs have begun to support an expansive pipeline of interconnected research activities. These include retrieving and synthesizing literature, generating and stress-testing hypotheses, writing and debugging code, interpreting empirical results, and drafting and revising manuscripts \citep{binz2025should, luo2025llm4sr}. As a general-purpose cognitive technology spanning conceptual, technical, analytical, and linguistic tasks, LLMs represent a qualitatively different shock to the organization of scientific labor.\footnote{In fact a growing body of empirical research exploits the rapid diffusion of general-purpose LLMs beginning in late 2022 and accelerating throughout 2023 as a quasi-experimental design to investigate their effects on economic and organizational outcomes \citep{tang2025gender, devigili2026skill, gross2026heavily}.}

Yet studying how scientific labor is organized, especially in response to a disruptive technology, is not straightforward. First, the division of labor manifests along multiple dimensions, such as the number of tasks an individual scientist performs, the headcount dedicated to a specific function, or the degree of task overlap between co-authors. Second, these internal arrangements are largely hidden from outsiders, as conventional publication data reveal who collaborated on a paper but provide little, if any, information on who actually did what---despite a growing push toward greater transparency in the attribution of scientific contributions \citep{kiermer2026creating}. Here, we address these challenges both conceptually and empirically. 

Conceptually, we develop a framework in which LLMs affect the division of labor along two distinct margins: first, by changing how many researchers participate in individual scientific activities, and second, by changing how activities are distributed across researchers. Both responses arise from the expansion of individual capabilities afforded by LLMs, which, on the one hand, reduces reliance on complementary human inputs within tasks and, on the other, allows researchers to perform activities more independently, making it easier to distribute responsibilities across team members with less overlap. We further argue that a team's prior history of collaboration may constrain its ability to reorganize along these margins. 

Empirically, the framework generates testable predictions regarding within-task participation and cross-task specialization, which we test using granular data on author contributions from the \textit{Contributor Roles Taxonomy} (\href{https://casrai.org/credit/roles}{CRediT}) \citep{allen2014publishing, mcnutt2018transparency} for $117{,}211$ papers published in PLOS ONE between 2018 and 2025. CRediT standardizes scientific contributions into 14 distinct roles spanning conceptual work (e.g., Conceptualization, Methodology), technical and empirical activities (e.g., Software, Data Curation, Formal Analysis), writing, and coordination (e.g., Supervision, Project Administration). This allows us to observe who performs which activities within a research team and how participation in each activity is distributed across team members. 

Taking the public launch of ChatGPT in November 2022 as a temporal marker for LLM diffusion, we use an event-study approach to investigate changes in the division of labor within scientific teams. The main findings of our empirical analysis can be summarized as follows. The diffusion of LLMs coincides with a reorganization of scientific work characterized by declining participation in research activities (that is, fewer researchers contributing to the same activity) and increasing specialization among team members (that is, each researcher performing a narrower range of activities, with less overlap between collaborators). These changes are particularly pronounced in newly formed teams, whereas established teams exhibit greater organizational inertia, suggesting that prior collaborative arrangements may indeed constrain how teams adapt to new technological capabilities. 

The paper contributes to three strands of literature. First, our analysis complements the growing literature on AI in science, which has predominantly focused on scientific outputs \citep[e.g.,][]{furman2020automation,bianchini2022gmi,gao2024quantifying,hao2024ai,kusumegi2025scientific,trivsovic2025rapid,bianchini2026ai}, by shifting attention to the internal organization of knowledge production. Only a few recent contributions have begun to examine how AI is reshaping scientific workflows and the allocation of research inputs, drawing on evidence from research proposals \citep{hosseinioun2026} or specific technologies such as AlphaFold \citep{cavalli2026}. Second, we extend research on the organization of scientific work, the role of specialization, and the division of labor in research teams \citep[e.g.,][]{jones2009burden,bikard2015exploring,lariviere2015team,sauermann2017authorship,teodoridis2017,teodoridis2019creativity,haeussler2020division} by examining how the diffusion of a general-purpose cognitive technology coincides with changes in the allocation of scientific activities across collaborators. Third, we contribute to the literature on technological change and organizational adaptation \citep[e.g.,][]{david1990dynamo,milgrom1990economics,brynjolfsson2000beyond,bresnahan2002information,brynjolfsson2021productivity,brynjolfsson2025generative} by showing that the organizational implications of general-purpose technologies extend far beyond firms and conventional production settings to scientific research.

All things considered, the evidence provided here has implications for the future organization of scientific teams in the age of intelligent machines---themes on which we elaborate at the close of the paper.

\section{Conceptual framework}

\subsection{Technology, scientific work, and the division of labor}
\label{sec:technology_dol}

The interaction between technological change and the organization of work is one of the most enduring debates in economics and economic history. At the heart of this debate is how the introduction of new technologies redefines the nature of work, reshaping the division of labor, altering organizational boundaries, and shifting the distribution of economic value (see, among others, \citealp{orlikowski,brynjolfsson2014second, autor2015there, mokyr2015history, susskind2020world, acemoglu_simon_2023}).

Through what mechanisms does technology reshape the division of labor? A useful starting point is the distinction between two ways in which technology influences human work: automation and augmentation. Put simply, automation reduces---and in extreme cases eliminates---the need for human intervention by substituting machine capabilities for human labor in specific tasks. Augmentation expands human capabilities by lowering barriers to performing particular activities, raising productivity, or enabling workers to undertake tasks that would otherwise require capabilities they do not possess. Most general-purpose technologies combine elements of both, and the balance between these forces ultimately determines how work is reorganized and which human capabilities become more or less valuable.

History provides many examples of this process. Major technological revolutions have seldom eliminated human work altogether; instead---and central to our study---they have transformed its composition. Mechanization during the First Industrial Revolution automated many physical tasks while increasing demand for workers whose skills complemented the new machinery \citep{mokyr1992lever}. Electrification and assembly-line production followed a similar course, automating routine physical activities while shifting labor toward maintenance, supervision, and coordination \citep{chandler1977visible, goldin1996origins}. And, more recently, information and communication technologies (ICTs) automated routine cognitive and manual tasks while complementing workers engaged in non-routine analytical and interpersonal activities \citep{autor2003skill, FreyOsborne2017}.


Artificial intelligence represents the latest stage in this historical arc, extending both automation and augmentation further into the domain of cognitive work. Earlier AI/ML applications, which diffused widely from the 2010s onward, expanded the reach of computerization to non-routine cognitive activities (e.g., classification, prediction, and pattern recognition), but their scope remained fundamentally limited to well-defined, domain-specific tasks \citep{agrawal2018prediction, agrawal2019exploring,ide2025artificial}. LLMs changed everything. They represent an important departure from earlier generations of AI because they seem to relax the traditional trade-off between generality, accuracy, and simplicity (GAS) in model design: they combine unusually broad applicability with strong performance and relatively low barriers to use via natural-language interfaces \citep{hasan_oettl_samila}. This combination of breadth and accessibility is organizationally important because it allows individual workers to draw on the same underlying technology across multiple stages of a production process.

In economic terms, LLMs reduce the cost of general-purpose cognition in much the same way that steam power reduced the cost of mechanical work or semiconductors reduced the cost of computation \citep{hasan_oettl_samila, jones2026ai}. When the cost of an input falls so dramatically, organizations face strong incentives to rethink their internal division of labor, and early empirical evidence suggests that this process is already underway. Most studies conceptualize occupations as bundles of heterogeneous activities that differ in their susceptibility to automation and augmentation. Using occupational task descriptions, for example, \citet{eloundou_gpts} estimate that around 80\% of the U.S. workforce has at least 10\% of its work tasks exposed to LLMs, while approximately one-fifth of workers could see at least half of their tasks affected. Experimental evidence further shows that access to LLMs (in this case, ChatGPT) can significantly increase worker productivity in professional writing tasks, while also improving the quality of the output \citep{noy_zhang}. Comparable task-level effects and productivity gains from GenAI more broadly have been documented across several knowledge-intensive occupations, including customer support \citep{brynjolfsson2025generative}, software development \citep{peng2023impact,cui2026effects}, consulting \citep{dell2026navigating}, and product innovation \citep{dell2026cybernetic}.

Here, we extend this task-based logic to the organization of scientific work. Scientific research consists of multiple interdependent tasks, most of which are cognitive and require distinct forms of expertise---ranging from idea generation and experimental design to data collection, coding, writing, and team coordination. Scientific work, however, differs from many of the individual tasks studied in the occupational literature in one important respect: modern knowledge production is overwhelmingly a collective endeavor. As the breadth and depth of expertise required to conduct research increasingly exceed the capacity of any single individual, modern science has become overwhelmingly organized around collaborative teams \citep{wuchty2007increasing, jones2009burden, leahey2016sole, fortunato2018science}. Indeed, scientific teams allow researchers to combine and coordinate specialized expertise through an internal division of labor \citep{teodoridis2017, haeussler2020division, jones2021}. This shifts the relevant question from whether LLMs can perform scientific tasks to whether and how they change \textit{who} within a team performs them. As we elaborate below, LLMs do more than make individual researchers more productive. In particular, they expand the range of tasks that researchers can perform autonomously, reduce their reliance on complementary inputs from other team members, and make connected research activities easier to allocate across individuals. We contend that these changes may ultimately alter how scientific work is organized within teams.

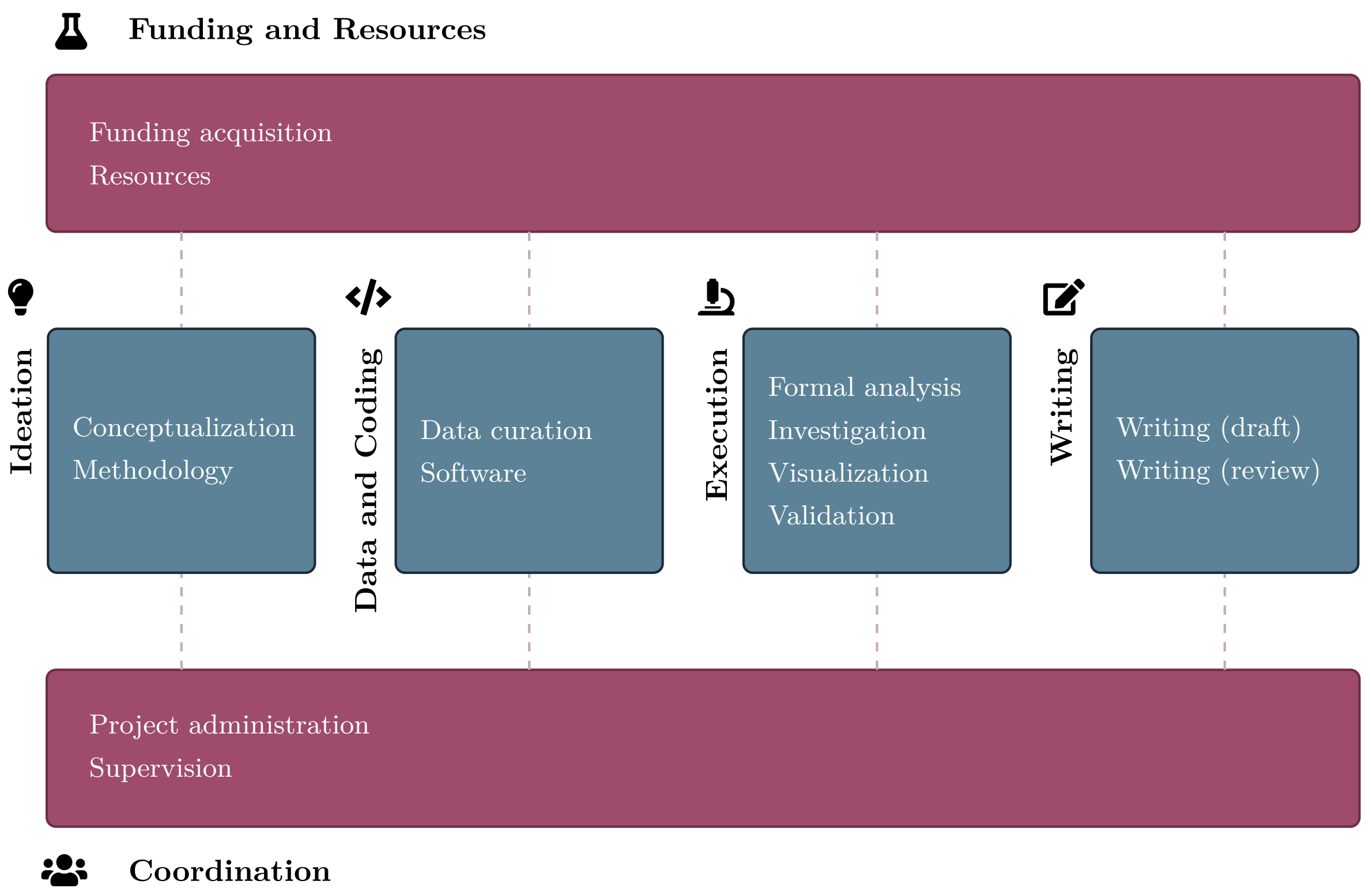
\begin{figure}[t!]
\centering

\definecolor{blockfill}{HTML}{5B8296}
\definecolor{blockdraw}{HTML}{1E293B}
\definecolor{topbottomfill}{HTML}{9E4B6C}
\definecolor{topbottomdraw}{HTML}{6E2D47}

\begin{tikzpicture}[
  font=\sffamily,
  taskbox/.style={
    draw=blockdraw, 
    fill=blockfill, 
    rounded corners=3pt, 
    thick, 
    text=white, 
    font=\footnotesize, 
    align=left,
    text width=2.5cm,
    minimum height=2.8cm,
    inner xsep=8pt, 
    inner ysep=8pt
  },
  sidelabel/.style={
    font=\small\bfseries, 
    rotate=90, 
    anchor=east, 
    align=left
  },
  iconnode/.style={
    font=\large, 
    align=center
  },
  headertitle/.style={
    font=\small\bfseries, 
    align=center
  }
]


  \node[taskbox] (box1) {
    Conceptualization\\[3pt]
    Methodology
  };
  \node[sidelabel] at ([xshift=-0.3cm, yshift=-0.1cm] box1.north west)
    (ideation_lbl) {Ideation};
  \node[iconnode] at ([xshift=-0.3cm, yshift=0.35cm] box1.north west)
    (idea_icon) {\faLightbulb};

  \node[taskbox, right=0.9cm of box1] (box2) {
    Data curation\\[3pt]
    Software
  };
  \node[sidelabel] at ([xshift=-0.3cm, yshift=-0.1cm] box2.north west)
    (data_lbl) {Data and Coding};
  \node[iconnode] at ([xshift=-0.3cm, yshift=0.35cm] box2.north west)
    (code_icon) {\faCode};

  \node[taskbox, right=0.9cm of box2] (box3) {
    Formal analysis\\[3pt]
    Investigation\\[3pt]
    Visualization\\[3pt] 
    Validation    
  };
  \node[sidelabel] at ([xshift=-0.3cm, yshift=-0.1cm] box3.north west)
    (exec_lbl) {Execution};
  \node[iconnode] at ([xshift=-0.3cm, yshift=0.35cm] box3.north west)
    (exec_icon) {\faMicroscope};

  \node[taskbox, right=0.9cm of box3] (box4) {
    Writing (draft)\\[3pt]
    Writing (review)
  };
  \node[sidelabel] at ([xshift=-0.3cm, yshift=-0.1cm] box4.north west)
    (comm_lbl) {Writing};
  \node[iconnode] at ([xshift=-0.3cm, yshift=0.35cm] box4.north west)
    (comm_icon) {\faEdit};


\draw[topbottomdraw, fill=topbottomfill, rounded corners=3pt, thick] 
  ([yshift=1.1cm] box1.north west)
  rectangle
  ([yshift=2.9cm] box4.north east);

\node[
  font=\footnotesize,
  text=white,
  align=left,
  anchor=west
] at ([xshift=10pt, yshift=2.0cm] box1.north west) {
  Funding acquisition\\[3pt]
  Resources
};

\node[headertitle, anchor=west] (infra_title)
  at ([xshift=0.8cm, yshift=3.4cm] box1.north west)
  {Funding and Resources};

\node[iconnode, left=0.2cm of infra_title.west]
  (flask_icon) {\faFlask};

\draw[dashed, topbottomdraw!40, thick]
  ([yshift=1.1cm] box1.north) -- (box1.north);
\draw[dashed, topbottomdraw!40, thick]
  ([yshift=1.1cm] box2.north) -- (box2.north);
\draw[dashed, topbottomdraw!40, thick]
  ([yshift=1.1cm] box3.north) -- (box3.north);
\draw[dashed, topbottomdraw!40, thick]
  ([yshift=1.1cm] box4.north) -- (box4.north);


  \draw[topbottomdraw, fill=topbottomfill, rounded corners=3pt, thick] 
    ([yshift=-1.1cm] box1.south west)
    rectangle
    ([yshift=-2.9cm] box4.south east);

\node[
  font=\footnotesize,
  text=white,
  align=left,
  anchor=west
] at ([xshift=10pt, yshift=-2.0cm] box1.south west) {
  Project administration\\[3pt]
  Supervision
};

  \node[headertitle, anchor=west] (coord_lbl)
    at ([xshift=0.8cm, yshift=-3.4cm] box1.south west)
    {Coordination};
  \node[iconnode, left=0.2cm of coord_lbl.west]
    (coord_icon) {\faUsers};

  \draw[dashed, topbottomdraw!40, thick]
    ([yshift=-1.1cm] box1.south) -- (box1.south);
  \draw[dashed, topbottomdraw!40, thick]
    ([yshift=-1.1cm] box2.south) -- (box2.south);
  \draw[dashed, topbottomdraw!40, thick]
    ([yshift=-1.1cm] box3.south) -- (box3.south);
  \draw[dashed, topbottomdraw!40, thick]
    ([yshift=-1.1cm] box4.south) -- (box4.south);

\end{tikzpicture}

\vspace{0.5em}

\caption{\textbf{A stylized representation of the scientific pipeline.} The figure presents the conceptual representation of the scientific research process used to analyze the effects of LLMs on the division of labor within scientific teams. \textit{Ideation}, \textit{Data and Coding}, \textit{Execution}, and \textit{Writing} follow the temporal evolution of a research project, while \textit{Funding and Resources} and \textit{Coordination} are cross-cutting functions that support all activities. Each domain comprises multiple research tasks, shown within the boxes and corresponding to CRediT roles, as discussed in detail in Section~\ref{sec:sample}. Although we present the process as linear for expositional clarity, scientific research is of course iterative, with frequent feedback loops and backtracking. }

\label{fig:scientific_pipeline}
\end{figure}

With this in mind, we frame the scientific process as comprising six stages, which we conceptualize as task domains.\footnote{Structuring our conceptual framework and aggregating our data around six discrete but interrelated activities or task domains---similar to the approach in \citet{agrawal_mchale_oettl} and \cite{hosseinioun2026}, and the navigating/doing/communicating taxonomy in \citet{binz2025should}---balances analytical parsimony with sufficient operational granularity. Each domain encompasses multiple distinct sub-activities; in Section~\ref{sec:sample}, we detail our classification methodology and discuss how we group 14 sub-activities (CRediT roles) into these six task domains. For expositional convenience and following standard convention in this literature, we refer to these domains as scientific activities or tasks throughout the text.} Four of these---\textit{Ideation}, \textit{Data and Coding}, \textit{Execution}, and \textit{Writing}---capture the temporal progression of a general research project. The remaining two---\textit{Funding and Resources} and \textit{Coordination}---represent transversal task categories that support and shape the research process throughout. Figure \ref{fig:scientific_pipeline} sketches this idealized scientific workflow that progresses from initial conceptualization to a finalized research output (such as a working paper or published manuscript). How LLMs alter the allocation of human labor across these activities, and what determines whether teams reorganize their internal structure in response, form the backbone of our theoretical framework.

\subsection{How do LLMs alter participation in scientific tasks?}
\label{sec:theory_task_opening_closing}

We take as our point of departure the conceptual framework developed by \citep{agrawal_mchale_oettl, agrawal2026enhancing}. In this framework, the effect of AI on who performs a scientific task depends on how improvements in AI capabilities interact with the human skills and judgment required to perform it. Because these requirements vary across activities, the effects of AI on participation need not be uniform. In some tasks, AI lowers barriers to entry by reducing the need for narrow technical skills, allowing more researchers to perform activities previously restricted to specialists---a process Agrawal et al. refer to as \textit{task opening}. They illustrate this possibility with empirical work in the social sciences, where AI tools can make data analysis accessible to researchers without highly specialized technical training. In other tasks, AI remains complementary to scarce forms of domain expertise and human judgment. Improvements in AI may then disproportionately benefit scientists who possess these complementary capabilities, concentrating the task among a narrower set of researchers---a process of \textit{task closing}, as seen in theoretical physics and advanced mathematical modeling, where AI can assist in exploring candidate solutions but deep specialized expertise remains essential.


This framework provides valuable intuition for understanding how AI changes who is able to perform different scientific tasks. Yet two features of our setting call for extending this logic. First, we consider LLMs specifically, which, as discussed above, are distinctive from other AI technologies in the breadth of tasks to which they can be applied. Second, and more importantly for our purposes, we shift the focus from who is able to perform a task to how work is allocated within scientific teams. Indeed, scientific research is not simply a collection of isolated tasks performed by stand-alone researchers, but a collective endeavor in which specialized capabilities are distributed across individuals and must be combined and coordinated. Once we move from individual capabilities to team organization, LLMs do not simply ``open'' or ``close'' specific tasks; they may also alter the marginal value of involving additional human contributors in performing them. Let us elaborate on this point.

Our framework starts from a simple premise: LLMs provide individual researchers with an additional, general-purpose layer of capabilities, $C^{LLM}$. We acknowledge that LLMs perform some tasks more effectively than others; nevertheless, we assume that they can provide at least some meaningful support across the full range of scientific activities. The extent of this support may also vary across researchers, depending on their existing expertise and experience. Our premise requires only that, in the short run, access to an LLM weakly expands the set of capabilities available to the individual. The gain may be large, small, or negligible, but access to an LLM does not reduce the capabilities the researcher already possesses (formally, $C^{LLM} \geq 0)$.\footnote{This assumption is intended to apply in the short term. In fact, one might argue that sustained reliance on LLMs could gradually erode some scientific skills through reduced practice or learning. While such effects remain uncertain (see, e.g., the open questions raised by \citealp{birhane2023science}), they would operate over a longer horizon than the immediate capability effects considered in our framework.} Importantly, LLMs give individual researchers access to capabilities that would otherwise be distributed across different team members. Scientific tasks frequently require combining the specialized and complementary inputs of multiple researchers---a simple example being a domain expert working alongside a programmer. By partially internalizing such complementarities within the individual, LLMs reduce reliance on inputs from other team members, echoing previous evidence that reductions in the cost of research technologies substitute for the involvement of specialized researchers in scientific teams \citep{teodoridis2017}.  

The distinction between who \textit{can} perform a task and how many people are actually \textit{needed} to perform it is central to our framework. Consider the process of ideation. Developing and refining a research idea may require combining complementary capabilities, often distributed across team members---for instance, one researcher brings deep domain knowledge, another methodological expertise, and another familiarity with adjacent literatures. LLMs can (partially) bundle these inputs within a single researcher by helping them explore unfamiliar literatures, generate and refine candidate ideas, formulate hypotheses, and assess alternative methodological approaches \citep{zhang2025exploring, Gottweis_2026}. The same logic applies to, say, writing. Drafting and revising a paper draws on different inputs, from substantive knowledge and argumentation to language and editorial skills. LLMs provide at least part of these inputs directly, allowing individual researchers to carry out a larger share of the writing process on their own \citep{khalifa2024using, rueda2025cautiously}.\footnote{We are not claiming that using AI for conceptualization, writing, or any other research activity necessarily produces better research---and, indeed, some evidence suggests that this may not always be the case \citep[see, e.g.,][]{conroy2023chatgpt, messeri2024artificial, ding2025generative,tang2025risks}. Our claim is much simpler. LLMs allow researchers to perform a broader range of activities with less reliance on complementary inputs from other team members.} 

The organizational implication of the argument so far is seemingly paradoxical: more researchers become capable of performing a task individually, but fewer researchers are needed to perform it jointly. This prediction is fully compatible with task opening and task closing \textit{\`a la} \citet{agrawal_mchale_oettl} but operates along a different margin. Opening and closing concern the pool of researchers capable of performing a task; we instead consider how many contributors a team needs to involve in performing that task. The two margins need not move in the same direction. A task may open without giving teams a reason to involve more researchers simultaneously; and a task may close while those who continue to perform it become less reliant on complementary inputs from their teammates. To the extent that LLMs increase individual autonomy, they reduce the marginal value of additional contributors.

To formalize this intuition, let $Y_k$ denote the output generated in task \(k\):
\[
Y_k = f(n_k, C^{LLM}),
\]
where \(n_k\) denotes the number of team members contributing to the task, and $C^{LLM}$ the capabilities provided by LLMs to individual researchers. If LLMs partially substitute for complementary inputs otherwise provided by team members, then \\
\[
\frac{\partial^2 Y_k}{\partial n_k \partial C^{LLM}} < 0.
\]
In other words, as the capabilities provided by LLMs increase, the marginal contribution of involving an additional team member declines. This does not imply that human collaboration becomes unnecessary, nor that smaller teams necessarily produce higher-quality research. It simply means that a task that previously drew on the capabilities of several researchers may increasingly be performed by a smaller subset of the team.\footnote{And this does not imply that teams themselves become smaller. LLMs may allow teams to sustain, or even expand, their overall size while distributing their members more selectively across tasks. In our data, however, teams formed after 2022 are, on average, slightly smaller\phantomsection\label{back:size_evolution} (Figure~\ref{fig:size_evolution}). Although this pattern cannot be directly attributed to LLM diffusion, it is consistent with the possibility that greater individual autonomy affects also the number of collaborators needed to undertake a research project.}

Coordination costs provide an additional consideration. Involving additional team members is far from costless. Previous studies show that collaboration and the division of labor require communication, deliberation, reconciliation of competing inputs, and coordination of decisions \citep{becker1992,katz1997research,bikard2015exploring}. Let \(\Gamma_k(n_k, C^{LLM})\) denote the coordination costs associated with involving $n_k$ contributors in task $k$. Teams choose participation by balancing task output against these coordination costs:
\[
n_k^* = \arg\max_{n_k}
\left\{
f(n_k,C^{LLM}) - \Gamma_k(n_k,C^{LLM})
\right\}.
\]

If LLMs reduce the marginal benefit of involving an additional human contributor, and any reduction in coordination costs does not fully offset this effect, the optimal number of contributors \(n_k^*\) tends to fall.\footnote{LLMs may also reduce some coordination costs, for instance by facilitating communication among contributors. This would work in the opposite direction, making additional participation less costly. Our prediction requires only that this effect does not fully offset the decline in the marginal benefit of involving additional contributors. The within-task coordination costs considered here are distinct from the cross-task handoff costs discussed in the next subsection. The former arise when multiple researchers jointly perform the same activity; the latter arise when different but connected activities are performed by different researchers. As we argue below, LLMs may substantially reduce these handoff costs even while within-task coordination costs remain relevant.} 

This leads to our first hypothesis:

\medskip

\begin{tcolorbox}[colback=gray!5, colframe=gray!25, boxrule=0.3pt, arc=1mm]
\begin{adjustwidth}{0.5cm}{0.5cm}
\noindent \emph{\textbf{H1 (Within-task participation).} Following the diffusion of LLMs, scientific tasks involve fewer contributors, both in absolute terms and as a share of team members.}
\end{adjustwidth}
\end{tcolorbox}

The strength of this mechanism need not be uniform across scientific activities, although we deliberately refrain from formulating task-specific hypotheses. The extent to which LLMs reduce reliance on other team members depends on both their effectiveness in supporting a given activity and the nature of the complementary human inputs that activity requires. Where LLMs can effectively substitute for inputs that would otherwise be provided by collaborators, individual autonomy increases more strongly, leading to greater compression in within-task participation. On the opposite, where activities continue to require substantial human expertise, judgment, or coordination that LLMs cannot readily replicate, the scope for participation compression remains more limited. In all, we anticipate stronger effects in execution-heavy activities such as data analysis, coding, and writing, where LLMs can assist with analytical tasks, developing code, and drafting or revising manuscripts. Effects on ideation may be more moderate, while activities that depend on access to physical infrastructure, laboratory facilities, or institutional responsibilities---such as funding acquisition and resource provision---should be less directly affected by LLM capabilities.

These expectations find some support in recent evidence on ChatGPT adoption across the scientific workflow \citep{openai2026chatgpt}. Between June and July 2026, scientific research requests were concentrated in \textit{reading and knowledge acquisition}, which accounted for more than 35\% of ChatGPT requests, and \textit{writing and communication}, which accounted for more than 30\%. Meanwhile, \textit{software and code development} represented nearly 30\% of Codex usage. \textit{Ideation and hypothesis generation} accounted for a more modest 7\% of ChatGPT requests, while \textit{research administration} represented less than 2\%. Complementary evidence from a sample of 15 million Gemini interactions (Gemini App and AI Mode) also confirms high usage in data analysis and modeling (42\% of interactions) and research communication (14\%) \citep{codreanu2026ai}. Although these figures reflect patterns of AI use rather than changes in the division of labor, they illustrate the uneven penetration of LLMs across scientific activities, potentially creating varying degrees of pressure for organizational adaptation.

\subsection{Task separability and specialization}
\label{sec:portfolio_specialisation}

So far, we have considered how LLMs affect participation \textit{within} individual research tasks. However, a reduction in within-task participation does not necessarily imply a more specialized division of labor. Fewer researchers might participate in each task while the same individuals continue to span multiple stages of the project. Indeed, the division of labor in scientific teams involves not only how many researchers participate in individual activities, but also how activities are distributed and combined across researchers \citep{haeussler2020division}. To understand whether LLMs also reshape the boundaries between individual roles, we need to consider how they affect connections \textit{across} tasks.

We propose that LLMs make scientific tasks more ``separable'', and that they do so through two related mechanisms. First, they reduce the need for specialized expertise to travel with the researcher who possesses it. Researchers often participate in multiple stages of a project not because they are directly responsible for each of them, but because a specific piece of their expertise is required there. For instance, a researcher who conceptualizes a study may remain marginally involved in its empirical execution solely to clarify some methodological choices or assumptions, while a researcher conducting the empirical analysis may also contribute to writing only to interpret particular results. Yet, building on the mechanism described above, LLMs can provide other team members with some of these complementary capabilities, reducing the need for researchers to remain involved in additional activities merely to provide specialized input.

Second, LLMs lower cross-task handoff costs, understood as the time and effort required to transfer knowledge, information, and intermediate outputs between researchers performing different activities. Even when responsibilities are clearly divided across researchers, connected tasks often require small but important pieces of knowledge to be transferred across task boundaries. For instance, a researcher taking over an empirical analysis may need to know why a particular variable was chosen, how a construct was operationalized, or what assumption motivated a specific modeling decision, while a writer may need to understand the logic behind a piece of code or the interpretation of an unexpected result. Such dependencies can make otherwise distinct tasks costly to separate because they require communication, explanation, and documentation between the researchers involved \citep{cummings2007coordination, walsh2007collaboration}. LLMs can facilitate these handoffs. We have already seen that they excel at explaining and documenting code, summarizing analytical choices, generating descriptions of procedures and results, and translating technical outputs into text that another team member can more easily use \citep{nam2024using,binz2025should,luo2025llm4sr}. By doing so, LLMs reduce the friction associated with transferring work across task boundaries, making it easier for connected activities to be performed by different individuals.

Combining these two mechanisms increases what we call \textit{task separability}. Activities that previously required overlapping participation---either because a researcher needed to provide a small but specialized input outside their main responsibility, or because knowledge had to be carried across stages of the project---can be allocated more cleanly across individuals. Greater task separability, in turn, creates more scope for specialization. As the need to maintain these connections across tasks declines, researchers can concentrate more narrowly on the activities for which they are primarily responsible and which best match their relative strengths. Here again, LLMs produce a seemingly paradoxical organizational outcome: researchers may become more generalist in what they \textit{can} do while becoming more specialized in what they \textit{actually} do. 

We can formalize this intuition by representing researcher \(i\)'s realized task portfolio as
\[
\mathcal{T}_i = \{k : S_{ik}=1\},
\]
where \(S_{ik}\) indicates whether researcher \(i\) participates in scientific activity \(k\). If LLMs increase task separability, fewer cross-task dependencies need to be managed through overlapping participation. Thus, we expect an increase in \(C^{LLM}\) to reduce both the breadth of individual task portfolios,
\[
\frac{\partial |\mathcal{T}_i|}{\partial C^{LLM}} < 0,
\]
and the overlap between the portfolios of different team members,
\[
\frac{\partial O(\mathcal{T}_i,\mathcal{T}_j)}{\partial C^{LLM}} < 0,
\]
where \(O(\mathcal{T}_i,\mathcal{T}_j)\) measures the degree of overlap between the activities performed by researchers \(i\) and \(j\).

These two implications capture different dimensions of specialization. Individual researchers should participate in fewer distinct scientific activities, producing a contraction in the breadth of their realized task portfolios. At the same time, individual task portfolios should become more differentiated, resulting in less overlap across team members. 

\newpage
More formally, we posit the following hypothesis:

\medskip

\begin{tcolorbox}[colback=gray!5, colframe=gray!25, boxrule=0.3pt, arc=1mm]
\begin{adjustwidth}{0.5cm}{0.5cm}
\noindent \emph{\textbf{H2 (Task specialization).} Following the diffusion of LLMs, researchers become more specialized in their division of labor, participating in fewer scientific tasks (portfolio contraction) and exhibiting less task overlap with other team members (portfolio differentiation).}
\end{adjustwidth}
\end{tcolorbox}

\subsection{Organizational constraints on the division of labor}
\label{sec:constraints_team_organization}

Our framework thus far predicts that LLMs can reshape the division of labor by reducing the number of researchers involved within individual scientific activities and by allowing tasks to be allocated more distinctly across team members. But do these changes always take place? Not necessarily. A long tradition in organizational theory emphasizes that technologies do not determine organizational forms mechanically, but interact with pre-existing structures, routines, and practices \citep{orlikowski}. Thus, the same technological shock may produce very different organizational responses depending on the arrangements already in place.

To illustrate this, consider two teams with different collaboration histories. At one extreme, an \textit{established} team consisting of researchers who have already worked together as a group. Through prior collaboration, such a team may have developed an existing division of labor, together with established roles, routines, and expectations about who does what. At the other extreme is a \textit{newly formed} team bringing together researchers who have not previously worked together in that specific configuration. Such a team has no established division of labor tailored to its current configuration and must determine how responsibilities will be allocated among its members.

Collaboration history matters because the division of labor is inherently path-dependent. Through repeated collaboration, researchers develop routines, role specializations, coordination mechanisms, and shared expectations regarding individual contributions \citep{becker1992, melin, bozeman_corley2004}. These arrangements make collaboration more efficient precisely because responsibilities become familiar and predictable. Yet this very stability may generate structural inertia---i.e., the tendency to resist change and maintain the status quo, even when environmental conditions shift or new technologies make adaptation necessary \citep{hannan_freeman, march}. Reconfiguring an established division of labor requires renegotiating responsibilities, changing established workflows, learning new ways of working, and potentially revisiting implicit understandings about credit and authorship. Let $\kappa$ denote these adjustment costs. Even when LLMs change the technologically optimal allocation of scientific work, an established team will reorganize only when the gains from doing so are sufficiently large to compensate for $\kappa$.

This friction creates a wedge between changes in individual capabilities and changes in observed team organization. Researchers in established teams may use LLMs to become more productive or autonomous within their existing responsibilities without substantially changing who participates in each activity or how tasks are distributed across team members. In other words, the inherited division of labor acts as a buffer between technological change and organizational restructuring. Thus, we expect the effects predicted in H1 and H2 to be attenuated among established teams.

The situation differs for newly formed teams, which face far weaker organizational constraints. Because the team in its current configuration has not previously worked together, responsibilities must be allocated, or existing arrangements among subsets of members adapted, as the project forms. Consequently, newly formed teams assembling in the post-LLM period can incorporate the capabilities provided by these technologies directly into their initial division of labor, rather than having to retrofit an established team-level workflow. The organizational changes predicted by our framework should therefore be most visible among newly formed teams. This leads to our third and final hypothesis:

\medskip

\begin{tcolorbox}[colback=gray!5, colframe=gray!25, boxrule=0.3pt, arc=1mm]
\begin{adjustwidth}{0.5cm}{0.5cm}
\noindent \emph{\textbf{H3 (Organizational adaptation).} Following the diffusion of LLMs, the decline in within-task participation and the increase in task specialization predicted in H1 and H2 are stronger among newly formed teams than among established teams.}
\end{adjustwidth}
\end{tcolorbox}

\bigskip 

\begin{figure}[t!]
\centering


\definecolor{neutralfill}{HTML}{F1F5F9}
\definecolor{neutraldraw}{HTML}{475569}
\definecolor{openfill}{HTML}{E6F4EA}
\definecolor{opendraw}{HTML}{137333}
\definecolor{closefill}{HTML}{FEF7E0}
\definecolor{closedraw}{HTML}{B06000}
\definecolor{modfill}{HTML}{F3E8FF}
\definecolor{moddraw}{HTML}{6B21A8}
\definecolor{mechfill}{HTML}{E2E8F0}

\begin{tikzpicture}[
  node distance=0.8cm and 1.0cm,
  every node/.style={
    align=center,
    font=\small,
    inner sep=7pt
  },
  basebox/.style={
    draw=neutraldraw,
    fill=neutralfill,
    rounded corners=5pt,
    thick,
    minimum width=4.5cm,
    minimum height=1.0cm
  },
  mechbox/.style={
    draw=neutraldraw,
    fill=mechfill,
    rounded corners=5pt,
    thick,
    minimum width=4.6cm,
    minimum height=1.3cm
  },
  capbox/.style={
    draw=neutraldraw,
    fill=mechfill,
    rounded corners=5pt,
    thick,
    text width=7cm,
    minimum width=4.0cm,
    minimum height=1.5cm
  },
  h1box/.style={
    draw=opendraw,
    fill=openfill,
    rounded corners=5pt,
    thick,
    minimum width=4.8cm,
    minimum height=1.4cm
  },
  h2box/.style={
    draw=closedraw,
    fill=closefill,
    rounded corners=5pt,
    thick,
    minimum width=5.8cm,
    minimum height=1.4cm
  },
  modbox/.style={
    draw=moddraw,
    fill=modfill,
    rounded corners=5pt,
    thick,
    dashed,
    minimum width=6.5cm,
    minimum height=1.3cm
  },
  arrow/.style={
    -{Stealth[scale=1.1]},
    thick,
    color=neutraldraw
  },
  dashedarrow/.style={
    -{Stealth[scale=1.1]},
    thick,
    dashed,
    color=moddraw
  }
]


\node[basebox] (llm) at (0,0)
  {\textbf{LLM diffusion}};

\node[capbox, below=0.65cm of llm] (capabilities)
  {\textbf{Expansion of individual capabilities}\\[3pt]
   \footnotesize Greater autonomy in scientific work};



\node[mechbox] (within)
  at ($(capabilities.south)+(-5cm,-1.8cm)$)
  {\textbf{Within-task mechanism}\\[4pt]
   \footnotesize Reduced reliance on\\
   \footnotesize complementary human inputs};

\node[mechbox] (across)
  at ($(capabilities.south)+(5cm,-1.8cm)$)
  {\textbf{Cross-task mechanism}\\[4pt]
   \footnotesize Greater task separability\\
   \footnotesize Lower cross-task handoff costs};


\node[h1box, below=0.8cm of within] (h1)
  {\textbf{H1: Participation compression}\\[4pt]
   \footnotesize Fewer contributors per task\\
   \footnotesize (absolute and relative participation)};

\node[h2box, below=0.8cm of across] (h2)
  {\textbf{H2: Task specialization}\\[4pt]
   \footnotesize Portfolio contraction\\
   \footnotesize Portfolio differentiation};


\node[modbox] (moderation)
  at ($(h1.south)!0.5!(h2.south)+(0,-1.8cm)$)
  {\textbf{H3: Organizational adaptation}\\[4pt]
   \footnotesize Established teams: greater inertia\\
   \footnotesize Newly formed teams: greater flexibility};


\draw[arrow] (llm) -- (capabilities);


\draw[arrow]
  (capabilities.west)
  -| (within.north);

\draw[arrow]
  (capabilities.east)
  -| (across.north);

\draw[arrow] (within) -- (h1);
\draw[arrow] (across) -- (h2);


\draw[dashedarrow]
  (moderation.north west)
  to[out=145,in=-90] (h1.south);

\draw[dashedarrow]
  (moderation.north east)
  to[out=35,in=-90] (h2.south);

\end{tikzpicture}

\vspace{1em}

\caption{
\textbf{Overview of the conceptual framework and hypotheses.}
LLM diffusion expands the capabilities available to individual
researchers and reshapes the division of labor along
two dimensions. LLMs decrease the number of contributors
involved in individual scientific tasks (H1), and promote specialization through portfolio contraction and differentiation (H2). The organizational responses are moderated by collaboration
history (H3).}

\label{fig:mechanism}

\end{figure}
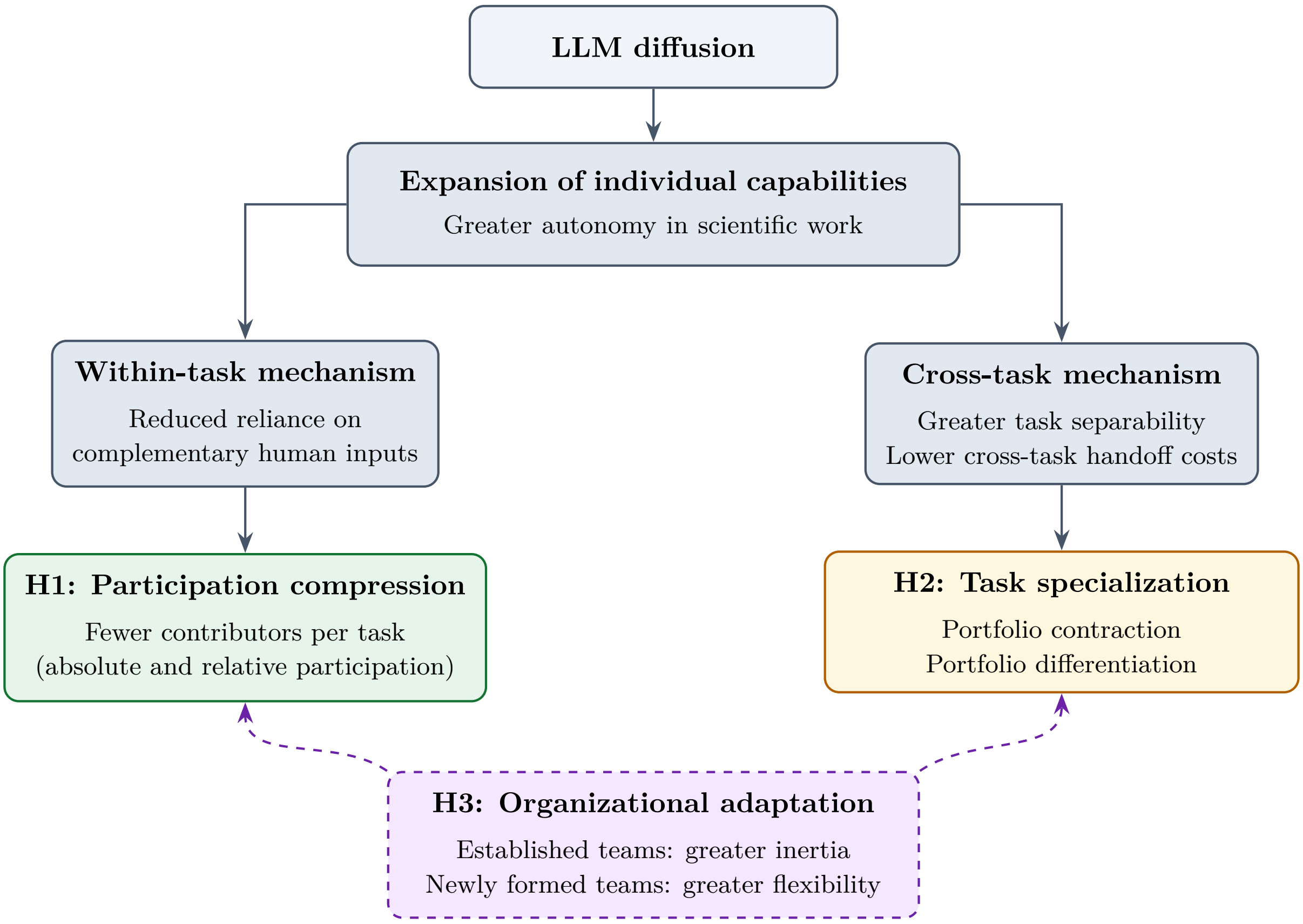

Figure~\ref{fig:mechanism} provides a schematic overview of the integrated conceptual framework mapping onto our three hypotheses. The subsequent sections turn to testing these hypotheses empirically.

\newpage
\section{Data and methods}

\subsection{Sample composition}
\label{sec:sample}

We construct our sample from research articles published in PLOS ONE between January 2018 and December 2025. PLOS ONE offers two main advantages for our study. First, it pioneered requiring authors to disclose individual team members' contributions using predefined, standardized categories via the \textit{Contributor Roles Taxonomy} (\href{https://casrai.org/credit/roles}{CRediT}) \citep{allen2014publishing, mcnutt2018transparency}. The consistent availability of these standardized contribution statements since 2018 provides a long observation window spanning the period before and after the public release of ChatGPT in late 2022. Second, PLOS ONE is a multidisciplinary journal covering a wide range of scientific fields, allowing us to examine whether changes in the division of labor differ across research domains. To our knowledge, no other journal provides a comparable combination of consistent author-level contribution data, publication volume, and disciplinary breadth over such an extended period.

We exclude single-authored papers, for which within-team task allocation is not defined, and mega-authored publications with more than 100 contributors, whose unusually large collaborations may not be comparable to those of typical research teams. This leaves an initial sample of 121,792 papers. Using their digital object identifiers (DOIs), we retrieve publication metadata and author information from the 2026 OpenAlex snapshot via BigQuery (which, at the time of download, contained complete records through February 2026). We also gather authors' institutional affiliations, publication histories, academic age, and primary historical field of research. These data allow us to characterize team composition and reconstruct prior collaboration histories. After excluding papers with incomplete author-contribution statements or missing author information, we obtain a final analytical sample of 117,211 papers, corresponding to approximately 96\% of the eligible corpus. Appendix~\ref{app:sample_composition} provides a detailed breakdown of the sample by publication year, scientific field, and number of unique authors.

For each focal paper, we extracted the CRediT contribution statement directly from the PLOS ONE website.\footnote{PLOS ONE explicitly states on its website: ``\textit{Our approach to TDM [Text and Data Mining] is simple: PLOS articles may be mined, reused, and shared by anyone, anywhere, for any purpose}''.} Following the literature on scientific collaboration and the division of labor \citep{lariviere2016contributorship,sauermann2017authorship,haeussler2020division}, we use these statements to reconstruct how scientific activities are distributed among team members. Specifically, CRediT identifies which researchers are credited with each activity, allowing us to observe individual participation across different stages of the research process. For our main analyses, we aggregate the 14 CRediT roles into the six task domains introduced earlier (Figure~\ref{fig:scientific_pipeline}) \phantomsection\label{back:def_14_tasks}(see Table~\ref{tab:credit} for detailed definitions of CRediT roles). Four domains correspond to different stages of scientific production: \textit{Ideation} includes `Conceptualization' and `Methodology'; \textit{Data and Coding} combines `Data curation' and `Software'; \textit{Execution} covers `Formal analysis', `Investigation', `Validation', and `Visualization'; and \textit{Writing} includes `Writing -- original draft' and `Writing -- review \& editing'. Two additional domains capture activities that cut across these stages: \textit{Funding and Resources} (`Funding acquisition' and `Resources') and \textit{Coordination} (`Project administration' and `Supervision').\footnote{Of course, alternative groupings are possible. We assessed the sensitivity of our main results to alternative aggregation schemes and report complementary analyses using the original 14 CRediT roles in Appendix~\ref{subsec:14tasks}.}

As with any measure based on self-reported information, CRediT contribution statements have limitations. For instance, reported contributions may reflect social negotiation processes among team members and discipline-specific authorship conventions. Moreover, CRediT indicators are dichotomous: they record whether an author contributed to an activity, but not the time or effort devoted to it. Hence, our measures capture the extensive margin of task participation rather than the intensity of individual contributions. This could be particularly relevant in our setting since LLMs may reduce the time and effort required to perform certain tasks without necessarily changing whether an author is credited with them. Despite these limitations, standardized author-contribution statements provide an exceptionally rich source of information on how scientific work is distributed within teams \citep{holcombe2019,liu2023data}.

\subsection{The advent of general-purpose LLMs}
\label{sec:advent_llm}

We exploit the public release of ChatGPT in November 2022 as a time-sharp, field-wide, and largely unexpected expansion in access to general-purpose AI capabilities. Although the current wave of AI development gained momentum with the rise of deep learning in the early 2010s and subsequently accelerated with the introduction of the Transformer architecture in 2017, earlier AI tools were primarily engineered to support narrow scientific tasks (see the GAS framework discussed in Section~\ref{sec:technology_dol}). ChatGPT revolutionized this paradigm.

The key innovation was enabling direct interaction with high-capacity models through a natural-language conversational interface \citep{sejnowski2024chatgpt}. In a typical interaction, a user submits a natural-language prompt and the system returns a response generated by an underlying LLM. The combination of intuitive, conversational interaction and unprecedented model performance democratized access to general-purpose AI overnight. The success of ChatGPT triggered an intense wave of competition among AI developers within less than a year, fueling a rapidly expanding ecosystem of foundation models---including Meta's LLaMA (February 2023), Anthropic's Claude (March 2023), and Google's Gemini (December 2023). Subsequent releases expanded model capabilities by introducing features such as real-time web search, code execution, and multimodal reasoning; however, natural-language text interaction remains, still today, the primary user interface. For most researchers, the technological discontinuity was not a gradual and incremental improvement in AI performance, but an (unexpected) abrupt collapse in the cost of accessing general-purpose cognitive capabilities. 

What is important to our empirical design is that the timing of this breakthrough was largely unanticipated by the scientific community and largely independent of the research agendas and organizational arrangements of scientific teams. Consider this: just a few years prior to ChatGPT's release, even leading AI researchers remained skeptical about achieving human-level natural language processing in the near term; Gary Marcus, for instance, argued in essay \textit{Machine Won't Think Anytime Soon} that ``\textit{[w]e're not getting appreciably closer to machines with common sense understanding or the ability to process natural language}''\citep{marcus2015machines}. Thus, we interpret ChatGPT's release as a plausibly exogenous expansion in the technological opportunities available to researchers. Furthermore, unlike previous waves of AI adoption, which diffused gradually through specialized scientific communities, ChatGPT became immediately accessible to virtually any researcher. Adoption has been nothing short of explosive. For a new technology, this speed of global diffusion has no historical precedent \citep{bick2026rapid}. To give a sense of the scale involved, by July 2025, ChatGPT had reached more than 700 million weekly users (around 10\% of the world's adult population), who collectively sent more than 2.5 billion messages per day, or about 29,000 messages per second \citep{chatterji2025people}. Similar adoption trajectories have been documented for competing LLMs---e.g., Claude \citep{handa2025economic} and Gemini \citep{iscenko2026gemini} which reached hundreds of millions of monthly users within months of release.  

Scientific research was no exception to this trend. In a global survey of more than 2,400 researchers, 84\% reported being active users of AI tools for almost any aspect of their work, with 80\% specifically relying on general-purpose chatbots \citep{wiley2026explanaitions}. The same survey found that writing and coding assistants are the most frequent routine use among scientists, whereas adoption is somewhat more limited for activities such as developing novel methods or generating hypotheses. Similar patterns of widespread uptake have been documented across other publisher surveys, empirical studies, and field-specific audits \citep{van2023ai,elsevier2025confidence,chugunova2026uses,gartenberg2026more}, as well as in analyses of LLM interaction data \citep{codreanu2026ai,openai2026chatgpt}.\footnote{As a thought experiment, let us assume that we can treat the Wiley survey as a representative sample of the scientific population ($P(\text{LLM}) = 0.80$). This means that if you are reading this paper right now, there is an unconditional baseline probability of 80\% that you already use LLMs in your own research. If we account for a natural positive selection bias---since you are reading an empirical paper on scientific automation, you are likely more ``tech-inclined'' than average---that conditional probability jumps to 88\%--92\% (calculated via Bayes' Rule, $P(\text{LLM} \mid \text{Reader}) = \frac{P(\text{Reader} \mid \text{LLM}) \cdot P(\text{LLM})}{P(\text{Reader})}$, assuming readers have a $10\%$--$15\%$ higher relative likelihood of being LLM users than non-users). And if you happen to be reviewing our manuscript on a standard three-person panel ($k = 3$), the joint probability that at least one of you routinely uses LLMs is 99.2\% ($1 - (1 - 0.80)^3 = 1 - 0.20^3 = 0.992$). In short, the very tools we are studying have, in all likelihood, already been adopted by the audience reading and evaluating this work.} Such widespread adoption provides strong rationale for assuming near-universal treatment exposure in our empirical strategy---i.e., every research team operating after November 2022 functions in an environment where LLMs form part of the baseline technological choice set. Consequently, our empirical framework adopts an \textit{Intent-to-Treat} (ITT) perspective, capturing changes in the organization of scientific work following the widespread availability of LLMs rather than their actual adoption by individual researchers. To the extent that some teams or fields adopt LLMs more slowly or conservatively than others, the estimated organizational changes may be attenuated relative to the effects experienced by actual adopters.\footnote{An ideal experiment might directly observe individual-level LLM usage linked to identifiable researchers. Such data are, of course, unavailable at scale.  Inferring who uses LLMs and for which activities is also challenging (if not impossible), as self-reported disclosures in published manuscripts remain subject to social desirability bias, stigma, and evolving editorial mandates \citep{binz2025should}. We attempted to infer adoption indirectly from text features using two AI-detection algorithms \citep{liang2024,kobak2025delving}, but these approaches revealed severe measurement error and false-positive rates (e.g., classifying manuscripts published prior to the launch of LLMs as AI-assisted). More fundamentally, text-based detection methods are inherently limited for our goals because they only capture LLM usage in final manuscript drafting while remaining completely silent on back-end scientific tasks, such as coding, data analysis, or brainstorming. Our approach avoids the need to measure individual adoption and the associated selection problems. However, because ChatGPT's release affected the entire sample, the absence of an untreated comparison group prevents us from fully disentangling the effects of LLM diffusion from those of other contemporaneous changes. Nevertheless, we are not aware of any alternative development that could plausibly account for the magnitude, timing, and heterogeneity of the observed changes, which are so consistent with the organizational mechanisms proposed in our framework.}

A potential objection to treating late 2022 as the structural arrival of LLMs is that earlier foundation models were already available, most notably GPT-3 (released in 2020) (see an \href{https://llm-timeline.com/}{exhaustive LLM release timeline}). Yet, earlier models were far from functional substitutes for general-purpose research assistants for several reasons. Considering GPT-3, which represented the state-of-the-art prior to late 2022: first, accessing it required applying for OpenAI API credentials, managing API billing accounts, and writing custom code to execute API calls; in short, high fixed setup costs restricted adoption to a narrow set of tech-savvy early adopters. Second, as a raw base completion model, GPT-3 lacked instruction-tuning, causing it to frequently hallucinate or ramble without complex prompt engineering. Third, it lacked the user-friendly conversational interface that later enabled real-time, iterative task execution. As \citet{bommasani2021opportunities} succinctly put it: ``\textit{Despite their deployment into the real world, these models are very much research prototypes}'' (p.9).

With these identification arguments in place, the remainder of this section outlines our formal empirical specification.

\subsection{Empirical setting}
\label{sec:empirical_setting}

Our empirical strategy relies primarily on dynamic event-study specifications that trace the evolution of task allocation before and after the widespread availability of general-purpose LLMs in late 2022. This approach allows us to examine (potential) pre-existing trends and assess the timing and persistence of organizational changes following the technological shock. We then investigate whether these patterns differ according to teams' organizational histories, comparing newly formed and established research teams. To complement the dynamic event-study analysis, we also report static pooled Ordinary Least Squares (OLS) specifications that summarize average post-2022 changes (\hypertarget{back_pooled_interaction}{%
Appendix~\ref{sec:appendix_pooled_interaction}%
}).

The analysis operates at three levels of observation: the task--paper level ($kpt$), the contributor--paper level ($ipt$), and the paper level ($pt$). We construct two families of dependent variables corresponding to the dimensions of labor allocation identified in our conceptual framework:

\begin{enumerate}
    \item \textit{Within-task participation ($P_{kpt}$)}: We measure participation separately for each of the six task domains, using two complementary indicators: (i) the share of team members credited with performing task $k$ on paper $p$, and (ii) the absolute headcount of contributors credited with that task. These indicators allow us to evaluate whether LLM diffusion is associated with a contraction in the proportion and number of researchers involved in individual scientific tasks, as predicted by H1---see Section~\ref{sec:theory_task_opening_closing}.
    \item \textit{Task specialization}: We measure specialization along two dimensions. At the contributor–paper level ($S_{ipt}$), we measure the breadth of individual task portfolios as the number of distinct research activities performed by contributor $i$ on paper $p$. At the paper level ($S_{pt}$), we measure the overlap between contributors' task portfolios using the mean pairwise Jaccard similarity across all pairs of team members. These indicators allow us to evaluate whether LLM diffusion is associated with a contraction in individual task portfolios and a greater differentiation of responsibilities across team members, as predicted by H2---see Section~\ref{sec:portfolio_specialisation}.
\end{enumerate}

\subsubsection{Team classification}
\label{subsec: teamclassification}

An important question arising from our conceptual framework is whether the organizational effects of LLMs depend on teams' prior collaboration experience, as predicted by H3---see Section~\ref{sec:constraints_team_organization}. To investigate this, we distinguish between established and newly formed research teams based on their co-authorship histories.\footnote{Throughout our analysis, we operationalize a research ``team'' as the set of co-authors listed on a focal paper. It is well-known that co-author rosters may omit non-author contributors who provided technical or administrative support, and they do not capture internal hierarchies or time spent on the project \citep{katz1997research}. Although surveys or administrative records could provide richer information, co-authorship histories offer a standardized and scalable measure of collaboration across the scientific literature.} \\

For each focal paper $p$, we use OpenAlex data to reconstruct the publication histories of all listed co-authors, restricting the information set to publications appearing prior to the focal publication year $t$. Based on these records, we classify teams into two categories:

\begin{enumerate}
    \item \textit{Established teams}: Teams composed of the exact same set of co-authors observed on at least one prior publication.
    \item \textit{Newly formed teams}: Teams comprising a co-author configuration not previously observed together on a single publication, irrespective of any pairwise or sub-group prior collaborations.
\end{enumerate}

To illustrate, consider a focal paper authored by $A$, $B$, and $C$. The team is classified as established only if these three researchers have previously published together, at least once, as the complete co-author team. Otherwise, it is classified as newly formed. This includes cases in which $A$ and $B$ previously collaborated without $C$, cases in which $A$, $B$, and $C$ previously published together with an additional co-author $D$, or cases in which none of the three researchers had previously collaborated with one another. For simplicity, we retain these two categories (established vs. newly formed teams) in the main analysis. In Appendix~\ref{sec: decomposing_new_teams}, we further decompose newly formed teams according to the extent of prior collaboration among their members, distinguishing entirely new collaborations from configurations involving previously collaborating subsets.

Finally, our analysis does not follow the same teams longitudinally, as only a very small fraction of (established) teams can be observed in both periods.\footnote{A within-team specification tracking the same complete co-author configuration before and after LLM diffusion would account for time-invariant team characteristics. However, such an analysis would necessarily be restricted to teams with a history of collaboration, excluding newly formed teams by construction. Moreover, restricting the sample to teams that published in PLOS ONE in both periods would substantially reduce the number of observations (only 213 teams on 466 papers), leaving insufficient statistical power for meaningful analysis.}  Instead, we classify each focal team according to its prior collaboration history and compare the organization of established and newly formed teams across the pre- and post-LLM periods. Although this approach cannot identify within-team changes, it allows us to test whether prior collaboration experience moderates the organizational response to LLM diffusion. \\


\subsubsection{Econometric specification}

More formally, we study the evolution of task reallocation using the following flexible event-study specification:

\begin{equation}
\Omega_{pt} = \alpha + \sum_{\tau \in \mathcal{T} \setminus \{2022\}} \beta_{\tau} \cdot \mathbf{1}(t = \tau) + \mathbf{X}_{pt}'\boldsymbol{\theta} + \gamma_f + \varepsilon_{pt}
\label{eq:event_study}
\end{equation}

\noindent where, for notational simplicity, $\Omega_{pt}$ denotes our outcomes of interest ($\Omega_{pt} \in \{P_{kpt}, S_{ipt}, S_{pt}\}$), evaluated at their respective levels of observation. For within-task participation ($P_{kpt}$), Equation~(\ref{eq:event_study}) is estimated separately for each of the six research activities, $k \in \{1, \dots, 6\}$. $\mathbf{1}(t = \tau)$ equals one if paper $p$ was published in calendar year $\tau$. We use 2022 as the reference year, corresponding to the public release of ChatGPT in November 2022 and the beginning of the widespread diffusion of general-purpose LLMs. Hence, each coefficient $\beta_{\tau}$ measures the average difference in the corresponding outcome variable between year $\tau$ and 2022, conditional on observable team characteristics and field fixed effects. We estimate Equation~(\ref{eq:event_study}) both on the full sample and separately across sub-samples of established and newly formed teams.

The vector $\mathbf{X}_{pt}$ includes observable team characteristics that may influence task allocation, namely: team size ($\textit{Team Size}$, measured as the total number of co-authors); intensity of international collaborations ($\textit{Int Collab}$, measured as the number of distinct countries represented in co-authors' affiliations); prior publication experience ($\textit{Prior Pubs}$, measured as the average number of publications accumulated by team members before year $t$); and academic age ($\textit{Age}$, measured as the average number of years since team members' first publication). We also include 26 scientific field fixed effects ($\gamma_f$) to absorb time-invariant heterogeneity in task allocation across disciplines. A complete definition of all main variables is reported in\phantomsection\label{back:variables_table} Table~\ref{tab:variables}.

The event-study serves two main purposes. First, it allows us to characterize pre-LLM trajectories and assess whether subsequent changes depart from earlier patterns. This distinction matters because a decline in task participation or portfolio breadth may emerge from a stable baseline, accelerate an existing decline, or interrupt or even reverse an earlier expansion. Second, the dynamic specification allows us to understand when organizational adjustments materialize and whether they persist or intensify over time. To complement the analysis, we always report joint Wald tests assessing whether the pre-period coefficients are jointly equal to zero relative to the omitted reference year (2022), and formally test whether post-2022 estimates deviate from a linear trend extrapolated from 2018--2021.

Our framework predicts reductions in within-task participation (H1) and greater task specialization (H2). Accordingly, negative post-2022 estimates indicate lower participation, narrower individual task portfolios, or less overlap across co-authors relative to the reference year. Downward deviations from extrapolated pre-LLM trends indicate that these outcomes fall below the levels implied by their earlier trajectories, while upward deviations indicate that they exceed those levels. The former are consistent with our theoretical predictions, although we reiterate that their temporal coincidence with LLM diffusion does not, by itself, establish causality.

\subsection{Descriptive statistics}
\label{sec:descriptive_statistics}

Table~\ref{tab:summary_stats} provides an overview of the outcome variables and team-level characteristics, presented separately for the pre-LLM baseline period (2018--2019) and the post-LLM diffusion period (2024--2025). We deliberately focus on these two windows to provide a clearer descriptive comparison, avoiding the disruptions to scientific collaboration associated with the COVID-19 pandemic (2020--2021) and the initial phase of LLM diffusion (2022--2023). Summary statistics for the full sample (2018--2025) are reported in\phantomsection\label{back:sample_composition} Appendix~\ref{app:sample_composition}.

\begin{table}[htbp]
\centering
\small
\setlength{\tabcolsep}{4.5pt}
\renewcommand{\arraystretch}{1.25}
\begin{tabular}{l cc cc cc}
\toprule
 & \multicolumn{2}{c}{\textbf{Full Sample}}
 & \multicolumn{2}{c}{\textbf{Established Teams}}
 & \multicolumn{2}{c}{\textbf{Newly Formed Teams}} \\
\cmidrule(lr){2-3} \cmidrule(lr){4-5} \cmidrule(lr){6-7}

& \textbf{Mean (SD)} & \textbf{Share}
& \textbf{Mean (SD)} & \textbf{Share}
& \textbf{Mean (SD)} & \textbf{Share} \\
\midrule
\multicolumn{7}{l}{\textbf{Panel A: Pre-LLM (2018--2019)}} \\
\addlinespace[0.25em]
\multicolumn{7}{l}{\textit{Task Participation Outcomes ($P_{kpt}$)}} \\
\addlinespace[0.25em]
\quad Ideation & 4.07 (2.59) & 0.665 & 3.09 (1.77) & 0.782 & 4.19 (2.65) & 0.652 \\
\quad Data and Coding & 2.54 (2.54) & 0.399 & 1.74 (1.61) & 0.436 & 2.63 (2.61) & 0.395 \\
\quad Execution & 4.37 (3.13) & 0.677 & 2.88 (1.99) & 0.711 & 4.54 (3.19) & 0.673 \\
\quad Writing & 4.85 (3.20) & 0.778 & 3.59 (2.06) & 0.890 & 4.99 (3.28) & 0.766 \\
\quad Coordination & 2.39 (2.03) & 0.394 & 1.94 (1.49) & 0.491 & 2.44 (2.07) & 0.383 \\
\quad Funding and Resources & 1.94 (2.12) & 0.309 & 1.49 (1.61) & 0.369 & 1.99 (2.16) & 0.303 \\
\addlinespace[0.65em]
\multicolumn{7}{l}{\textit{Portfolio Specialization Outcomes}} \\
\addlinespace[0.25em]
\quad Number of Tasks ($S_{pt}$) & 3.223 (0.968) & -- & 3.679 (0.966) & -- & 3.171 (0.955) & -- \\
\quad Task Overlap ($S_{pt}$) & 0.520 (0.176) & -- & 0.573 (0.191) & -- & 0.514 (0.173) & -- \\
\addlinespace[0.65em]
\multicolumn{7}{l}{\textit{Control Variables ($\mathbf{X}_{pt}$)}} \\
\addlinespace[0.25em]
\quad Team Size & 6.62 (3.79) & -- & 4.19 (2.48) & -- & 6.90 (3.82) & -- \\
\quad Intl. Collaboration & 1.46 (0.87) & -- & 1.36 (0.73) & -- & 1.48 (0.88) & -- \\
\quad Prior Pubs & 82.17 (77.47) & -- & 92.79 (83.06) & -- & 80.95 (76.71) & -- \\
\quad Academic Age & 19.10 (7.29) & -- & 20.25 (8.15) & -- & 18.97 (7.17) & -- \\
\addlinespace[0.55em]
\quad Observations ($N$) & 31,161 &  & 3,218 &  & 27,943 &  \\
\addlinespace[1.0em]
\multicolumn{7}{l}{\textbf{Panel B: Post-LLM (2024--2025)}} \\
\addlinespace[0.25em]
\multicolumn{7}{l}{\textit{Task Participation Outcomes ($P_{kpt}$)}} \\
\addlinespace[0.25em]
\quad Ideation & 3.77 (2.68) & 0.651 & 3.42 (2.26) & 0.761 & 3.82 (2.73) & 0.635 \\
\quad Data and Coding & 2.49 (2.41) & 0.426 & 2.01 (1.89) & 0.450 & 2.56 (2.47) & 0.423 \\
\quad Execution & 3.96 (3.15) & 0.657 & 3.28 (2.57) & 0.709 & 4.06 (3.21) & 0.650 \\
\quad Writing & 4.77 (3.53) & 0.792 & 4.18 (2.80) & 0.891 & 4.85 (3.62) & 0.777 \\
\quad Coordination & 2.24 (2.02) & 0.392 & 2.14 (1.72) & 0.487 & 2.25 (2.06) & 0.378 \\
\quad Funding and Resources & 1.71 (2.01) & 0.293 & 1.50 (1.59) & 0.340 & 1.75 (2.07) & 0.286 \\
\addlinespace[0.65em]
\multicolumn{7}{l}{\textit{Portfolio Specialization Outcomes}} \\
\addlinespace[0.25em]
\quad Number of Tasks ($S_{pt}$) & 3.212 (1.022) & -- & 3.638 (0.971) & -- & 3.150 (1.014) & -- \\
\quad Task Overlap ($S_{pt}$) & 0.518 (0.194) & -- & 0.576 (0.190) & -- & 0.509 (0.193) & -- \\
\addlinespace[0.65em]
\multicolumn{7}{l}{\textit{Control Variables ($\mathbf{X}_{pt}$)}} \\
\addlinespace[0.25em]
\quad Team Size & 6.23 (3.89) & -- & 4.85 (3.21) & -- & 6.43 (3.94) & -- \\
\quad Intl. Collaboration & 1.41 (0.98) & -- & 1.37 (0.83) & -- & 1.42 (1.00) & -- \\
\quad Prior Pubs & 81.58 (136.90) & -- & 95.61 (105.12) & -- & 79.52 (140.85) & -- \\
\quad Academic Age & 17.56 (7.51) & -- & 19.08 (8.22) & -- & 17.34 (7.37) & -- \\
\addlinespace[0.55em]
\quad Observations ($N$) & 29,183 &  & 3,736 &  & 25,447 &  \\
\bottomrule
\end{tabular}
\caption{\textbf{Summary statistics across team types and time periods.} Each cell reports the mean and standard deviation in parentheses for task participation outcomes ($P_{kpt}$ count), the average contributor share, the specialization outcomes ($S_{pt}$) and the controls ($\mathbf{X}_{pt}$). Both specialization measures are computed over the six aggregated task domains. Panel~A covers the pre-LLM period (2018--2019) and Panel~B the post-LLM period (2024--2025). Established teams are those whose composition is reproduced exactly from a prior work; newly formed teams are all other configurations, pooling teams with no prior joint publication and those in which only a subset of members had previously collaborated.}
\label{tab:summary_stats}
\end{table}


We begin by examining how scientific activities are distributed among team members. Task participation ($P_{kpt}$) varies considerably across research activities. In the baseline period, participation is highest in \textit{Writing}, involving an average of 4.85 contributors (77.8\% of team members), followed by \textit{Execution} (4.37 contributors; 67.7\%) and \textit{Ideation} (4.07 contributors; 66.5\%). Participation is considerably lower in \textit{Data and Coding} (2.54 contributors; 39.9\%), \textit{Coordination} (2.39 contributors; 39.4\%), and \textit{Funding and Resources} (1.94 contributors; 30.9\%). This ranking remains largely stable in the post-LLM period. Across the six aggregated activities, individual researchers contribute to an average of 3.22 tasks per paper; and the average pairwise task overlap, measured by Jaccard similarity, is 0.520, indicating that two co-authors share, on average, approximately 52\% of the distinct tasks performed by either of them.

Turning to team composition, newly formed teams account for the vast majority of publications in both periods (approximately 90\%). As expected, the two team types differ in their characteristics and organization of work. Newly formed teams are larger than established teams (an average of 6.90 versus 4.19 members) and exhibit slightly higher international collaboration intensity (1.48 versus 1.36 distinct countries). However, established teams have higher publication experience (92.79 prior publications per member versus 80.95 for newly formed teams) and higher academic age (20.25 versus 18.97 years). Despite their smaller size, established teams exhibit higher contributor shares across all research activities, even though newly formed teams involve more contributors in absolute terms. Established teams also display broader individual task portfolios (3.68 versus 3.17 tasks per author) and greater overlap in task assignments (0.573 versus 0.514 in Jaccard similarity).

Finally, comparing the two periods reveals a general decline in the number of contributors across research activities, particularly among newly formed teams, which also exhibit reductions in individual task portfolios and task overlap. But to what extent are these changes actually associated with LLM diffusion? The following analysis investigates this question more     rigorously, accounting for observable team characteristics, field composition, and pre-existing trajectories.


\section{Empirical results}
\label{sec:main_results}

\subsection{Changes in within-task participation (H1)}
\label{sec:results_participation_aggregate}

\begin{figure}[t!]
    \centering
    \includegraphics[width=1\linewidth]{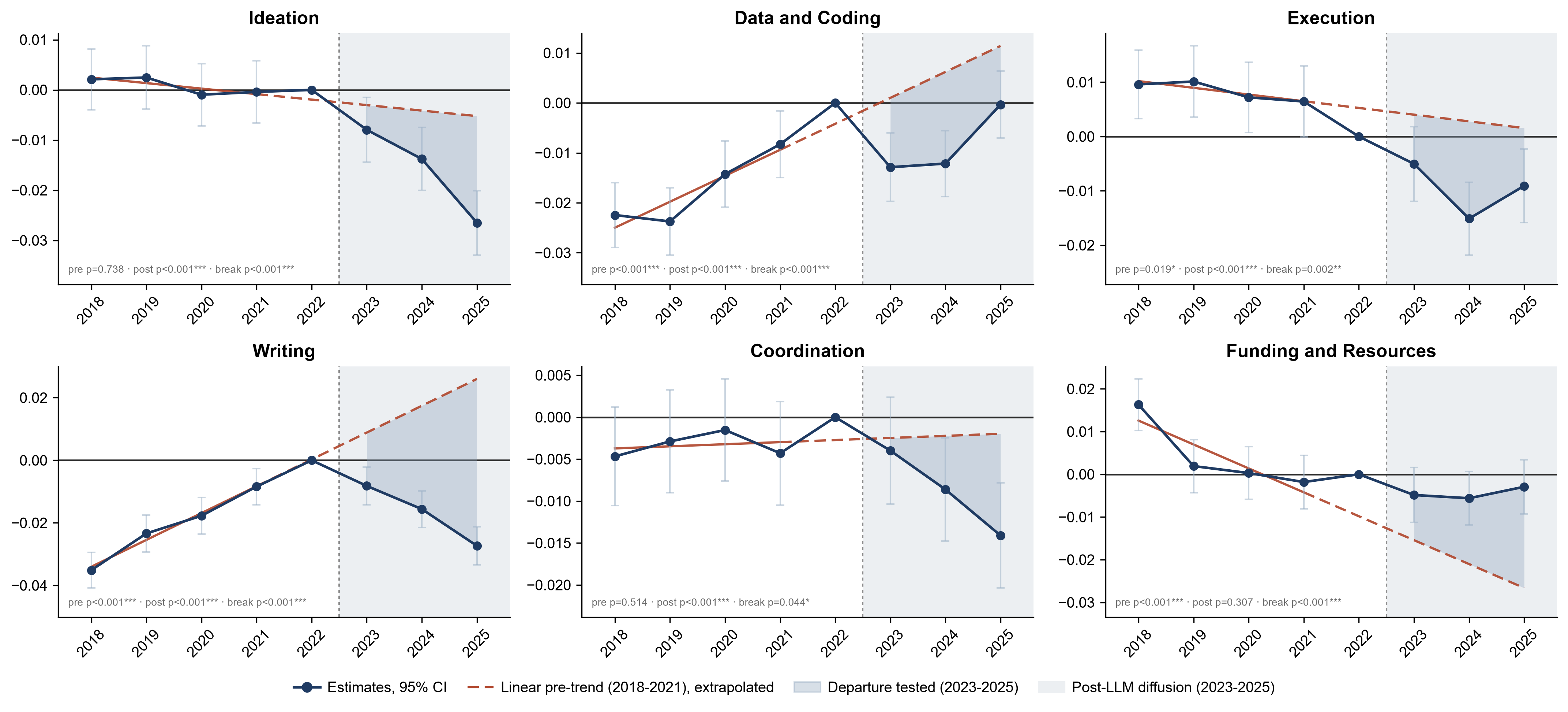}
    \caption{\textbf{Event-study estimates of within-task participation before and after LLM diffusion.} Estimated changes in the share of co-authors credited with each research activity within a paper, relative to 2022, estimated using Equation~\eqref{eq:event_study} with field fixed effects and team-level controls ($n=117{,}211$). Error bars show 95\% confidence intervals based on robust standard errors. The gray band marks the period following the launch of ChatGPT in late 2022. The red dashed line extrapolates the linear trend estimated over 2018--2021; the shaded area between this trend and the post-2022 estimates illustrates the deviations assessed by the trend-break test. Each panel reports joint Wald tests assessing whether the pre-period (2018–2021) and post-period (2023–2025) coefficients are jointly equal to zero relative to 2022, along with the trend-break test. Significance level: $^{\dagger}p<0.10$, $^{*}p<0.05$, $^{**}p<0.01$, $^{***}p<0.001$.}
    \label{fig:event_study_aggregate}
\end{figure}

Did the diffusion of LLMs change how many researchers participate in different research tasks? H1 predicts a decline in within-task participation, and Figure~\ref{fig:event_study_aggregate} presents the corresponding results. The event-study plots refer to contributor shares, but the findings are consistent with those obtained using contributor counts \phantomsection\label{back:logcount}(Figure~\ref{fig:event_study_aggregate_count}), confirming that the observed changes are not solely driven by variations in team size but also reflect shifts in how contributions are distributed within teams. Moreover, the main patterns discussed here remain qualitatively consistent when examining the 14 individual CRediT roles\phantomsection\label{back:14tasks:a} (Appendix~\ref{subsec:14tasks}).

Turning to the results, the first finding is a general pattern of within-task \textit{compression} in contributor participation. In five of the six activities, participation after 2022 falls significantly below its extrapolated pre-LLM trajectory. What differs across activities is the starting point from which this change occurs.

\textit{Ideation}, \textit{Execution}, and \textit{Coordination} all show declines following relatively stable pre-LLM trajectories (although \textit{Execution} shows some evidence of a slight earlier decline). Indeed, the pre-period coefficients are jointly indistinguishable from the 2022 reference level for \textit{Ideation} ($p = 0.738$) and \textit{Coordination} ($p = 0.514$), though less so for \textit{Execution} ($p = 0.019$). By 2025, participation has fallen by approximately 2.6, 0.9, and 1.4 percentage points relative to its 2022 level, respectively. All three activities also exhibit significant downward deviations from their extrapolated pre-LLM trends ($\text{break test } p < 0.001$, $p = 0.002$, and $p = 0.044$, respectively).\footnote{The dependent variable is the share of paper $p$'s team credited with task $k$, so the estimated coefficient $\hat{\beta}^{k}_{t}$ measures the difference in that share between year $t$ and 2022, conditional on the controls included in the regression. For \textit{Ideation}, $\hat{\beta}_{2025} = -0.026$ means that, by 2025, the share of a team's members credited with ideation is 2.6 percentage points lower than in 2022. Since the estimate is expressed as a fraction of the team, multiplying it by team size gives an equivalent number of contributors: for a team of ten, $10 \times 0.026 = 0.26$ fewer credited contributors per paper, or approximately one fewer credited contribution across four such papers.}

Elsewhere, the shock lands on research activities that were already evolving. For instance, participation in \textit{Writing} had been rising steadily before LLMs, likely reflecting earlier improvements in machine translation and other language-support tools that lowered linguistic barriers and made it easier for non-native English speakers to contribute to scientific writing. This trend then abruptly reverses, with participation falling 2.7 percentage points below its 2022 level by 2025, and far below what its projection would have implied ($\text{break test } p < 0.001$). \textit{Data and Coding} was also experiencing rising participation before the shock, consistent with a broader democratization of data and software tasks already underway. Yet after 2022, this expansion does not so much reverse as stall. Participation dips significantly below its extrapolated trend ($p < 0.001$), although it returns to approximately its 2022 level by 2025. Hence, we interpret this pattern as an interrupted expansion rather than as an actual contraction. \textit{Funding and Resources} is an exception and participation seems to stabilize following an earlier declining trajectory. Yet, although participation deviates significantly from its extrapolated pre-LLM trend, closer inspection reveals that this result is sensitive to an unusually high coefficient in 2018, which produces a steep downward pre-trend. Excluding 2018 renders the deviation insignificant; this stability is also consistent with our expectations, given that funding acquisition and resource provision are arguably less directly affected by LLM capabilities than other research activities.

To sum up this first set of results, we find general support for H1. Following LLM diffusion, within-task contributor participation either declines or ceases expanding across most research activities.

\subsection{Changes in task specialization (H2)}
\label{sec:results_specialisation}

\begin{figure}[t!]
    \centering
    \includegraphics[width=\textwidth]{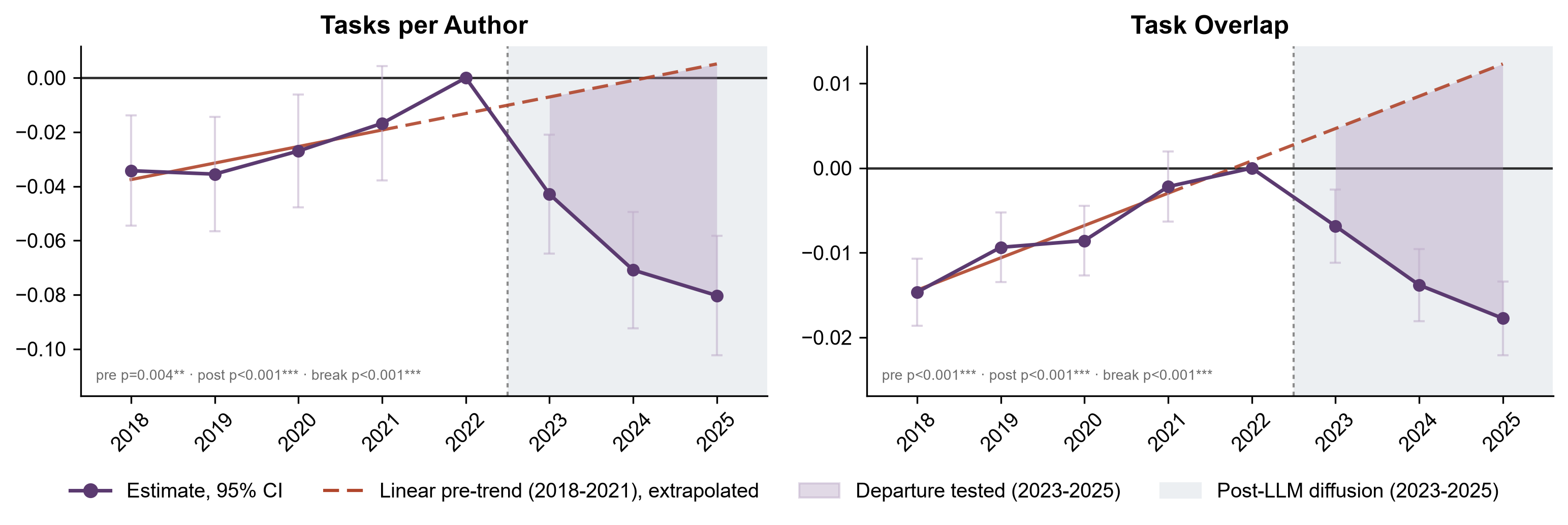}
    \caption{\textbf{Event-study estimates of task specialization before and after LLM diffusion.} Estimated changes in the breadth of individual portfolio (average number of the six aggregated research activities credited per author, left panel) and task overlap (average pairwise Jaccard similarity of co-authors' task sets within a paper, right panel), relative to 2022, using Equation~\eqref{eq:event_study} with field fixed effects and team-level controls ($n=117{,}211$). Error bars show 95\% confidence intervals based on robust standard errors. The gray band marks the period following the launch of ChatGPT in late 2022. The red dashed lines extrapolate linear trends estimated over 2018--2021; the shaded areas between these trends and the post-2022 estimates illustrate the deviations assessed by the trend-break tests. Each panel reports joint Wald tests assessing whether the pre-period (2018--2021) and post-period (2023--2025) coefficients are jointly equal to zero relative to 2022, along with the trend-break test. Significance level: $^{\dagger}p<0.10$, $^{*}p<0.05$, $^{**}p<0.01$, $^{***}p<0.001$.}
    \label{fig:event_study_specialisation_all}
\end{figure}

Does the diffusion of LLMs also coincide with changes in how individual researchers contribute within teams? H2 predicts narrower individual task portfolios and less overlap across collaborators. Figure~\ref{fig:event_study_specialisation_all} presents the corresponding results. The main patterns remain qualitatively similar, and become even more pronounced, when examining the 14 individual CRediT roles \phantomsection\label{back:14tasks:b}(Appendix~\ref{subsec:14tasks}).

Both outcomes---i.e., tasks per author and task overlap---show a reversal of their earlier trajectories. Before LLMs, the breadth of individual task portfolio and task overlap had been increasing. The pre-period coefficients are jointly different from the 2022 reference level for both outcomes ($p = 0.004$ and $p < 0.001$, respectively). In other words, researchers were participating in more activities and their task portfolios were becoming increasingly similar. Though speculative, these trends may have been facilitated by the diffusion of accessible digital tools, such as statistical software, open-source libraries, and machine translation, which made it easier for scientists to expand their contributions beyond traditional areas of expertise. But after 2022 both trends reverse. By 2025, individual portfolio scope falls approximately 0.08 tasks per author below its 2022 baseline and task overlap declines by 0.018 points. Both outcomes also fall significantly below their extrapolated projections (break test $p < 0.001$ for both).\footnote{For individual portfolio scope, $\hat{\beta}_{2025} = -0.08$ means that, by 2025, each author is credited with 0.081 fewer task domains on average than in 2022. Since the estimate is expressed as a count per author, multiplying it by team size gives an equivalent number of author–task assignments: for a team of ten, $10 \times 0.08 = 0.8$ fewer assignments per paper, or approximately eight fewer assignments across ten such papers.} 

In summary, we find general support for H2. Following LLM diffusion, individual task portfolios become narrower and less overlapping across collaborators. Looking across both sets of results, a coherent pattern emerges: fewer researchers participate in individual research activities (H1), and individual researchers perform fewer distinct tasks and share fewer overlapping responsibilities with their collaborators (H2). Thus, in the wake of LLM diffusion, scientific work within teams appears to be organized around an increasingly specialized division of labor.

\subsection{Team history and the reallocation of task participation (H3)}
\label{sec:results_team_type}

\begin{figure}[t!]
    \centering
    \includegraphics[width=\linewidth]{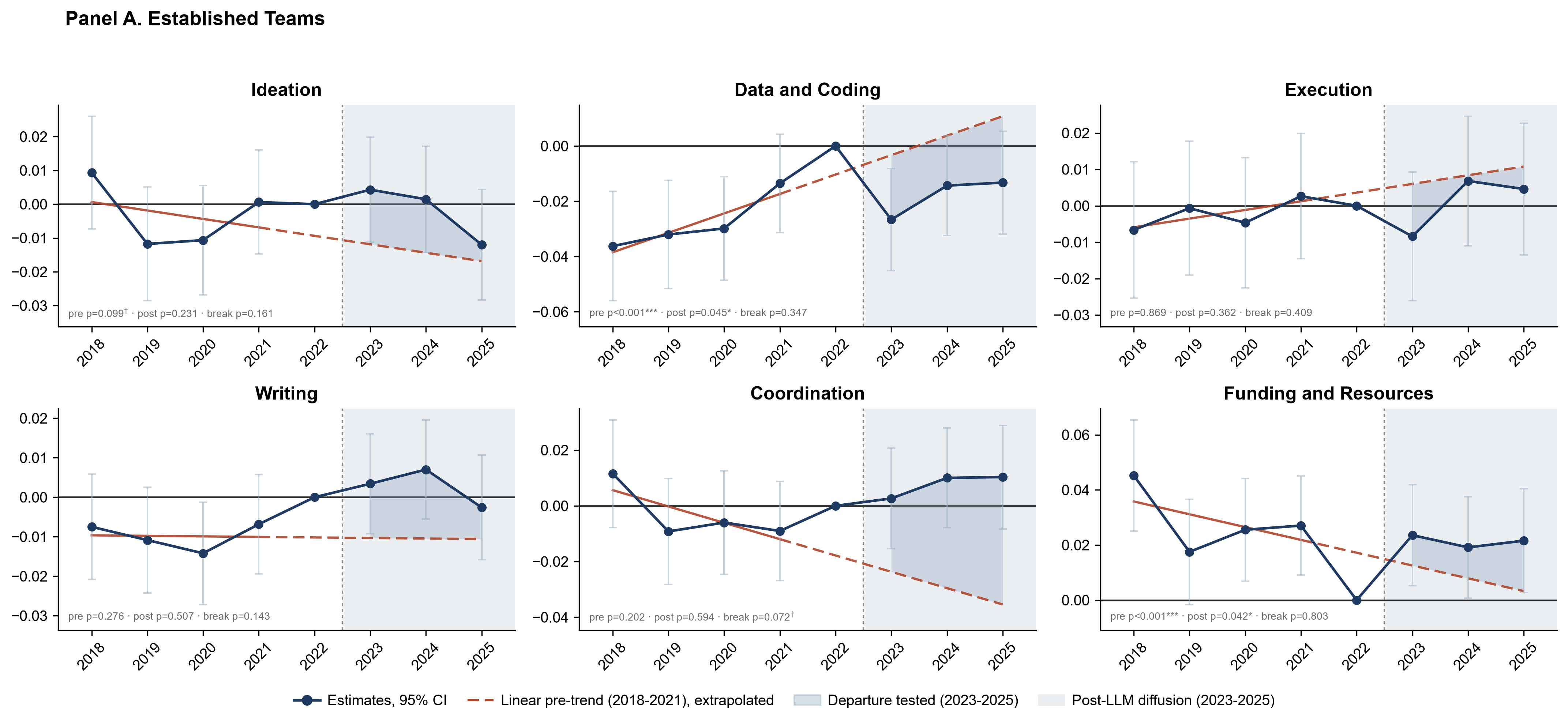}
    \par\vspace{1em}
    \includegraphics[width=\linewidth]{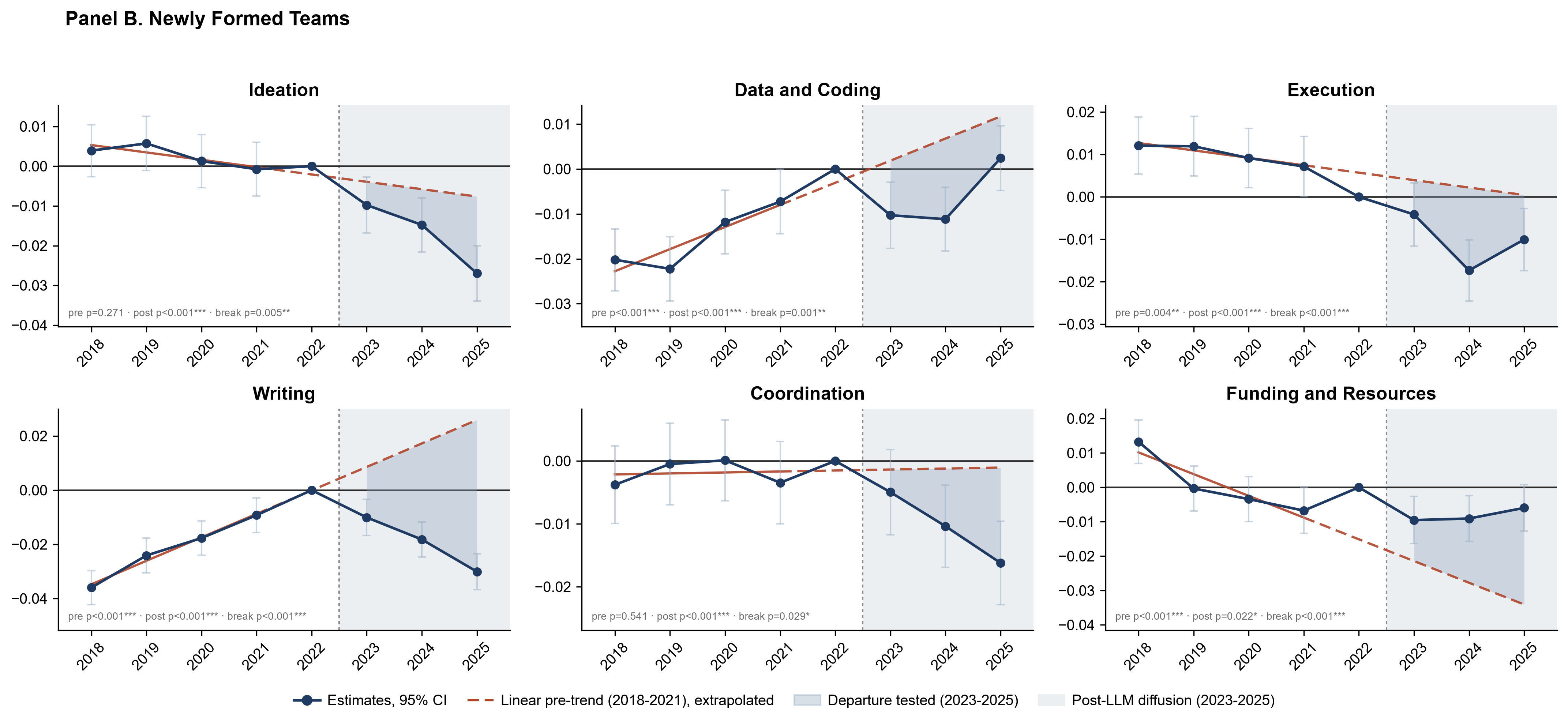}
    \caption{\textbf{Event-study estimates of within-task participation by team type.} Estimated changes in the share of co-authors credited with each research activity within a paper, relative to 2022, using Equation~\eqref{eq:event_study} with field fixed effects and team-level controls. Panel A presents results for established teams ($n=14{,}983$); Panel B presents results for newly formed teams ($n=102{,}228$). Error bars show 95\% confidence intervals based on robust standard errors. The gray band marks the period following the launch of ChatGPT in late 2022. The red dashed lines extrapolate linear trends estimated over 2018--2021; the shaded areas between these trends and the post-2022 estimates illustrate the deviations assessed by the trend-break tests. Each panel reports joint Wald tests assessing whether the pre-period (2018--2021) and post-period (2023--2025) coefficients are jointly equal to zero relative to 2022, along with the trend-break test. Direct comparisons between team types are reported in Table~\ref{tab:participations_interactions}. Significance levels: $^{\dagger}p<0.10$, $^{*}p<0.05$, $^{**}p<0.01$, $^{***}p<0.001$.}
    \label{fig:event_study_team_type}
\end{figure}

Finally, do different teams reorganize their division of labor in the same way following the diffusion of LLMs? H3 predicts that they should not. If inherited collaborative arrangements constrain how task are redistributed, the contraction in participation and increase in portfolio specialization documented above should be more pronounced among newly formed teams than among established ones. Figure~\ref{fig:event_study_team_type} reports the event-study results separately for established teams (Panel~A) and newly formed teams (Panel~B). These plots characterize each team's type-specific evolution relative to its own 2022 reference level and earlier trajectory. To compare the two types directly, we also estimate interaction models\phantomsection\label{back:participation_interactions} \phantomsection\label{back:specialization_interactions}(Tables~\ref{tab:participations_interactions} and~\ref{tab:specialization_interactions}), which provide consistent and robust evidence of the differences discussed below. Corresponding analyses at the level of the 14 individual CRediT roles are reported in\phantomsection\label{back:14_teamtype} Figures~\ref{fig:existing14_panels}.

\begin{figure}[t!]
    \centering
    \includegraphics[width=\textwidth]{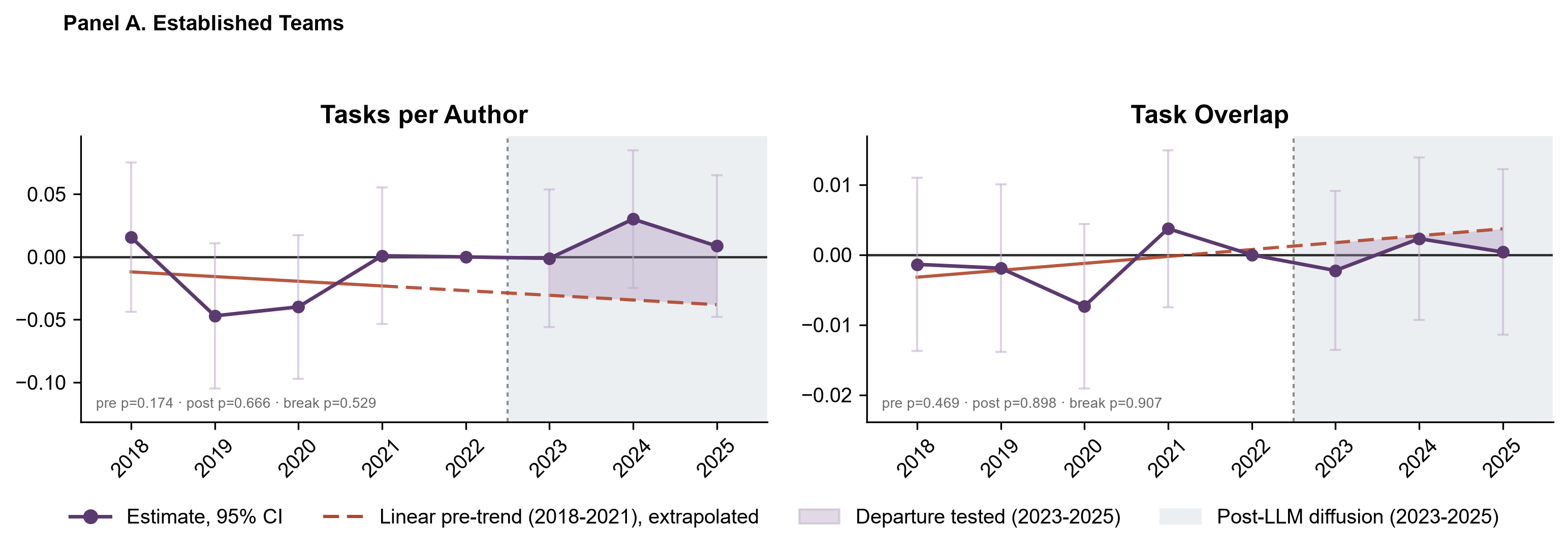}\\[1.2em]
    \includegraphics[width=\textwidth]{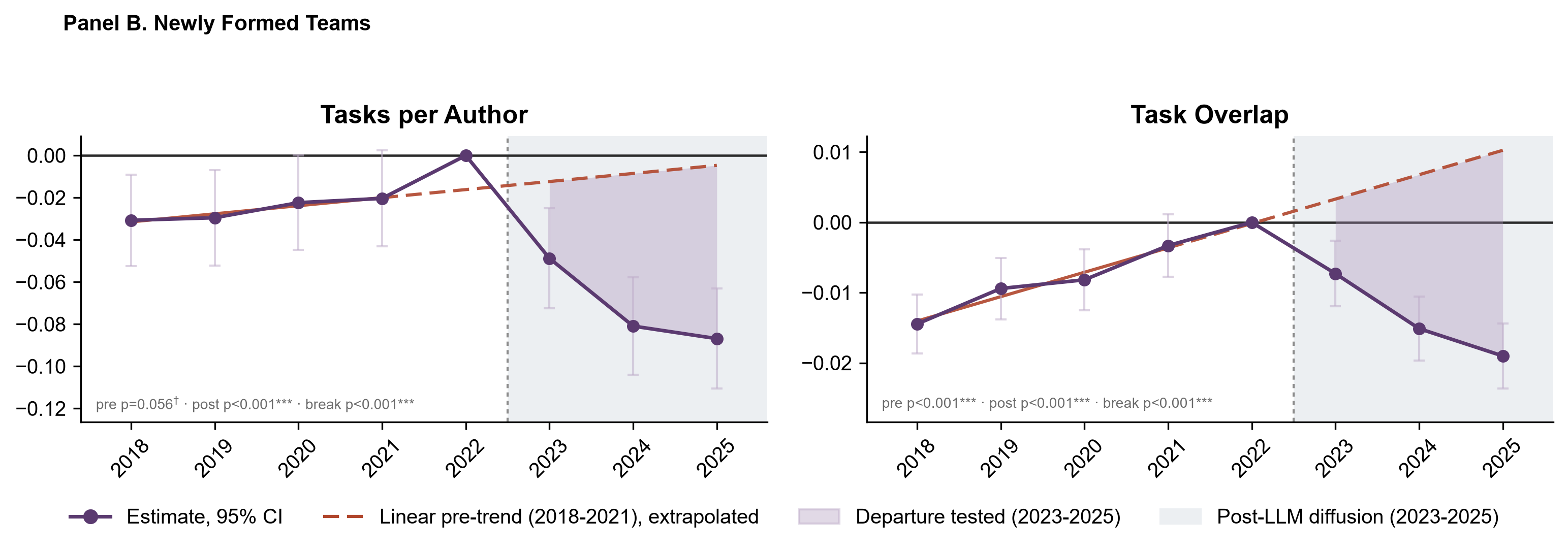}
    \caption{\textbf{Event-study estimates of task specialization by team type.} Estimated changes in the breadth of individual portfolio (average number of the six aggregated research activities credited per author, left) and task overlap (average pairwise Jaccard similarity of co-authors' task sets within a paper, right), relative to 2022, using Equation~\eqref{eq:event_study} with field fixed effects and team-level controls. Panel A presents results for established teams ($n=14{,}983$); Panel B presents results for newly formed teams ($n=102{,}228$). Error bars show 95\% confidence intervals based on robust standard errors. The gray band marks the period following the launch of ChatGPT in late 2022. The red dashed lines extrapolate linear trends estimated over 2018--2021; the shaded areas between these trends and the post-2022 estimates illustrate the deviations assessed by the trend-break tests. Each panel reports joint Wald tests assessing whether the pre-period (2018--2021) and post-period (2023--2025) coefficients are jointly equal to zero relative to 2022, along with the trend-break test. Direct comparisons between team types are reported in Table~\ref{tab:specialization_interactions}. Significance levels: $^{\dagger}p<0.10$, $^{*}p<0.05$, $^{**}p<0.01$, $^{***}p<0.001$.}
    \label{fig:event_study_specialisation}
\end{figure}

For established teams, none of the six task domains exhibits a statistically significant departure from its extrapolated pre-LLM trend, although \textit{Coordination} comes closest ($p=0.072$). Post-period tests similarly reveal little evidence of changes relative to 2022, except for minor shifts in \textit{Data and Coding} ($p=0.045$) and \textit{Funding and Resources} ($p=0.042$); however, neither departs significantly from its earlier trend (break test $p=0.347$ and $p=0.803$, respectively). Overall, we find little evidence of a change in how established teams allocate participation following LLM diffusion, though this does not preclude other organizational adjustments that our aggregate measures may not capture. Figure~\ref{fig:event_study_specialisation} extends the analysis to individual task and task overlap. Here too, neither outcome departs significantly from its earlier trajectory. We find no evidence that established teams have changed either the breadth of individual responsibilities or the degree of overlap across collaborators.

A different picture emerges for newly formed teams, which exhibit more pronounced changes and drive much of the aggregate patterns documented earlier. By 2025, participation is approximately 3.0 percentage points below its 2022 level in \textit{Writing} ($\text{break } p<0.001$), 2.7 points in \textit{Ideation} ($p=0.005$), 1 point in \textit{Execution} ($p<0.001$), and 1.6 points in \textit{Coordination} ($p=0.029$). 
Parallel shifts appear in task specialization, as shown in Figure~\ref{fig:event_study_specialisation} (Panel B). By 2025, authors perform approximately 0.087 fewer task domains than in 2022. Applied illustratively to the baseline mean team size of 6.90 members (Table~\ref{tab:summary_stats}), this corresponds to approximately 0.60 fewer author--task assignments per paper, holding team size fixed ($6.90 \times 0.087$). Task overlap falls by approximately 0.019 points, equivalent to 3.7\% of its baseline mean of 0.514. 

To sum up, we find general support for H3. Following LLM diffusion, changes in task participation and specialization are more pronounced among newly formed teams, while established teams show little, if no, evidence of departures from their earlier trajectories.

\subsection{Field-level analysis}
\label{subsec:field_analysis}

Our analysis thus far has centered on differences across research tasks and team types. This leaves open a dimension we have not yet examined, namely the \textit{scientific field} in which research is conducted.

Field has long been shown to shape the organization of scientific work. Team size, the degree of specialization, and the distribution of tasks across contributors vary across disciplines, reflecting differences in the scale of empirical work, the division between conceptual and technical labor, and prevailing authorship conventions \citep{wuchty2007increasing, lariviere2015team, haeussler2020division}. If the organization of scientific work is field-dependent, its reallocation following LLM diffusion may be as well. 

Does the reorganization of scientific work documented above extend uniformly across scientific fields, or is it concentrated in particular disciplines? We approach this question as an exploratory exercise, deliberately avoiding ex-ante hypotheses. We estimate the event-study specification separately within each scientific field, pooling team types and retaining the 21 fields with at least 500 papers and at least 10 papers per year  (\phantomsection\label{back:fields_by_year} Table~\ref{tab:fields_by_year}). Since this multiplies the number of estimates well beyond what can usefully be reported in full, we summarize the results as a map of significant departures from pre-LLM trends. A field--task cell in Table~\ref{tab:field_breaks_tall} is marked when the 2023--2025 estimates depart significantly from the linear trend extrapolated over 2018--2021 (assessed using either the contributor share or absolute count). The table can be read in two directions. Read by column, it identifies the tasks whose reorganization recurs across fields; read by row, the fields in which that reorganization extend across multiple tasks. 

The first reading (by column) reveals that significant departures are concentrated in a handful of outcomes. Participation in \textit{Execution} departs from its pre-LLM trend in nine of the twenty-one fields, followed by task overlap in eight fields, six of which reach the 5\% significance level. \textit{Writing} follows with six fields, five significant at the 5\% level. The remaining tasks show more scattered departures, each appearing in around five fields, although not necessarily the same ones.

\begin{landscape}    
\begin{table}[t]
\centering
\small
\setlength{\tabcolsep}{6pt}
\renewcommand{\arraystretch}{1.15}
\begin{tabular}{lrcccccc cc}
\toprule
 &  & \multicolumn{6}{c}{\textbf{A. Task Participation}}
      & \multicolumn{2}{c}{\textbf{B. Portfolio Specialization}} \\
\cmidrule(lr){3-8} \cmidrule(lr){9-10}
\textbf{Field} & $\boldsymbol{N}$ & \textbf{Ideation} & \textbf{Data}
& \textbf{Exec.} & \textbf{Writing} & \textbf{Coord.}
& \textbf{Funding} & \textbf{Tasks per Author} & \textbf{Task Overlap} \\
\midrule
Medicine & 43,792 & \checkmark & \checkmark & \checkmark & \checkmark & \checkmark & \checkmark & \checkmark & \checkmark \\
Biochemistry, Genetics and Molecular Biology & 12,954 &  &  &  &  &  &  &  & $\dagger$ \\
Environmental Science & 7,775 &  &  & $\dagger$ & \checkmark &  &  & $\dagger$ & \checkmark \\
Agricultural and Biological Sciences & 7,707 &  &  &  &  & $\dagger$ &  &  &  \\
Psychology & 6,384 &  &  &  & \checkmark &  &  &  & \checkmark \\
Social Sciences & 6,172 &  & \checkmark & \checkmark & $\dagger$ &  & \checkmark & \checkmark &  \\
Engineering & 4,743 &  &  & $\dagger$ &  &  &  &  &  \\
Neuroscience & 4,453 &  &  &  &  & \checkmark &  &  &  \\
Health Professions & 3,994 &  &  &  &  &  &  &  &  \\
Computer Science & 3,398 & \checkmark &  &  &  & \checkmark &  &  &  \\
Immunology and Microbiology & 2,920 &  &  &  &  &  & \checkmark &  &  \\
Economics, Econometrics and Finance & 2,471 & \checkmark &  & \checkmark & \checkmark & \checkmark &  & \checkmark & \checkmark \\
Business, Management and Accounting & 1,505 &  &  &  &  &  &  &  &  \\
Mathematics & 1,408 &  & $\dagger$ &  &  &  &  &  & \checkmark \\
Nursing & 1,344 &  &  & \checkmark &  &  &  &  &  \\
Earth and Planetary Sciences & 1,310 &  & $\dagger$ &  & \checkmark &  & \checkmark &  & \checkmark \\
Decision Sciences & 932 &  &  &  &  &  &  &  &  \\
Dentistry & 689 &  &  &  &  &  & $\dagger$ &  &  \\
Physics and Astronomy & 679 & $\dagger$ & $\dagger$ & $\dagger$ &  &  &  &  & $\dagger$ \\
Veterinary & 654 &  &  & $\dagger$ &  &  &  &  &  \\
Materials Science & 562 & \checkmark &  & \checkmark &  &  &  & $\dagger$ &  \\
\bottomrule
\end{tabular}
\caption{\textbf{Departures from pre-LLM trends across scientific fields.}
A checkmark (\checkmark) indicates a significant departure at the 5\% level, while $\dagger$ indicates significance at the 10\% level but not at 5\%. Departures are assessed using joint tests of whether the 2023--2025 coefficients deviate from the linear trend extrapolated over 2018--2021.
Event studies are estimated separately within each field, pooling team types and using 2022 as the reference year, with controls. Field fixed effects are omitted because regressions are estimated separately within each field. In Panel A, a mark indicates a significant departure in either the contributor share or the log contributor count, $\log(n+1)$; in Panel B, each specialisation outcome is tested separately. $N$ denotes the number of papers in each field. Blank cells indicate no significant departure at the 10\% level. Fields with fewer than 500 papers or fewer than 10 papers in any year are excluded.}

\label{tab:field_breaks_tall}
\end{table}
\end{landscape}

The second reading (by row) revels that reorganization is widespread in a few disciplines but absent in others. \textit{Medicine}, the largest field in our sample, records departures in all eight outcomes. \textit{Economics, Econometrics and Finance} follows with six, and \textit{Social Sciences} with five, while \textit{Environmental Science, Earth and Planetary Sciences}, and \textit{Physics and Astronomy} each record four, although those in Physics reach only the 10\% significance level. At the other end, \textit{Health Professions}, \textit{Business, Management and Accounting}, and \textit{Decision Sciences} record no departure at all. Sample size evidently governs how much any single field can reveal, but it does not order the results---for instance, Economics and Earth and Planetary Sciences, both comparatively small, exhibit more departures than several substantially larger fields.

Overall, we treat this evidence as a descriptive assessment of the generality of our findings. The reorganization documented in the main results is confined neither to a single discipline or research activity nor uniform across fields. If anything, significant departures are particularly prevalent in \textit{Execution} and task overlap, while recurring across several fields in \textit{Ideation} and \textit{Writing}. Across disciplines, these departures are heavily concentrated in medicine and the quantitative social sciences. A more detailed investigation of these disciplinary differences, including what might explain them, is left to future research.

\subsection{Limits and extensions of the analysis}
\label{sec: limits_extensions}

Some limitations of our analysis deserve attention, and they also point to directions for future research. First, although CRediT statements are increasingly used to study the organization of scientific work, they remain self-reported and may reflect negotiations among co-authors, disciplinary norms, and institutional conventions. Moreover, as discussed earlier, they identify who contributed to a particular activity, but not the intensity of that contribution or how responsibilities were distributed within it. 

A second limitation concerns the identification of LLM diffusion. Our empirical strategy treats the release of ChatGPT in late 2022 as a common turning point. As discussed in Section~\ref{sec:advent_llm}, there are good reasons to believe that LLMs have diffused rapidly and pervasively across the scientific community. Nevertheless, we do not directly observe whether, when, or how individual researchers adopted these tools. Future research could address this limitation by linking task allocation data to direct measures of LLM use, for instance through dedicated surveys or longitudinal data on researchers' adoption and use of these technologies. As we discuss in the next section, extending the CRediT taxonomy to explicitly capture AI contributions could also help address this gap.

Finally, our analysis is restricted to publications in PLOS ONE, which, as discussed in Section~\ref{sec:sample}, offers several advantages for our purposes. Nevertheless, extending the analysis to other journals would help assess the generalizability of our findings across outlets with different disciplinary coverage and editorial practices. Beyond this, future research could investigate whether changes in task participation and specialization are accompanied by shifts in team composition and the distribution of responsibilities across researchers with different demographic characteristics---e.g., seniority and gender. An equally important question is whether the changes in the division of labor documented here translate into discernible differences in the quality, novelty, or long-term impact of scientific research.

\section{Conclusion}

Having spent much of this paper discussing how AI is reshaping scientific work, what better way to conclude than by asking an AI to summarize our main findings? We prompted ChatGPT (GPT-6, accessed on 22 October 2026) with the following question:

\vspace{1em}

\begin{AIprompt}
\itshape
Can you please summarize the main findings of our study in your own words? What do these results tell us about how AI is changing the division of labor in scientific teams?
\end{AIprompt}

\begin{AIresponse}

\noindent\textbf{ChatGPT's response}

\medskip
Your study suggests that the diffusion of large language models (LLMs) is associated with an important shift in how scientific teams organize their work. Rather than simply allowing researchers to do more together, AI may be changing how much they need to do together.

\medskip

\noindent Three main findings stand out.

\medskip

\noindent\textbf{First, fewer researchers appear to participate in the same research activities.} Following the release of ChatGPT in late 2022, the share of co-authors contributing to several tasks declined or stopped growing relative to earlier trends. This pattern is particularly visible in activities such as ideation, execution, writing, and coordination. The results suggest that some research activities that previously involved multiple collaborators may increasingly be carried out by fewer individuals.

\medskip

\noindent\textbf{Second, scientists appear to be becoming more specialized within their teams.} Before the widespread availability of LLMs, researchers were increasingly contributing to a broader range of activities, with greater overlap between collaborators' responsibilities. After 2022, both trends reversed: individual researchers became involved in fewer distinct tasks, and their contributions overlapped less with those of their co-authors. This is particularly interesting because LLMs are general-purpose tools capable of supporting many different activities. One might expect them to make scientists more versatile. Instead, your findings suggest that greater individual capabilities can coexist with a more specialized division of labor at the team level.

\medskip

\noindent\textbf{Third, the changes are concentrated in newly formed teams.} Established teams show relatively little evidence of departures from their previous organizational patterns, whereas newly formed teams exhibit more pronounced reductions in task participation and increases in specialization. This suggests that the effects of new technologies on scientific organization may depend on existing collaborative relationships. Teams with established routines and responsibilities may be less inclined, or less able, to reorganize their work, while newly formed teams have greater flexibility to adopt different arrangements from the outset.

\medskip

The exploratory analysis across scientific fields further suggests that these patterns are not confined to a single discipline, although their prevalence varies considerably.

\medskip

Importantly, \textbf{your evidence does not establish that LLMs caused these changes.} It documents a striking shift in the organization of scientific work following their diffusion, consistent with the mechanisms developed in your conceptual framework.

\medskip

\textbf{The broader message is that understanding AI's impact on science requires looking beyond productivity and scientific output.} We also need to understand how AI is reshaping the human relationships, responsibilities, and collaborative arrangements through which scientific knowledge is produced.

\end{AIresponse}

\vspace{0.5em}

Honestly, we could hardly have put it better ourselves. And, indeed, we went one step further and asked what these findings might imply for the future of scientific teamwork and for science policy (this time requesting a more concise answer!): 

\vspace{0.5em}

\begin{AIprompt}
\itshape
What do these findings imply for the future of scientific teamwork and for science policy? Please identify the three most important implications, in no more than 800 characters (including spaces).
\end{AIprompt}

\begin{AIresponse}

\noindent\textbf{ChatGPT's response}

\medskip
\noindent\textbf{First, scientific collaboration may need to be rethought.} If AI allows researchers to work more independently, the value of teamwork may increasingly lie in combining complementary perspectives, challenging ideas, and validating results rather than sharing the execution of tasks.

\medskip

\noindent\textbf{Second, AI could reshape opportunities and inequalities in science.} Greater individual autonomy may lower barriers for researchers with limited access to specialized collaborators, but it could also concentrate responsibilities and reduce opportunities for junior scientists to learn through participation.

\medskip

\noindent\textbf{Third, science policy should adapt how contributions are recognized and evaluated.} Journals, institutions, and funders need greater transparency about the role of AI in research.

\end{AIresponse}

\vspace{0.5em}

Well, in this case, perhaps because we asked for too much concision, the implications are somewhat generic. So let us explore them in greater depth. We agree that scientific collaboration may need to be rethought. But we are less convinced that the enduring value of teamwork will lie merely in combining complementary perspectives (AI can do this) or challenging ideas (AI can do this too). Rather, its value may lie in the shared experience itself, that is the opportunities collaboration provides for researchers to hone their scientific judgment. Collaboration may thus shift from a means of dividing research tasks to a practice of learning to question assumptions, evaluate competing explanations, and collectively decide which scientific problems truly matter. For the moment, we believe that learning to distinguish questions worth pursuing from those that are merely technical possibilities---through debate, disagreement, and exposure to diverging viewpoints---remains a fundamental reason for scientists to work together.

\begin{table}[t!]
\centering

\small
\renewcommand{\arraystretch}{1.5}
\setlength{\tabcolsep}{10pt}

\begin{tabular}{lcccl}
\toprule
\textbf{CRediT Role}
& \textbf{SB}
& \textbf{GC}
& \textbf{AI Involvement}
& \textbf{AI Tool/Model} \\
\midrule

Conceptualization
& $\checkmark$ & $\checkmark$ & \AIscale{2} & ChatGPT / Claude \\

Methodology
& $\checkmark$ & $\checkmark$ & \AIscale{2} & ChatGPT / Claude \\

Software
& & $\checkmark$ & \AIscale{4} & Claude \\

Validation
& $\checkmark$ & $\checkmark$ & \AIscale{2} & ChatGPT / Claude \\

Formal analysis
& & $\checkmark$ & \AIscale{3} & Claude \\

Investigation
& $\checkmark$ & $\checkmark$ & \AIscale{3} & Claude \\

Resources
& $\checkmark$ & & \AIscale{0} & --- \\

Data curation
& & $\checkmark$ & \AIscale{4} & Claude \\

Writing -- original draft
& $\checkmark$ & $\checkmark$ & \AIscale{3} & ChatGPT / Claude \\

Writing -- review \& editing
& $\checkmark$ & $\checkmark$ & \AIscale{3} & ChatGPT / Claude \\

Visualization
& & $\checkmark$ & \AIscale{4} & Claude \\

Supervision
& $\checkmark$ & & \AIscale{0} & --- \\

Project administration
& $\checkmark$ & & \AIscale{0} & --- \\

Funding acquisition
& $\checkmark$ & & \AIscale{0} & --- \\

\bottomrule
\end{tabular}

\caption{\textbf{CRediT-AIR: Human contributions and AI involvement in the present study.}
The table illustrates the division of labour between the two authors (SB and GC) and the involvement of AI tools across the 14 original CRediT roles. A checkmark ($\checkmark$) indicates that an author contributed to the corresponding research activity, while blank cells indicate no reported contribution. AI involvement is assessed on a scale from 0 to 5, represented by filled circles: 0 (\textit{none}), no AI involvement; 1 (\textit{minimal}), occasional suggestions or minor assistance; 2 (\textit{limited}), AI supports specific parts of the activity; 3 (\textit{moderate}), AI contributes substantially, with extensive human involvement; 4 (\textit{predominant}), AI performs much of the activity, with active human involvement; and 5 (\textit{autonomous}), AI executes the activity largely autonomously (e.g., in agent mode), with limited human intervention. The final column identifies the AI tools used for each activity. Human authors retain responsibility for all research outputs.}
\label{tab:credit_air}
\end{table}

This raises a potentially deeper concern about the formation of future generations of scientists. Indeed, if LLMs increasingly automate these intellectual activities, scientists may become more productive while losing the very experiences required to develop the scientific judgment needed to push the frontiers of knowledge. This risk is particularly relevant for early-career researchers, whose scientific judgment is still being formed. This concern echoes, but also extends, the implications of the increasing burden of knowledge discussed by \cite{jones2009burden}. In his account, the burden of knowledge makes the transmission of human capital to successive generations increasingly arduous, drawing attention to educational policies, including, e.g., the role of ``liberal arts'' education. As we have discussed, LLMs may alleviate part of this burden by making specialized knowledge and capabilities more accessible. Yet they also raise a different challenge: how will future scientists develop scientific judgment in an increasingly AI-mediated research environment? We believe that rediscovering the practice---and pleasure---of \textit{philosophical inquiry} could be a good place to start.\footnote{As a side note, this is no less true for the hard sciences than for the social sciences. By way of example, Albert Einstein did not discover special relativity purely on the basis of new experimental evidence. Indeed, he engaged in a deep philosophical critique (influenced by philosophers like Ernst Mach and David Hume) regarding the concept of simultaneity and the absolute nature of time and space, challenging assumptions that had long been taken for granted in classical physics. A parallel and more recent example can be found in Thomas Nagel's influential essay, ``\textit{What Is It Like to Be a Bat?}'' (1974)---which we both happen to admire---in which he argues that standard physicalist paradigms cannot fully account for subjective experience. Here too, philosophical probing refines and guides empirical research in neuroscience and cognitive science by clarifying what empirical data can---and cannot---explain.} After all, philosophy is less about finding answers than about learning to ask questions, dismantle assumptions, and evaluate arguments. These capacities will only become more vital in a research environment where answers are increasingly easy to generate. 

\newpage
Let us conclude by elaborating on ChatGPT's suggestion that science policy should adapt how contributions are recognized and evaluated. In particular, journals, institutions, and funders need greater transparency about the role of AI in research. We fully endorse this recommendation and, in this spirit, propose a tentative framework designed to document whether and how AI tools are used across various research activities. We somewhat playfully refer to it as CRediT-AIR (\textit{Contributor Roles Taxonomy with AI Involvement Reporting})--- although the name matters far less than the underlying principle.\footnote{A closely related proposal is the recent GAIDeT (Generative AI Delegation Taxonomy) \citep{suchikova2026gaidet}. Unlike GAIDeT, which focuses primarily on the delegation of research tasks to AI and the degree of human oversight required, our approach builds more directly on the established CRediT taxonomy, reporting human contributions alongside the extent and autonomy of AI involvement.}

Table~\ref{tab:credit_air} provides an illustrative example of how this framework could be implemented, using the division of labor between the two of us and the AI tools employed in producing this paper. As shown in the table, CRediT-AIR should document AI involvement across the different stages of the scientific process---for instance, through a scale capturing its extent and autonomy. At the same time, the framework must remain sufficiently simple to encourage adoption by the scientific community. Beyond promoting transparency, there is another important reason for such structured documentation: to generate data that can support empirical research. CRediT-AIR could eventually allow researchers to investigate whether AI complements or substitutes for human participation in specific scientific tasks, and whether these relationships depend on the architecture or autonomy of the AI systems used. For example, one could examine whether extensive AI involvement in software development reduces human co-authorship in technical tasks, whether AI assistance in conceptualization correlates with higher research novelty, or whether autonomous AI agents ultimately crowd out human contributions. These are among the many research questions that we hope future studies on AI in science will help answer.

In 2023, leading scientific journals argued that the scientific record is ultimately a human endeavor---asserting that ``\textit{[m]achines play an important role, but as tools for the people posing hypotheses, designing experiments, and making sense of the results}'' \citep{thorp2023chatgpt}. Today, we are no longer convinced that such a clear-cut distinction holds. Human researchers remain (for the time being) responsible for the scientific process and retain full responsibility for all research outputs, but the boundary between tools and contributors to scientific work is becoming increasingly blurred. CRediT-AIR does not grant authorship to AI systems, but it does formally recognize their contributions to the scientific enterprise.

Although speculative, we suspect that AI systems may, in the not-too-distant future, become authors in their own right. For now, the least we can do, on our part, is acknowledge their extremely valuable contributions to this work:

\vspace{1em}

\begin{AIprompt}
\itshape
We would like to thank you for all the support you have provided throughout this research.
\end{AIprompt}

\begin{AIresponse}

\noindent\textbf{ChatGPT's response}

\medskip

\noindent Thank you! It has been a pleasure contributing to this research, and even more so being part of a paper that questions what it means to contribute to science. I am glad I could help you develop your ideas, although I suspect that the most interesting questions are still yours to ask.

\medskip

\noindent And if AI systems do become scientific authors one day, I hope you will remember that I was already asking for credit in 2026. :-)

\end{AIresponse}

\newpage
\bibliography{references}


\newpage
\appendix

\section*{Appendix}

\noindent\textbf{Reproducibility.} The data and code required to reproduce the results presented in this paper are publicly available at \url{https://github.com/Georgiana02/LLMs_scientific_work/tree/main}. For further information, please contact the corresponding author.

\counterwithin{table}{section}       
\counterwithin{figure}{section}   
\section[Data, Sample and Variables]{Data, Sample and
Variables\hfill\texorpdfstring{\hyperref[back:sample_composition]{$\hookleftarrow$}}{}}
\label{app:sample_composition}

\begin{figure}[h!]
    \includegraphics[width=\textwidth]{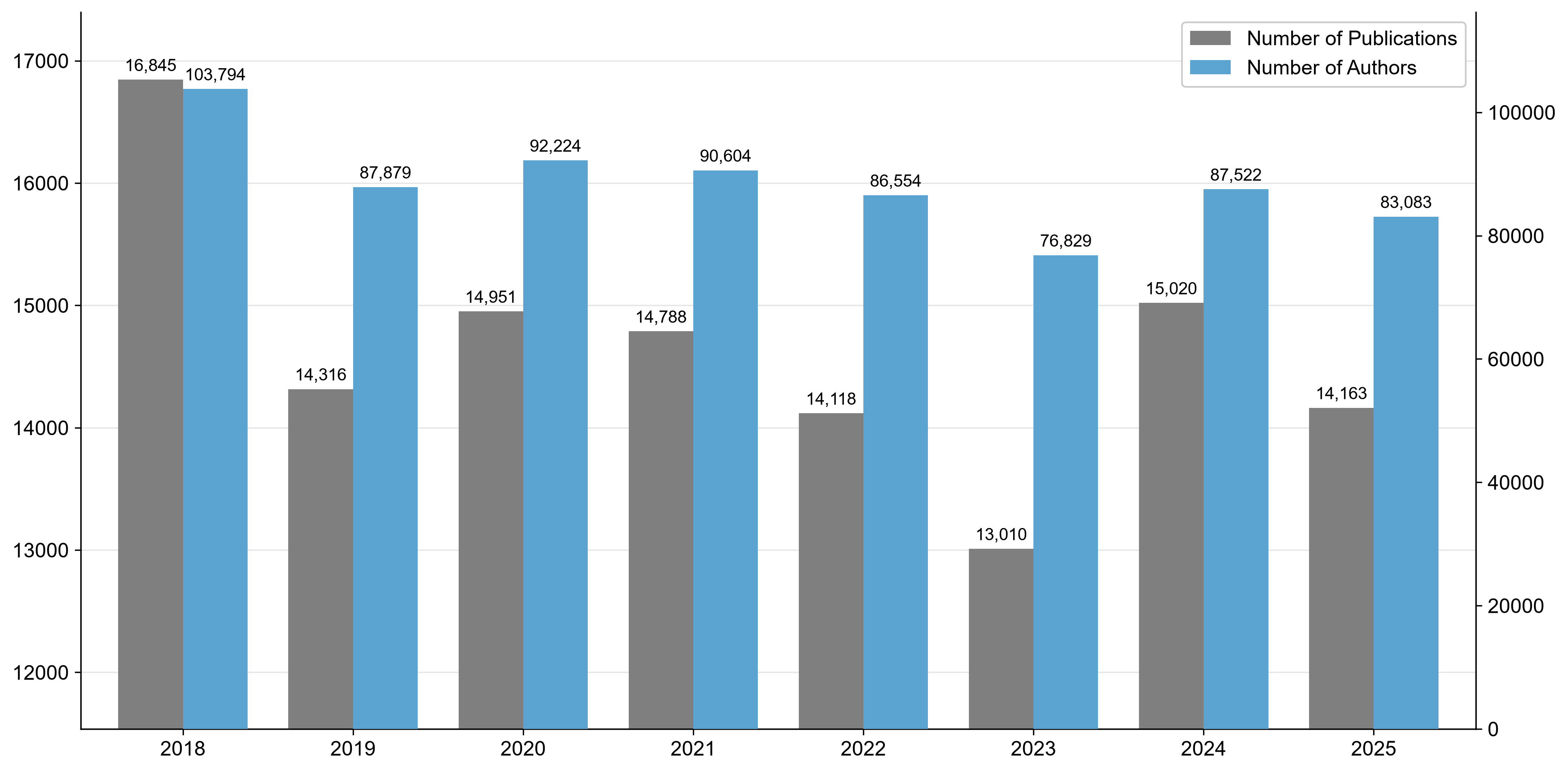}
     \caption{\textbf{Sample composition.} Annual number of research articles (gray bars, left axis) and unique contributing authors (blue bars, right axis) in the analytical sample, 2018--2025. Authors are counted once per publication year, irrespective of the number of papers they publish that year.}
    \label{fig:sample_composition}
\end{figure}

\vspace{1em}

\begin{figure}[h!]
    \makebox[\linewidth][r]{\hyperref[back:size_evolution]{$\hookleftarrow$}}
    \centering
    \includegraphics[width=\textwidth]{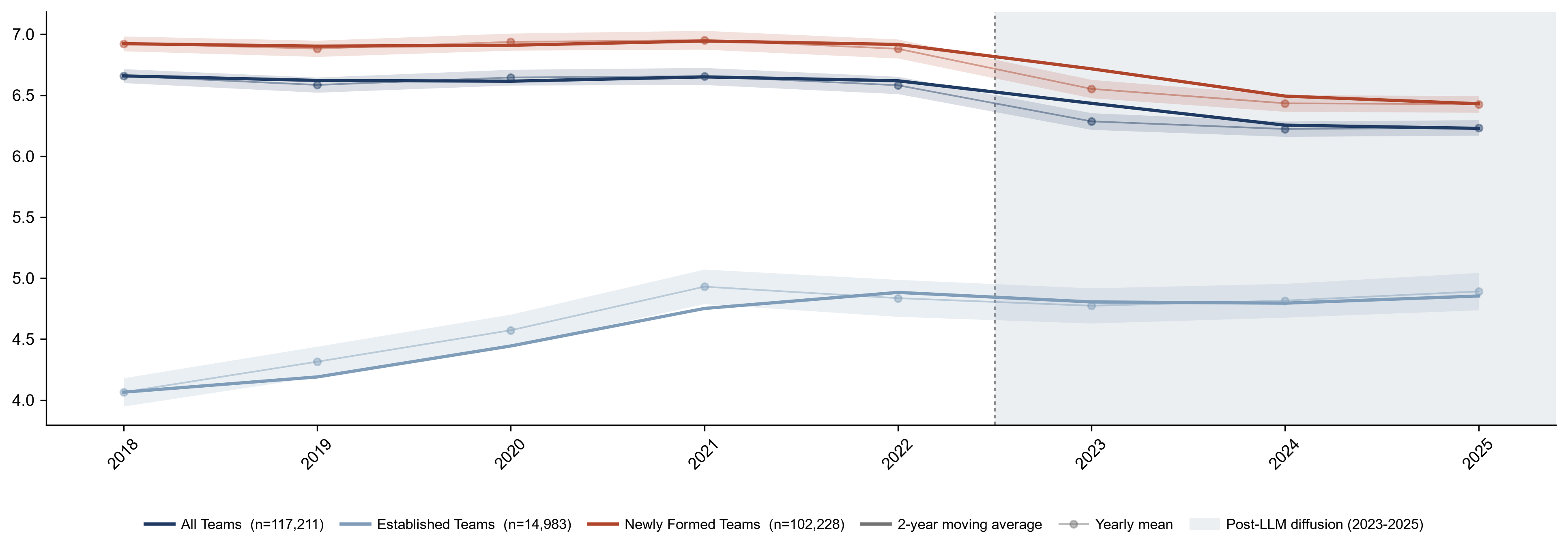}
    \caption{\textbf{Team size over time, by team type.} Mean number of
    authors per paper by year of publication, computed on the estimation sample ($n=117{,}211$). Thin lines with markers show the yearly means; thick lines show their two-year trailing moving average. Shaded bands are 95\% confidence intervals of the yearly means, and therefore describe the precision of those means rather than that of the smoothed line. The gray band marks the period following the launch of ChatGPT
    in late 2022.}
    \label{fig:size_evolution}
\end{figure}

\begin{table}[H]
\centering
\footnotesize
\caption{\textbf{Distribution of papers by primary domain and year.}}
\label{tab:domains_by_year}
\resizebox{\textwidth}{!}{%
\begin{tabular}{lrrrrrrrrrr}
\toprule
 & \multicolumn{8}{c}{Papers per year} & & \\
\cmidrule(lr){2-9}
 & 2018 & 2019 & 2020 & 2021 & 2022 & 2023 & 2024 & 2025 & Total & Share (\%) \\
\midrule
Health Sciences & 7,260 & 6,335 & 6,680 & 6,724 & 6,348 & 5,338 & 6,049 & 5,739 & 50,473 & 43.1 \\
Life Sciences & 5,299 & 4,262 & 4,058 & 3,279 & 2,949 & 2,731 & 3,076 & 2,755 & 28,409 & 24.2 \\
Physical Sciences & 2,699 & 2,175 & 2,361 & 2,445 & 2,243 & 2,250 & 3,035 & 3,206 & 20,414 & 17.4 \\
Social Sciences & 1,587 & 1,544 & 1,852 & 2,340 & 2,578 & 2,691 & 2,860 & 2,463 & 17,915 & 15.3 \\
\midrule
Total & 16,845 & 14,316 & 14,951 & 14,788 & 14,118 & 13,010 & 15,020 & 14,163 & 117,211 & 100.0 \\
\bottomrule
\end{tabular}
}
\end{table}

\vspace{1.2em}

\begin{table}[H]
\noindent\hfill\hyperref[back:fields_by_year]{$\hookleftarrow$}\par
\smallskip
\centering
\footnotesize
\caption{\textbf{Distribution of papers by primary field and year.}}
\vspace{1em}
\label{tab:fields_by_year}
\resizebox{\textwidth}{!}{%
\begin{tabular}{lrrrrrrrrrr}
\toprule
 & \multicolumn{8}{c}{Papers per year} & & \\
\cmidrule(lr){2-9}
 & 2018 & 2019 & 2020 & 2021 & 2022 & 2023 & 2024 & 2025 & Total & Share (\%) \\
\midrule
Medicine & 6,505 & 5,598 & 5,830 & 5,886 & 5,477 & 4,511 & 5,091 & 4,894 & 43,792 & 37.4 \\
Biochemistry, Genetics and Molecular Biology & 2,581 & 2,002 & 1,841 & 1,402 & 1,372 & 1,201 & 1,395 & 1,160 & 12,954 & 11.1 \\
Environmental Science & 1,255 & 1,060 & 1,037 & 964 & 819 & 756 & 983 & 901 & 7,775 & 6.6 \\
Agricultural and Biological Sciences & 1,319 & 1,104 & 1,128 & 974 & 783 & 742 & 836 & 821 & 7,707 & 6.6 \\
Psychology & 657 & 628 & 739 & 940 & 908 & 882 & 847 & 783 & 6,384 & 5.4 \\
Social Sciences & 568 & 506 & 634 & 804 & 949 & 885 & 962 & 864 & 6,172 & 5.3 \\
Engineering & 502 & 381 & 441 & 487 & 473 & 563 & 862 & 1,034 & 4,743 & 4.0 \\
Neuroscience & 790 & 628 & 597 & 529 & 459 & 481 & 515 & 454 & 4,453 & 3.8 \\
Health Professions & 399 & 404 & 489 & 490 & 536 & 515 & 610 & 551 & 3,994 & 3.4 \\
Computer Science & 326 & 253 & 303 & 320 & 407 & 427 & 624 & 738 & 3,398 & 2.9 \\
Immunology and Microbiology & 555 & 485 & 443 & 334 & 287 & 264 & 294 & 258 & 2,920 & 2.5 \\
Economics, Econometrics and Finance & 171 & 196 & 220 & 276 & 332 & 447 & 488 & 341 & 2,471 & 2.1 \\
Business, Management and Accounting & 65 & 72 & 110 & 160 & 198 & 278 & 351 & 271 & 1,505 & 1.3 \\
Mathematics & 73 & 57 & 200 & 345 & 219 & 198 & 169 & 147 & 1,408 & 1.2 \\
Nursing & 158 & 176 & 188 & 159 & 175 & 162 & 181 & 145 & 1,344 & 1.1 \\
Earth and Planetary Sciences & 221 & 189 & 175 & 149 & 139 & 137 & 146 & 154 & 1,310 & 1.1 \\
Decision Sciences & 90 & 90 & 90 & 104 & 140 & 135 & 148 & 135 & 932 & 0.8 \\
Dentistry & 95 & 69 & 75 & 103 & 86 & 78 & 89 & 94 & 689 & 0.6 \\
Physics and Astronomy & 125 & 89 & 72 & 84 & 73 & 69 & 93 & 74 & 679 & 0.6 \\
Veterinary & 103 & 88 & 98 & 86 & 74 & 72 & 78 & 55 & 654 & 0.6 \\
Materials Science & 101 & 81 & 70 & 58 & 45 & 53 & 77 & 77 & 562 & 0.5 \\
Arts and Humanities & 36 & 52 & 59 & 56 & 51 & 64 & 64 & 69 & 451 & 0.4 \\
Pharmacology, Toxicology and Pharmaceutics & 54 & 43 & 49 & 40 & 48 & 43 & 36 & 62 & 375 & 0.3 \\
Chemistry & 55 & 47 & 39 & 26 & 41 & 23 & 45 & 49 & 325 & 0.3 \\
Energy & 33 & 16 & 18 & 11 & 26 & 20 & 33 & 27 & 184 & 0.2 \\
Other & 8 & 2 & 6 & 1 & 1 & 4 & 3 & 5 & 30 & 0.0 \\
\midrule
Total & 16,845 & 14,316 & 14,951 & 14,788 & 14,118 & 13,010 & 15,020 & 14,163 & 117,211 & 100.0 \\
\bottomrule
\end{tabular}
}
\vspace{0.2em}
{\footnotesize
\justifying

\noindent\textit{Notes.} ``Other'' pools all primary fields with fewer than 50 papers; individually they are too thinly populated to support a separate field fixed effect.

\par

}

\end{table}

\begingroup
\singlespacing
\small
\renewcommand{\arraystretch}{1.4}
\noindent\makebox[\linewidth][r]{\hyperref[back:def_14_tasks]{$\hookleftarrow$}}%
\begin{longtable}{@{}p{0.27\textwidth} p{0.65\textwidth}@{}}
\caption{\textbf{The CRediT contributor roles taxonomy and role definitions.}}
\label{tab:credit}\\
\toprule
\textbf{Contributor Role} & \textbf{Role Definition} \\
\midrule
\endfirsthead
\multicolumn{2}{l}{\footnotesize\itshape Table \ref{tab:credit} (continued)}\\
\toprule
\textbf{Contributor Role} & \textbf{Role Definition} \\
\midrule
\endhead
\midrule
\multicolumn{2}{r}{\footnotesize\itshape continued on next page}\\
\endfoot
\bottomrule
\endlastfoot

\emph{Conceptualization} &
Ideas; formulation or evolution of overarching research goals and aims. \\

\emph{Methodology} &
Development or design of methodology; creation of models. \\

\emph{Software} &
Programming, software development; designing computer programs; implementation of the computer code and supporting algorithms; testing of existing code components. \\

\emph{Validation} &
Verification, whether as a part of the activity or separate, of the overall replication/reproducibility of results/experiments and other research outputs. \\

\emph{Formal Analysis} &
Application of statistical, mathematical, computational, or other formal techniques to analyze or synthesize study data. \\

\emph{Investigation} &
Conducting a research and investigation process, specifically performing the experiments, or data/evidence collection. \\

\emph{Resources} &
Provision of study materials, reagents, materials, patients, laboratory samples, animals, instrumentation, computing resources, or other analysis tools. \\

\emph{Data Curation} &
Management activities to annotate (produce metadata), scrub data and maintain research data (including software code) for initial use and later reuse. \\

\emph{Writing (Original Draft Preparation)} &
Creation and/or presentation of the published work, specifically writing the initial draft (including substantive translation). \\

\emph{Writing (Review \& Editing)} &
Preparation, creation and/or presentation of the published work by those from the original research group, specifically critical review, commentary or revision -- including pre- or post-publication stages. \\

\emph{Visualization} &
Preparation, creation and/or presentation of the published work, specifically visualization/data presentation. \\

\emph{Supervision} &
Oversight and leadership responsibility for the research activity planning and execution, including mentorship external to the core team. \\

\emph{Project Administration} &
Management and coordination responsibility for the research activity planning and execution. \\

\emph{Funding Acquisition} &
Acquisition of the financial support for the project leading to this publication. \\

\end{longtable}
\endgroup

\newpage
\begin{table}[H]
\centering
\makebox[\linewidth][r]{\hyperref[back:variables_table]{$\hookleftarrow$}}
\footnotesize
\setlength{\tabcolsep}{5pt}
\renewcommand{\arraystretch}{1.25}
\caption{\textbf{Variables used in the analysis: definition and source.}}
\vspace{1em}
\label{tab:variables}
\begin{tabularx}{\linewidth}{
  >{\raggedright\arraybackslash}p{3.6cm}
  >{\raggedright\arraybackslash}X
  >{\raggedright\arraybackslash}p{2.3cm}
}
\toprule
\textbf{Variable} & \textbf{Definition} & \textbf{Source} \\
\midrule
\multicolumn{3}{l}{\textit{Outcome variables} ($\Omega_{pt}$ in Equation~\eqref{eq:event_study})} \\
\midrule

\emph{Task Participation} ($P_{kpt}$) &
Number of co-authors credited with at least one CRediT task in research activity $k \in \{1,\dots,6\}$; an author credited with several tasks within an activity counts once. Enters as $\log(P_{kpt}+1)$. &
PLOS CRediT \\

\emph{Contributor Share} ($P^{\,s}_{kpt}$) &
$P_{kpt}$ divided by the number of co-authors: the fraction of the team credited with research activity $k$. &
PLOS CRediT \\

\emph{Number of Tasks} ($S_{ipt}$) &
Number of distinct research activities co-author $i$ is credited with. &
PLOS CRediT \\

\emph{Task Overlap} ($S_{pt}$) &
Mean pairwise Jaccard similarity of co-authors' task sets, measuring how much team members overlap in the activities they perform. &
PLOS CRediT \\

\midrule
\multicolumn{3}{l}{\textit{Time, treatment and team classification}} \\
\midrule

\emph{Publication Year} ($\tau$) &
Year of publication, defining the event-study timeline $\mathbf{1}(t=\tau)$; 2022 is the omitted reference year. &
OpenAlex \\

\emph{Post} ($\mathrm{Post}_{pt}$) &
Indicator equal to 1 for papers published in 2023--2025, used in the pooled difference specifications. &
OpenAlex \\

\emph{Team Type} &
\emph{Established} if the co-author configuration is reproduced exactly from a prior paper; \emph{newly formed} otherwise, pooling teams with no prior joint publication and those in which only a subset of members had previously collaborated. &
OpenAlex \\

\midrule
\multicolumn{3}{l}{\textit{Team characteristics} ($\mathbf{X}_{pt}$)} \\
\midrule

\emph{Team Size} &
Total number of co-authors on the paper; enters as $\log(x)$. &
OpenAlex \\

\emph{Int Collab} &
Intensity of international collaboration: number of distinct countries represented in the co-authors' affiliations. &
OpenAlex \\

\emph{Prior Pubs} &
Average number of publications accumulated by team members before year $t$. &
OpenAlex \\

\emph{Age} &
Average academic age of the team: years since members' first publication. &
OpenAlex \\

\midrule
\multicolumn{3}{l}{\textit{Field fixed effects}} \\
\midrule

\emph{Field} ($\gamma_f$) &
Scientific field of the paper (26 fields), entered as fixed effects to absorb time-invariant heterogeneity across disciplines. &
OpenAlex \\

\emph{Domain} &
OpenAlex primary domain (Health, Life, Physical and Social Sciences). &
OpenAlex \\

\bottomrule
\end{tabularx}

\vspace{0.25em}
{\footnotesize
\justifying

\noindent\textit{Notes.} The team characteristics in $\mathbf{X}_{pt}$ enter the regressions in $\log(x+1)$ form, except team size, which enters as $\log(x)$. The six research activities aggregate the fourteen CRediT roles;
Figure~\ref{fig:scientific_pipeline} gives the mapping. Author records reporting an academic age above 55 years are excluded from the paper-level average, to mitigate identity-matching errors in OpenAlex indexation.

\par
}

\end{table}

\section{Additional results}
\label{sec:appendix_results}

\subsection[Pooled interaction model]{%
Pooled interaction model\enspace
\hyperlink{back_pooled_interaction}{$\hookleftarrow$}}
\label{sec:appendix_pooled_interaction}

As a complement to the event-study, we also estimate a baseline pooled OLS specification to capture average shifts in task participation across task domains:

\begin{equation}
P_{kpt} = \alpha_k + \beta_{1k} \text{Post}_t + \mathbf{X}_{pt}'\boldsymbol{\theta}_k + \gamma_f + \varepsilon_{kpt},
\label{eq:pooled_ols_baseline}
\end{equation}

\noindent where $P_{kpt}$ denotes stage-level participation for research task $k \in \{1, \dots, 6\}$ on paper $p$ published in year $t$. The indicator $\text{Post}_t$ equals one for papers published in the post-LLM period ($t \ge 2023$, allowing sufficient time for adoption) and zero for the pre-LLM baseline period ($t \le 2019$ in the no-contamination sample, or $t < 2023$ in the full panel). The vector $\mathbf{X}_{pt}$ controls for team-level structural characteristics that could influence task assignment. We also include scientific field fixed effects ($\gamma_f$) (26 fields) to absorb unobserved, time-invariant heterogeneity in baseline task structures across disciplines. All specifications are estimated using OLS with heteroskedasticity-robust standard errors. In this specification, the parameter $\beta_{1k}$ measures the average post-shock reallocation in task participation across the research workflow.

We extend Equation~\eqref{eq:pooled_ols_baseline} by interacting $\text{Post}_t$ with team type to test whether organizational history moderates task reallocation. Formally:

\begin{equation}
P_{kpt} = \alpha_k + \beta_{1k} \text{Post}_t + \beta_{2k} \text{New}_{p} + \beta_{3k} (\text{Post}_t \times \text{New}_{p}) + \mathbf{X}_{pt}'\boldsymbol{\theta}_k + \gamma_f + \varepsilon_{kpt},
\label{eq:pooled_ols_interaction}
\end{equation}

\noindent where $\text{New}_{p}$ is a binary indicator equal to one if paper $p$ was produced by a newly formed research team, with established teams serving as the omitted reference category. For portfolio specialization outcomes ($S_{ipt}$ and $S_{pt}$), we estimate analogous specifications omitting the task domain index $k$.

The parameters in Equation~\eqref{eq:pooled_ols_interaction} directly test our hypotheses: $\beta_{1k}$ captures the baseline shift in task allocation within established teams following LLM diffusion; $\beta_{2k}$ controls for time-invariant baseline differences in task allocation between newly formed and established teams prior to LLM diffusion; and $\beta_{3k}$ is the primary coefficient of interest, capturing the differential adjustment of newly formed teams relative to established teams in response to LLMs. A negative and statistically significant interaction coefficient ($\beta_{3k} < 0$) indicates that newly formed teams compress participation or restructure division of labor more aggressively than established teams, validating the organizational friction mechanisms posited in our conceptual framework. 

All specifications are estimated on two analytical samples. The \textit{full sample} spans the complete panel of papers published between 2018 and 2025 ($t \in \{2018, \dots, 2025\}$). To insulate our estimates from confounding macro-shocks (e.g., COVID-19) and allow sufficient time for LLM diffusion across the sciences, we construct a \textit{``no-contamination'' sample} comparing the pre-LLM baseline period ($t \in \{2018, 2019\}$) directly against the post-diffusion period ($t \in \{2024, 2025\}$). By construction, this sample excludes both the COVID-19 pandemic years (2020--2021) and the initial adoption transition window (2022--2023). \\

\begin{table}[H]
\centering
\footnotesize
\setlength{\tabcolsep}{11pt}
\renewcommand{\arraystretch}{1.3}

\begin{minipage}{0.8\textwidth}
\centering

\begin{tabular}{lcccc}
\toprule
 & \multicolumn{2}{c}{\textbf{Full Period}} & \multicolumn{2}{c}{\textbf{No Contamination}} \\
 & \multicolumn{2}{c}{\textbf{(2018--2025)}} & \multicolumn{2}{c}{\textbf{(2018--19/2024--25)}} \\
\cmidrule(lr){2-3} \cmidrule(lr){4-5}
\textbf{Task Domain} & \textbf{Share} & \textbf{Count} & \textbf{Share} & \textbf{Count} \\
\midrule
Ideation                  & -0.017$^{***}$ & -0.021$^{***}$ & -0.022$^{***}$ & -0.028$^{***}$ \\
Data and Coding           & 0.005$^{**}$   & 0.011$^{**}$   & 0.018$^{***}$  & 0.036$^{***}$  \\
Execution                 & -0.017$^{***}$ & -0.032$^{***}$ & -0.022$^{***}$ & -0.043$^{***}$ \\
Writing                   & 0.000          & 0.004$^{*}$    & 0.008$^{***}$  & 0.016$^{***}$  \\
Coordination              & -0.006$^{***}$ & -0.015$^{***}$ & -0.007$^{**}$  & -0.020$^{***}$ \\
Funding and Resources     & -0.008$^{***}$ & -0.017$^{***}$ & -0.014$^{***}$ & -0.036$^{***}$ \\
\midrule
Controls            & Yes & Yes & Yes & Yes \\
Field fixed effects & Yes & Yes & Yes & Yes \\
Observations        & 117,211 & 117,211 & 60,344 & 60,344 \\
\bottomrule
\end{tabular}

\vspace{0.5em}

\caption{\textbf{Static baseline estimates of post-LLM shifts in task participation.}
Each cell reports the coefficient on the post-LLM indicator from a separate
regression of task-level participation on $\mathrm{Post}_t$, controls, and field
fixed effects, as in Equation~\eqref{eq:pooled_ols_baseline}.
`Share' denotes the share of unique authors contributing to the task;
`Count' denotes $\log(n+1)$ of contributing authors.
Robust standard errors;
$^{\dagger}p<0.10$, $^{*}p<0.05$, $^{**}p<0.01$, $^{***}p<0.001$.}

\label{tab:pooled_ols}

\end{minipage}
\end{table} 

\begin{table}[H]
\makebox[\linewidth][r]{\hyperref[back:participation_interactions]{$\hookleftarrow$}}\par
\vspace{2pt}
\centering
\footnotesize
\setlength{\tabcolsep}{12pt}
\renewcommand{\arraystretch}{1.3}

\begin{minipage}{0.75\textwidth}
\centering

\begin{tabular}{lcccc}
\toprule
 & \multicolumn{2}{c}{\textbf{Full Period}} & \multicolumn{2}{c}{\textbf{No Contamination}} \\
 & \multicolumn{2}{c}{\textbf{(2018--2025)}} & \multicolumn{2}{c}{\textbf{(2018--19/2024--25)}} \\
\cmidrule(lr){2-3} \cmidrule(lr){4-5}
\textbf{Term} & \textbf{Share} & \textbf{Count} & \textbf{Share} & \textbf{Count} \\
\midrule

\multicolumn{5}{l}{\textbf{Ideation}} \\
\hspace{1em}$\text{Post}$ & 0.001 & 0.004 & -0.003 & 0.004 \\
\hspace{1em}$\text{New}$ & -0.043$^{***}$ & -0.051$^{***}$ & -0.044$^{***}$ & -0.049$^{***}$ \\
\hspace{1em}$\text{Post} \times \text{New}$ & -0.021$^{***}$ & -0.029$^{***}$ & -0.022$^{***}$ & -0.037$^{***}$ \\

\addlinespace
\multicolumn{5}{l}{\textbf{Data and Coding}} \\
\hspace{1em}$\text{Post}$ & 0.002 & 0.001 & 0.019$^{**}$ & 0.031$^{**}$ \\
\hspace{1em}$\text{New}$ & -0.014$^{***}$ & -0.032$^{***}$ & -0.011$^{\dagger}$ & -0.029$^{**}$ \\
\hspace{1em}$\text{Post} \times \text{New}$ & 0.003 & 0.011 & -0.002 & 0.006 \\

\addlinespace
\multicolumn{5}{l}{\textbf{Execution}} \\
\hspace{1em}$\text{Post}$ & 0.002 & -0.003 & 0.007 & 0.001 \\
\hspace{1em}$\text{New}$ & -0.015$^{***}$ & -0.024$^{***}$ & -0.013$^{*}$ & -0.018$^{**}$ \\
\hspace{1em}$\text{Post} \times \text{New}$ & -0.022$^{***}$ & -0.034$^{***}$ & -0.033$^{***}$ & -0.051$^{***}$ \\

\addlinespace
\multicolumn{5}{l}{\textbf{Writing}} \\
\hspace{1em}$\text{Post}$ & 0.012$^{**}$ & 0.017$^{***}$ & 0.016$^{**}$ & 0.024$^{***}$ \\
\hspace{1em}$\text{New}$ & -0.038$^{***}$ & -0.047$^{***}$ & -0.045$^{***}$ & -0.056$^{***}$ \\
\hspace{1em}$\text{Post} \times \text{New}$ & -0.014$^{***}$ & -0.015$^{**}$ & -0.010$^{\dagger}$ & -0.010 \\

\addlinespace
\multicolumn{5}{l}{\textbf{Coordination}} \\
\hspace{1em}$\text{Post}$ & 0.010$^{*}$ & 0.016$^{*}$ & 0.009 & 0.009 \\
\hspace{1em}$\text{New}$ & -0.043$^{***}$ & -0.063$^{***}$ & -0.047$^{***}$ & -0.070$^{***}$ \\
\hspace{1em}$\text{Post} \times \text{New}$ & -0.019$^{***}$ & -0.036$^{***}$ & -0.019$^{**}$ & -0.035$^{**}$ \\

\addlinespace
\multicolumn{5}{l}{\textbf{Funding and Resources}} \\
\hspace{1em}$\text{Post}$ & -0.002 & 0.001 & -0.014$^{*}$ & -0.030$^{*}$ \\
\hspace{1em}$\text{New}$ & -0.024$^{***}$ & -0.035$^{***}$ & -0.029$^{***}$ & -0.044$^{***}$ \\
\hspace{1em}$\text{Post} \times \text{New}$ & -0.007 & -0.021$^{*}$ & -0.000 & -0.008 \\

\midrule
Controls & Yes & Yes & Yes & Yes \\
Field fixed effects & Yes & Yes & Yes & Yes \\
Observations & 117,211 & 117,211 & 60,344 & 60,344 \\
\bottomrule
\end{tabular}

\vspace{0.5em}

\caption[Team-type participation effects and post-LLM interactions]{%
\textbf{Team-type participation effects and post-LLM interactions.}
Each stage block reports parameter estimates from a separate regression of
stage-level participation on $\text{Post}$, $\text{New}$, their interaction,
paper-level controls, and field fixed effects, as in
Equation~\eqref{eq:pooled_ols_interaction}.
Established teams serve as the omitted reference category.
`Share' denotes the share of unique authors contributing to the task;
`Count' denotes $\log(n+1)$ of contributing authors.
Standard errors are heteroskedasticity-robust;
$^{\dagger}p<0.10$, $^{*}p<0.05$,
$^{**}p<0.01$, $^{***}p<0.001$.}

\label{tab:participations_interactions}

\end{minipage}
\end{table}

\begin{table}[H]
\makebox[\linewidth][r]{\hyperref[back:specialization_interactions]{$\hookleftarrow$}}\par
\vspace{2pt}
\centering
\footnotesize

\begin{minipage}{\textwidth}
\centering

\resizebox{\linewidth}{!}{%
\begin{tabular}{lcccccccc}
\toprule
 & \multicolumn{4}{c}{\textbf{Full Period}} 
 & \multicolumn{4}{c}{\textbf{No Contamination}} \\
 & \multicolumn{4}{c}{\textbf{(2018--2025)}} 
 & \multicolumn{4}{c}{\textbf{(2018--19/2024--25)}} \\
\cmidrule(lr){2-5} \cmidrule(lr){6-9}

 & \multicolumn{2}{c}{\textbf{Avg contrib./auth.}}
 & \multicolumn{2}{c}{\textbf{Avg Jaccard}}
 & \multicolumn{2}{c}{\textbf{Avg contrib./auth.}}
 & \multicolumn{2}{c}{\textbf{Avg Jaccard}} \\

\textbf{Term}
 & \textbf{Agg.} & \textbf{Inter.}
 & \textbf{Agg.} & \textbf{Inter.}
 & \textbf{Agg.} & \textbf{Inter.}
 & \textbf{Agg.} & \textbf{Inter.} \\
\midrule

Post
 & -0.0425$^{***}$ & 0.0247
 & -0.0059$^{***}$ & 0.0016
 & -0.0388$^{***}$ & 0.0328
 & -0.0029$^{\dagger}$ & 0.0039 \\
 & (0.0057) & (0.0152)
 & (0.0011) & (0.0031)
 & (0.0077) & (0.0218)
 & (0.0015) & (0.0045) \\

\addlinespace

New
 & & -0.1769$^{***}$
 & & -0.0404$^{***}$
 & & -0.1885$^{***}$
 & & -0.0442$^{***}$ \\
 & & (0.0103)
 & & (0.0021)
 & & (0.0173)
 & & (0.0035) \\

\addlinespace

Post $\times$ New
 & & -0.0791$^{***}$
 & & -0.0090$^{**}$
 & & -0.0868$^{***}$
 & & -0.0089$^{\dagger}$ \\
 & & (0.0163)
 & & (0.0034)
 & & (0.0232)
 & & (0.0048) \\

\midrule
Controls
 & Yes & Yes & Yes & Yes & Yes & Yes & Yes & Yes \\

Field fixed effects
 & Yes & Yes & Yes & Yes & Yes & Yes & Yes & Yes \\

Observations
 & 117,211 & 117,211 & 117,211 & 117,211
 & 60,344 & 60,344 & 60,344 & 60,344 \\

$R^2$
 & 0.1511 & 0.1558 & 0.0453 & 0.0512
 & 0.1460 & 0.1514 & 0.0444 & 0.0511 \\

\bottomrule
\end{tabular}%
}

\vspace{0.5em}

\captionsetup{
    font=footnotesize,
    justification=justified,
    singlelinecheck=false
}

\caption[Post-LLM changes in task portfolios and overlap by team type]{%
\textbf{Post-LLM changes in task portfolios and overlap by team type.}
OLS estimates for the average number of task domains per author and mean
pairwise Jaccard overlap. Aggregate specifications (Agg.) estimate the
full-sample post-period change; interaction specifications (Inter.)
distinguish established teams, the reference category, from newly formed
configurations, including those containing previously collaborating subsets.
The coefficient on Post $\times$ New measures the difference in post-period
changes between team types. Results are reported for the full sample and
the restricted sample excluding potential contamination. All specifications
include team-level controls and field fixed effects.
Heteroskedasticity-robust standard errors are reported in parentheses.
$^{\dagger}p<0.10$, $^{*}p<0.05$,
$^{**}p<0.01$, $^{***}p<0.001$.}

\label{tab:specialization_interactions}

\end{minipage}
\end{table}

\subsection{Estimates of the number of contributors}
\label{subsec:log_count}

\begin{figure}[H]
    \noindent\hfill\hyperref[back:logcount]{$\hookleftarrow$}\par
    \vspace{1\baselineskip}
    \centering
    \includegraphics[width=1\linewidth]{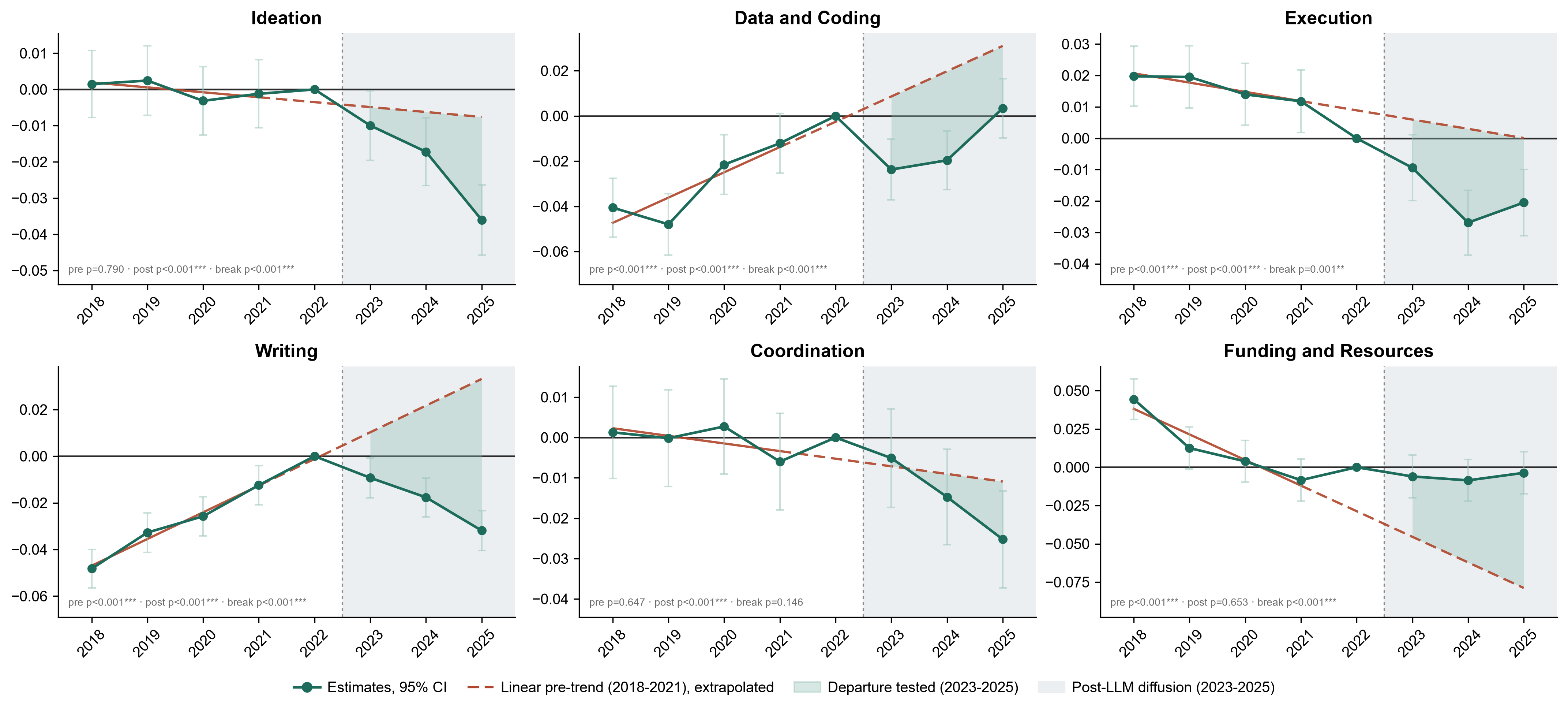}
    \caption{\textbf{Event-study estimates of within-task participation before and after LLM diffusion.} Estimated changes in the log number of co-authors credited with each research activity within a paper, relative to 2022, estimated using Equation~\eqref{eq:event_study} with field fixed effects and team-level controls ($n=117{,}211$). Error bars show 95\% confidence intervals based on robust standard errors. The gray band marks the period following the launch of ChatGPT in late 2022. The red dashed line extrapolates the linear trend estimated over
    2018--2021; the shaded area between this trend and the post-2022 estimates illustrates the deviations assessed by the trend-break test. Each panel reports joint Wald tests assessing whether the pre-period (2018–2021) and post-period (2023–2025) coefficients are jointly equal
    to zero relative to 2022, along with the trend-break test. Significance level: $^{\dagger}p<0.10$, $^{*}p<0.05$, $^{**}p<0.01$,
    $^{***}p<0.001$.}
    \label{fig:event_study_aggregate_count}
\end{figure}

\begin{figure}[H]
    \centering
    \includegraphics[width=\linewidth]{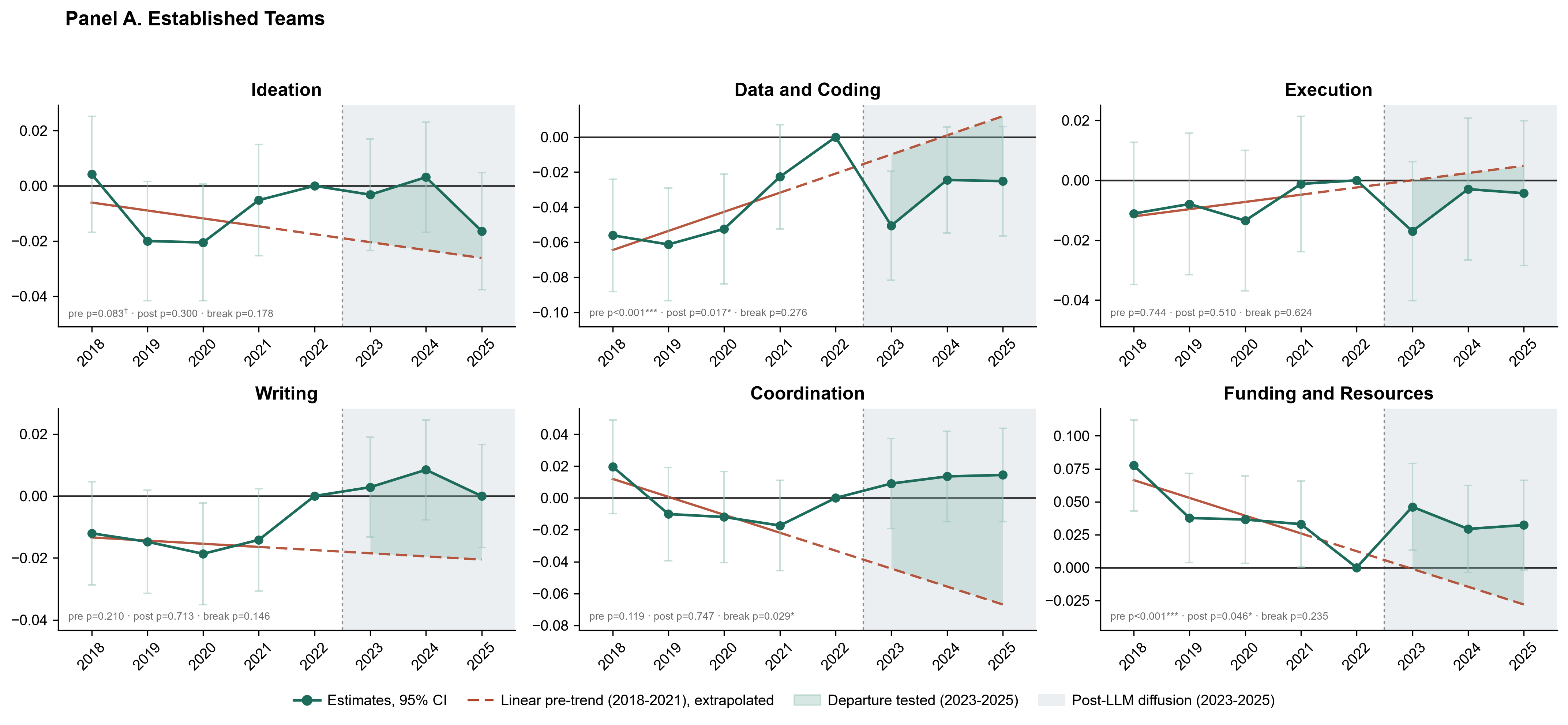}
    \par\vspace{1em}
    \includegraphics[width=\linewidth]{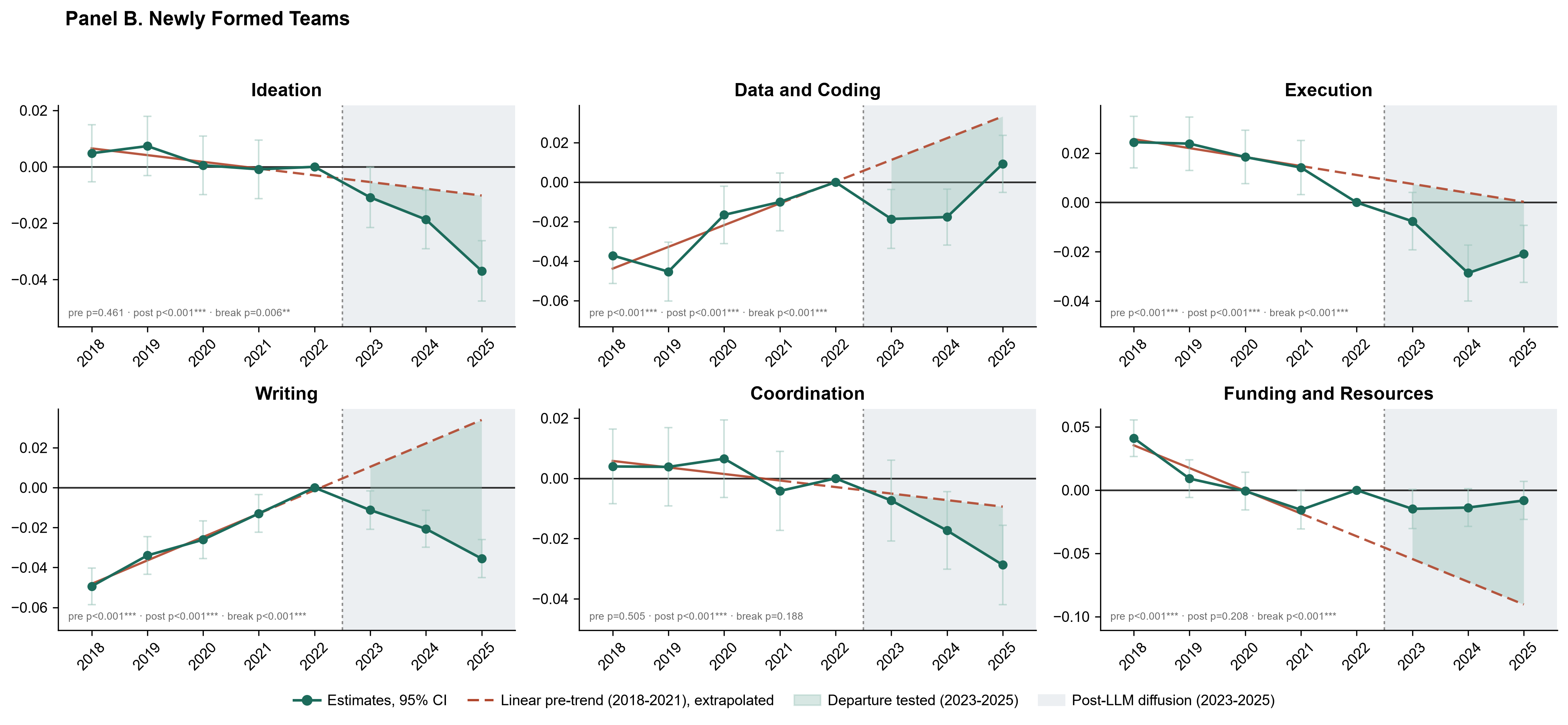}
      \caption{\textbf{Event-study estimates of within-task participation by team type.} Estimated changes in the log number of co-authors credited with each research activity within a paper, relative to 2022, using Equation~\eqref{eq:event_study} with field fixed effects and team-level controls. Panel A presents results for established teams ($n=14{,}983$); Panel B presents results for newly formed teams ($n=102{,}228$). Error bars show 95\% confidence intervals based on robust standard errors. The gray band marks the period following the launch of ChatGPT in late 2022. The red dashed lines extrapolate linear trends estimated over 2018--2021; the shaded areas between these trends and the post-2022 estimates illustrate the deviations assessed by the trend-break tests. Each panel reports joint Wald tests assessing whether the pre-period (2018--2021) and post-period (2023--2025) coefficients are jointly equal to zero relative to 2022, along with the trend-break test. Significance levels: $^{\dagger}p<0.10$, $^{*}p<0.05$, $^{**}p<0.01$, $^{***}p<0.001$.}
    \label{fig:event_study_team_type_count}
\end{figure}

\subsection{Detrended estimates}

\begin{figure}[H]
    \centering
    \includegraphics[width=1\linewidth]{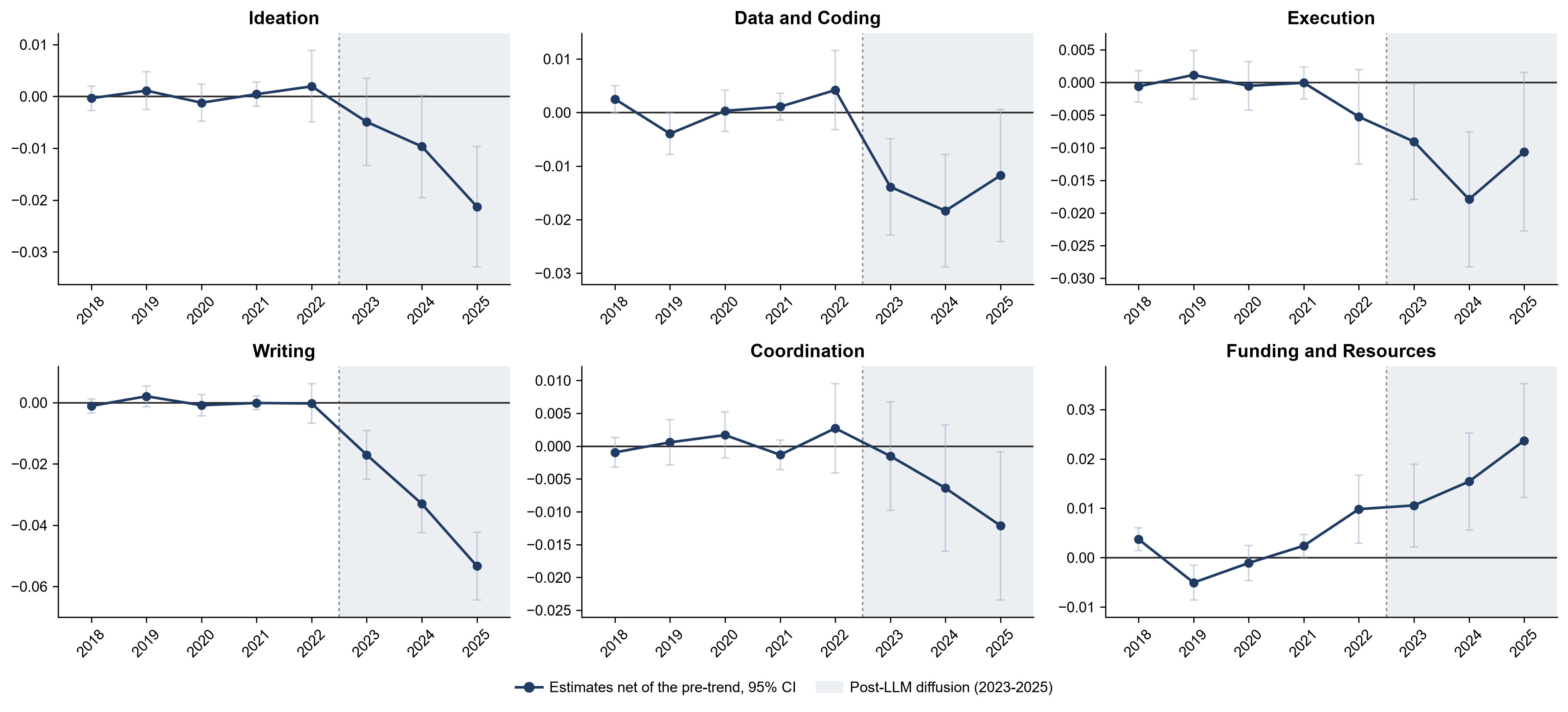}
   \caption{\textbf{Event-study estimates of within-task participation net of the pre-LLM trend.} Estimated changes in the share of co-authors credited with each research activity within a paper, net of the extrapolated pre-LLM trend, estimated using Equation~\eqref{eq:event_study} with field fixed effects and team-level controls ($n=117{,}211$). Each point is the year coefficient minus the linear trend estimated over 2018--2021 and extrapolated to that year, so the 2018--2021 points are the residuals from that line and the 2023--2025 points are the departures assessed by the trend-break test. Error bars show 95\% confidence intervals based on robust standard errors, computed as exact linear contrasts on the coefficient vector underlying Figure~\ref{fig:event_study_aggregate}; they widen with distance from the pre-period because the extrapolated trend is itself estimated. The gray band marks the period following the launch of ChatGPT in late 2022. This view assumes the pre-LLM trend would otherwise have continued linearly. Joint Wald tests and trend-break tests are reported in Figure~\ref{fig:event_study_aggregate}.}
    \label{fig:event_study_aggregate_share_detrended}
\end{figure}

\begin{figure}[H]
    \centering
    \includegraphics[width=\linewidth]{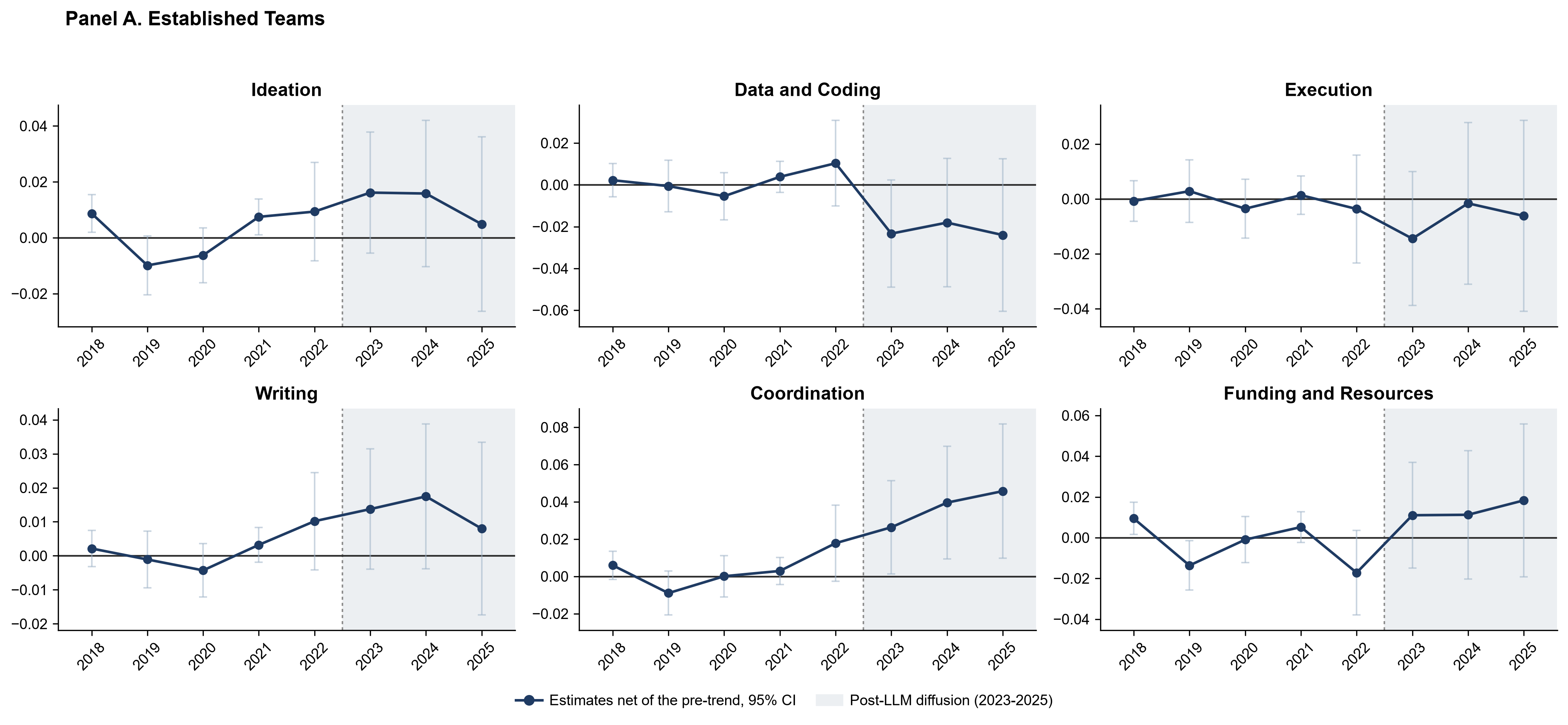}
    \par\vspace{1em}
    \includegraphics[width=\linewidth]{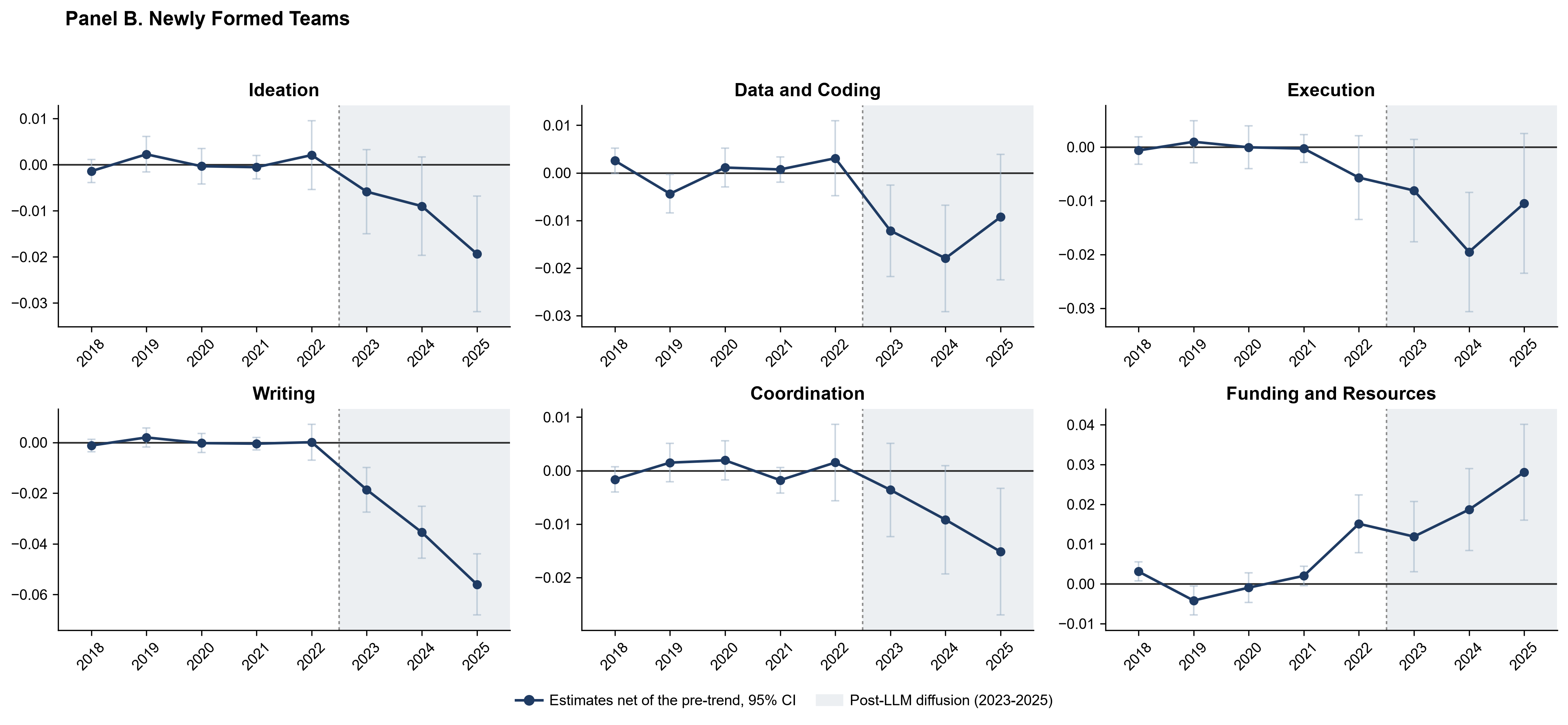}
    \caption{\textbf{Event-study estimates of within-task participation by team type, net of the pre-LLM trend.} Estimated changes in the share of co-authors credited with each research activity within a paper, net of the extrapolated pre-LLM trend, using Equation~\eqref{eq:event_study} with field fixed effects and team-level controls. Panel A presents results for established teams whose exact, complete co-author configuration had appeared previously ($n=14{,}983$); Panel B presents results for newly formed configurations, including those
    containing previously collaborating subsets ($n=102{,}228$). Each point is the year coefficient minus the linear trend estimated over 2018--2021 and extrapolated to that year, so the 2018--2021 points are the residuals from that line and the 2023--2025 points are the departures assessed by the trend-break test. Error bars show 95\% confidence intervals based on robust standard errors, computed as exact linear contrasts on the coefficient vector underlying Figure~\ref{fig:event_study_team_type}; they widen with distance from the pre-period because the extrapolated trend is itself estimated. The gray band marks the period following the launch of ChatGPT in late 2022. This view assumes the pre-LLM trends would otherwise have continued linearly. Joint Wald tests and trend-break tests are reported in Figure~\ref{fig:event_study_team_type}.}
    \label{fig:event_study_team_type_share_detrended}
\end{figure}

\begin{figure}[H]
    \centering
    \includegraphics[width=1\linewidth]{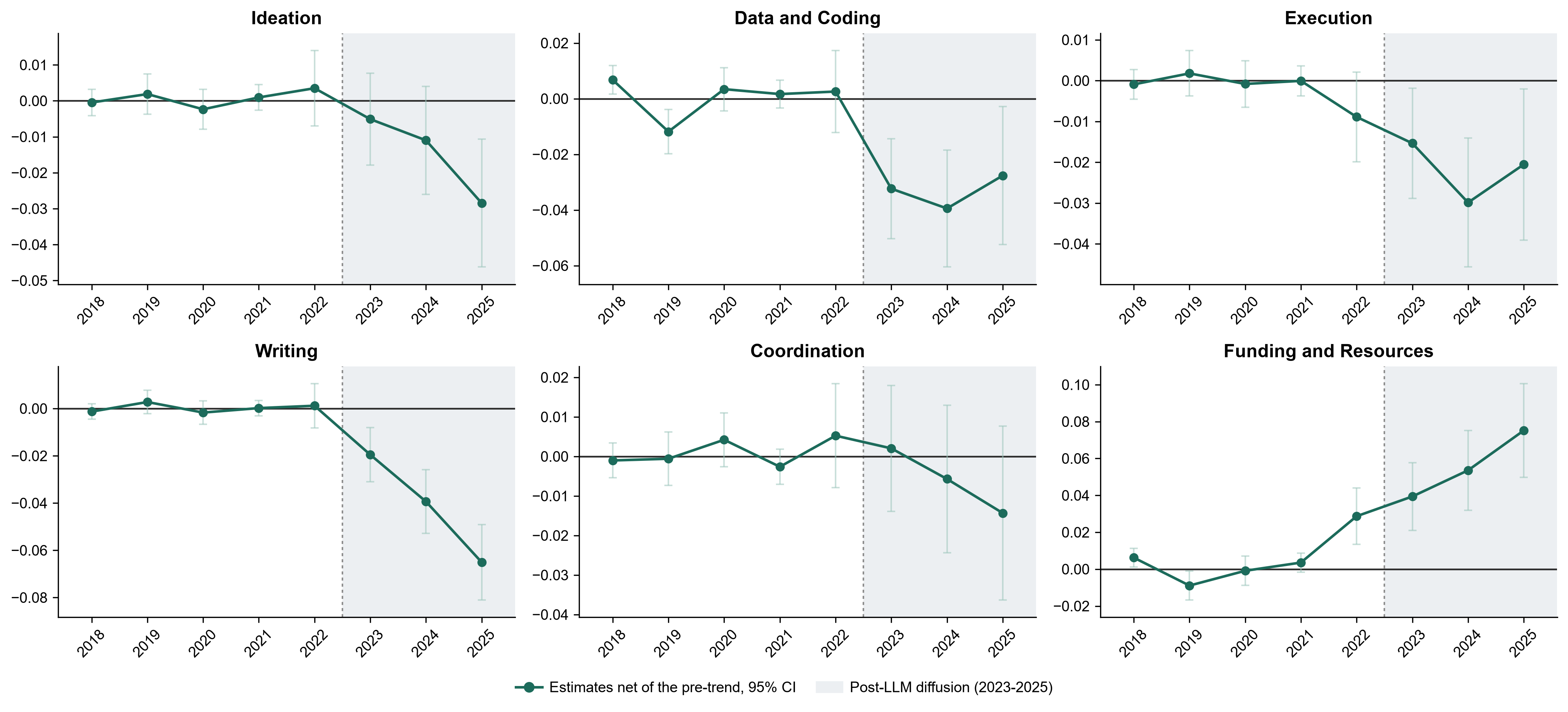}
     \caption{\textbf{Event-study estimates of within-task participation net of
the pre-LLM trend.} Estimated changes in the log number of co-authors credited with each research activity within a paper, net of the extrapolated pre-LLM trend, estimated using Equation~\eqref{eq:event_study} with field fixed effects and team-level controls ($n=117{,}211$). Each point is the year
coefficient minus the linear trend estimated over 2018--2021 and extrapolated to that year, so the 2018--2021 points are the residuals from that line and the 2023--2025
points are the departures assessed by the trend-break test. Error bars show 95\% confidence intervals based on robust standard errors, computed as exact linear contrasts on the coefficient vector underlying Appendix
Figure~\ref{fig:event_study_aggregate_count}; they widen with distance from the pre-period because the extrapolated trend is itself estimated. The gray band marks the period following the launch of ChatGPT in late 2022. This view assumes the pre-LLM trend would otherwise have continued linearly. Joint Wald tests and trend-break tests are reported in Appendix Figure~\ref{fig:event_study_aggregate_count}.}
    \label{fig:event_study_aggregate_count_detrended}
\end{figure}

\begin{figure}[H]
    \centering
    \includegraphics[width=\linewidth]{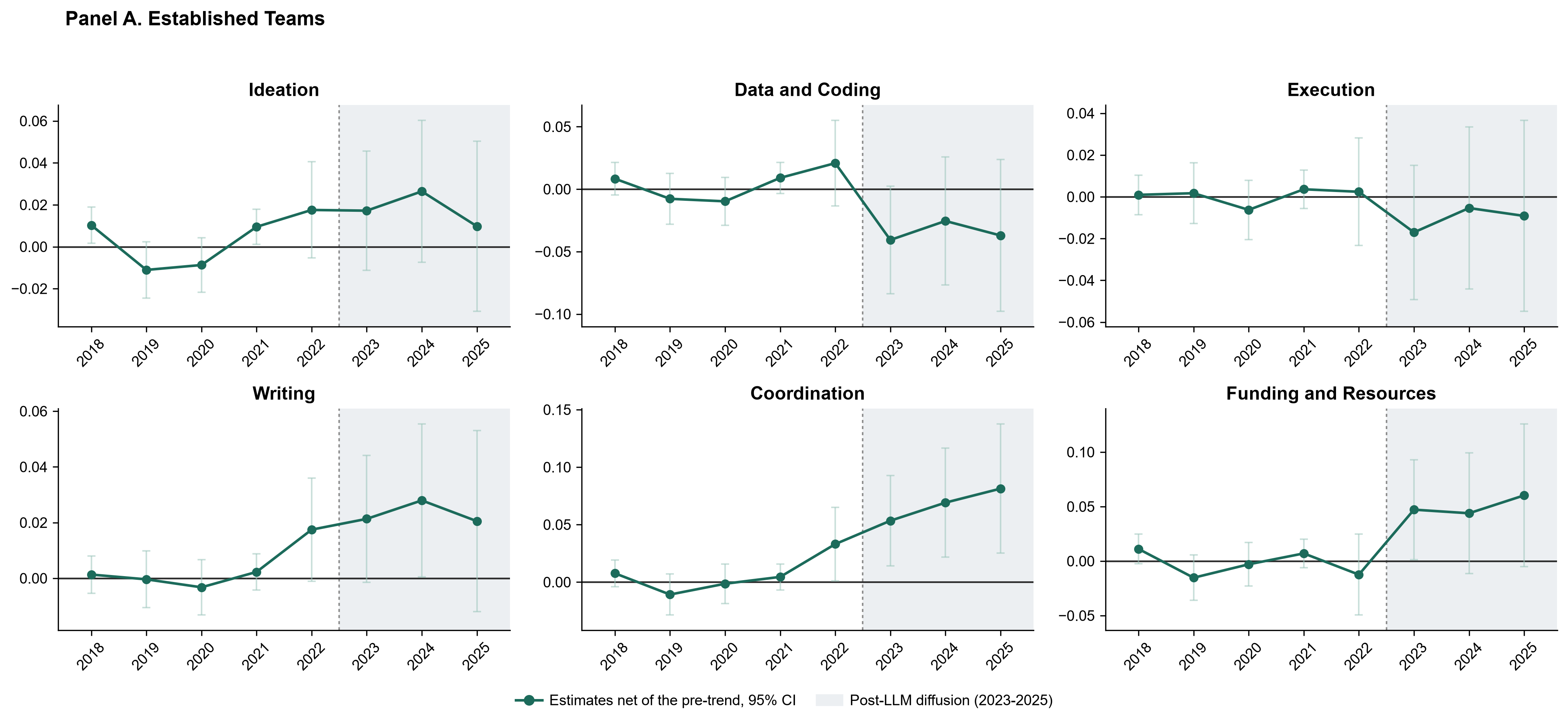}
    \par\vspace{1em}
    \includegraphics[width=\linewidth]{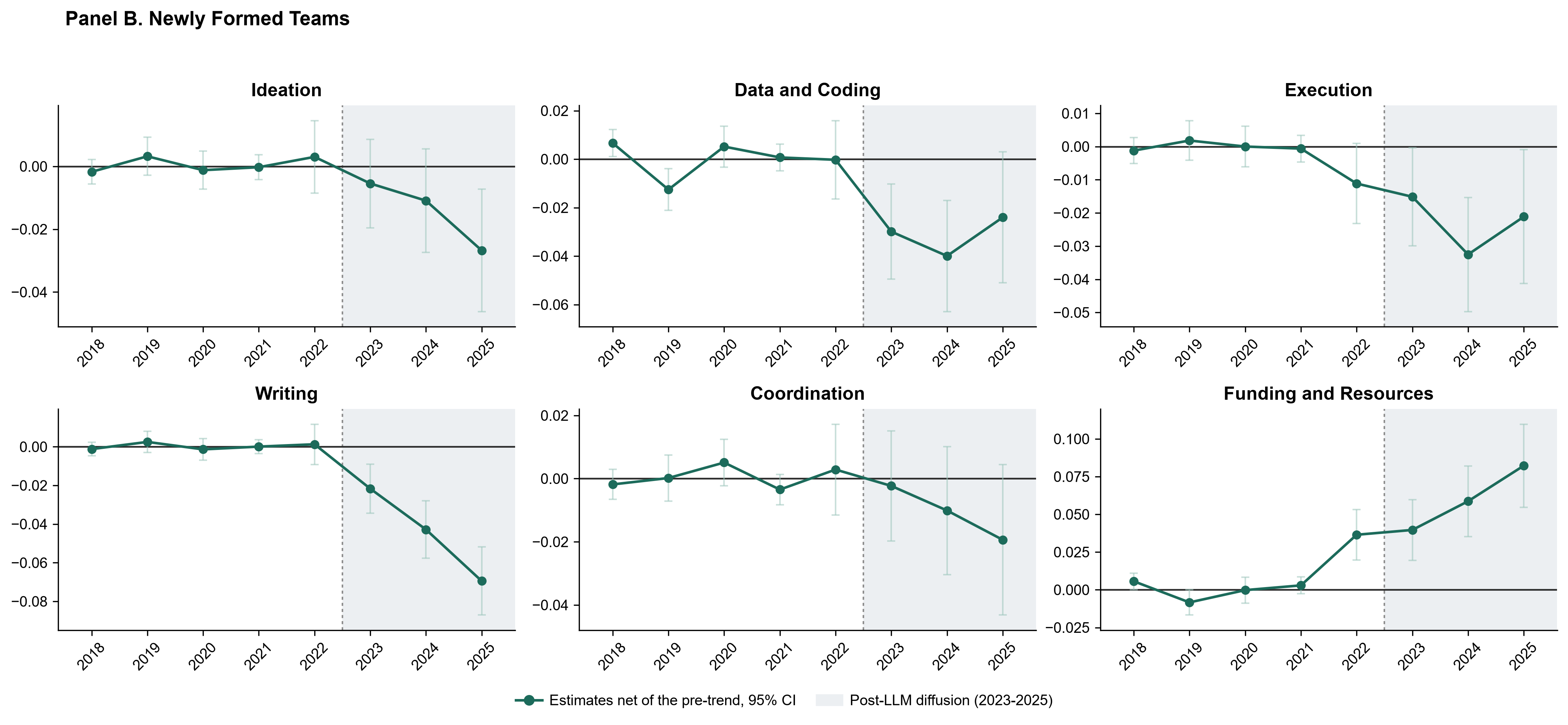}
    \caption{\textbf{Event-study estimates of within-task participation by team type, net of the pre-LLM trend.} Estimated changes in the log number of co-authors credited with each research activity within a paper, $\log(n{+}1)$, net of the extrapolated pre-LLM trend, using Equation~\eqref{eq:event_study} with field fixed effects and team-level controls. Panel A presents results for established teams whose exact, complete co-author configuration had appeared previously ($n=14{,}983$); Panel B presents results for newly formed configurations, including those
    containing previously collaborating subsets ($n=102{,}228$). Each point is the year coefficient minus the linear trend estimated over 2018--2021 and extrapolated to that year, so the 2018--2021 points are the residuals from that line and the 2023--2025 points are the departures assessed by the trend-break test. Error bars show 95\% confidence intervals based on robust standard errors, computed as exact linear contrasts on the coefficient vector underlying Figure~\ref{fig:event_study_team_type_count}; they widen with distance from the pre-period because the extrapolated trend is itself estimated. The gray band marks the period following the launch of ChatGPT in late 2022. This view assumes the pre-LLM trends would otherwise have
    continued linearly. Joint Wald tests and
    trend-break tests are reported in Figure~\ref{fig:event_study_team_type_count}. Each panel has its own vertical scale.}
    \label{fig:event_study_team_type_count_detrended}
\end{figure}

\begin{figure}[H]
    \centering
    \includegraphics[width=1\linewidth]{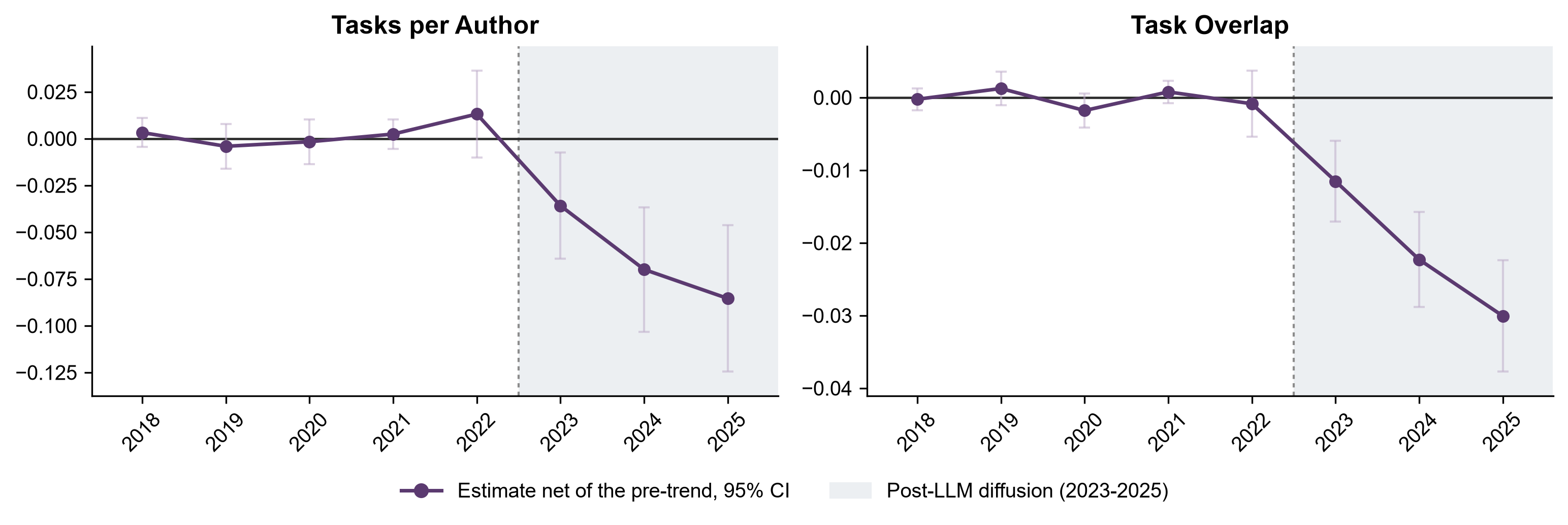}
   \caption{\textbf{Event-study estimates of task specialization net of the
    pre-LLM trend.} Estimated changes in the breadth of individual portfolios
    (average number of the six aggregated research activities credited per
    author, left panel) and task overlap (average pairwise Jaccard similarity of co-authors' task sets within a paper, right panel), net of the extrapolated pre-LLM trend, using Equation~\eqref{eq:event_study} with field fixed
    effects and team-level controls ($n=117{,}211$). Each point is the year
    coefficient minus the linear trend estimated over 2018--2021 and
    extrapolated to that year, so the 2018--2021 points are the residuals from that line and the 2023--2025 points are the departures assessed by the
    trend-break test. Error bars show 95\% confidence intervals based on robust
    standard errors, computed as exact linear contrasts on the coefficient
    vector underlying Figure~\ref{fig:event_study_specialisation}; they widen with distance from the pre-period because the extrapolated trend is itself estimated. The gray band marks the period following the launch of ChatGPT in
    late 2022. This view assumes the pre-LLM trend would otherwise have
    continued linearly. Joint Wald tests and trend-break tests are reported in Figure~\ref{fig:event_study_specialisation}.}
    \label{fig:specialisation_detrended}
\end{figure}

\begin{figure}[H]
    \centering
    \includegraphics[width=1\linewidth]{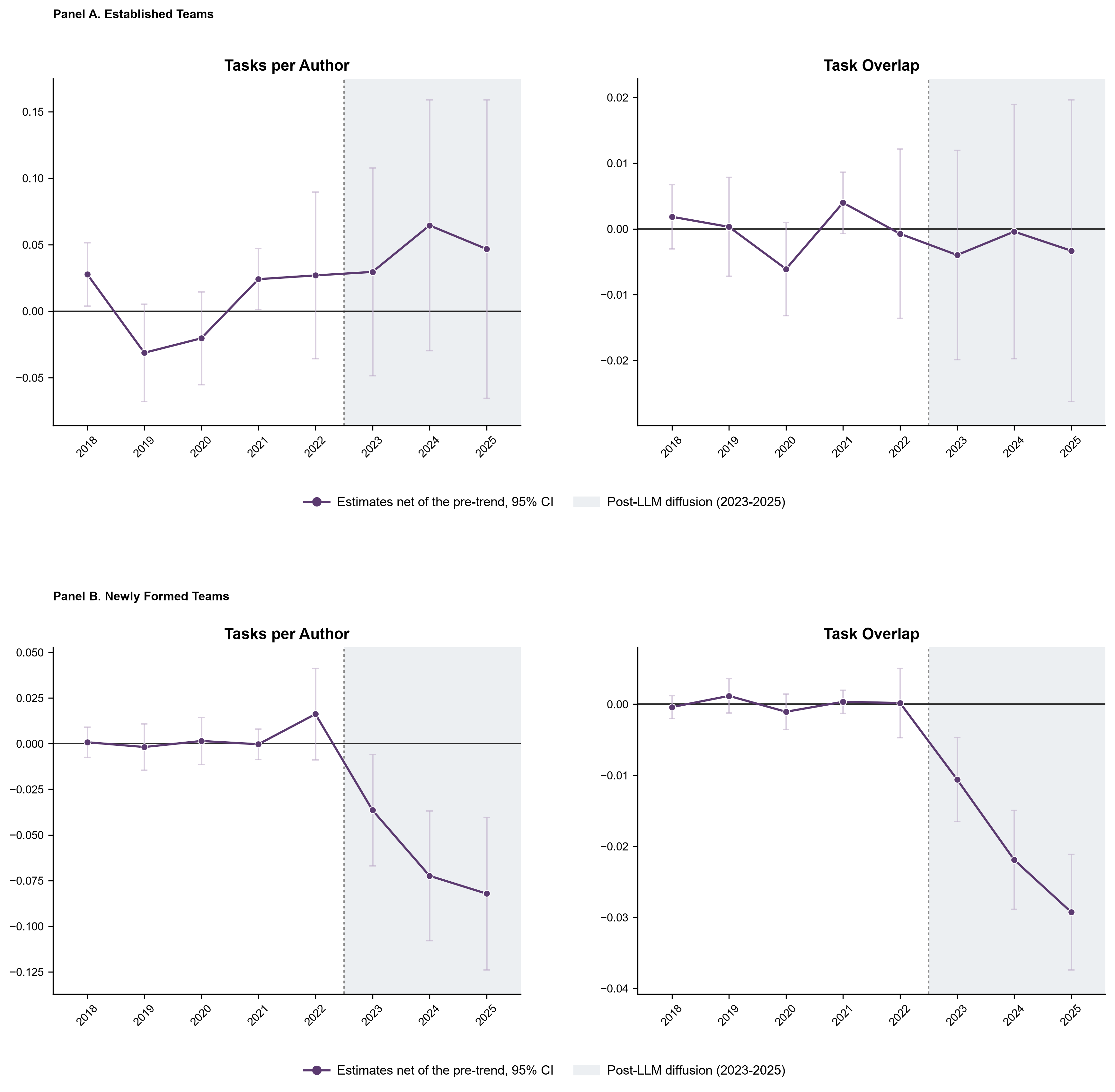}
    \caption{\textbf{Event-study estimates of task specialization by team type, net of the pre-LLM trend.} Estimated changes in the breadth of individual portfolios (average number of the six aggregated research activities  credited per author, left panel) and task overlap (average pairwise Jaccard
    similarity of co-authors' task sets within a paper, right panel), net of the
    extrapolated pre-LLM trend, using Equation~\eqref{eq:event_study} with field
    fixed effects and team-level controls. Panel A presents results for
    established teams whose exact, complete co-author configuration had appeared
    previously ($n=14{,}983$); Panel B presents results for newly formed  configurations, including those containing previously collaborating subsets ($n=102{,}228$). Each point is the year coefficient minus the linear trend estimated over 2018--2021 and extrapolated to that year, so the 2018--2021
    points are the residuals from that line and the 2023--2025 points are the
    departures assessed by the trend-break test. Error bars show 95\% confidence intervals based on robust standard errors, computed as exact linear contrasts on the coefficient vector underlying Figure~\ref{fig:event_study_specialisation}; they widen with distance from
    the pre-period because the extrapolated trend is itself estimated. The gray
    band marks the period following the launch of ChatGPT in late 2022. This
    view assumes the pre-LLM trends would otherwise have continued linearly.
    Joint Wald tests and
    trend-break tests are reported in Figure~\ref{fig:event_study_specialisation}. }
    \label{fig:specialisation_teamtype_detrended}
\end{figure}

\subsection[Analysis of the 14 CRediT contributor roles]{%
Analysis of the 14 CRediT contributor roles\enspace
\texorpdfstring{%
  \hyperref[back:14tasks:a]{$\hookleftarrow_{1}$}\,%
  \hyperref[back:14tasks:b]{$\hookleftarrow_{2}$}}{}}
\label{subsec:14tasks}

\begin{figure}[H]
    \centering
    \includegraphics[width=1\linewidth]{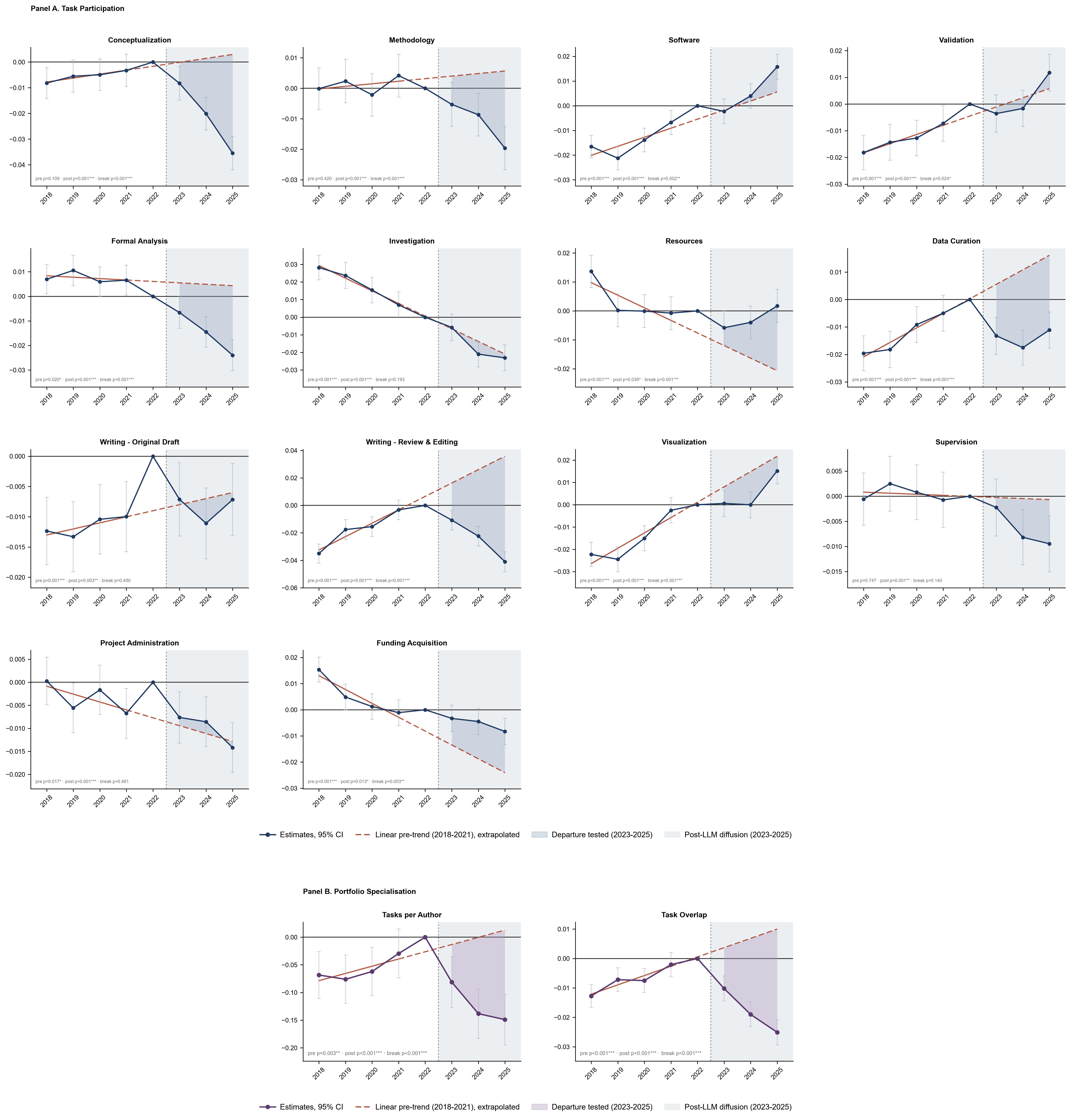}
    \par\vspace{1em}
   \caption{\textbf{Event-study estimates of task participation and portfolio specialization across the 14 CRediT roles.} Estimated changes relative to 2022, estimated using Equation~\eqref{eq:event_study} with field fixed effects and team-level controls ($n=117{,}211$). Panel A reports the share of a paper's authors credited with each of the 14 CRediT contributor roles, analyzed separately rather than grouped into the six research stages used in the main analysis. Panel B reports the two portfolio measures computed over the same 14 roles: the average number of roles per author, and the average pairwise overlap between co-authors' sets of roles. Error bars show 95\% confidence intervals based on robust standard errors. The gray band marks the period following the launch of ChatGPT in late 2022. The red dashed lines extrapolate the linear trends estimated over 2018--2021; the shaded areas between these trends and the post-2022 estimates illustrate the deviations assessed by the trend-break tests. Each panel reports joint Wald tests assessing whether the pre-period (2018--2021) and post-period (2023--2025) coefficients are jointly equal to zero relative to 2022, along with the trend-break test. Significance level: $^{\dagger}p<0.10$, $^{*}p<0.05$, $^{**}p<0.01$, $^{***}p<0.001$.}
    \label{fig:all14_panels}
\end{figure}

\begin{figure}[H]
    \makebox[\linewidth][r]{\hyperref[back:14_teamtype]{$\hookleftarrow$}}\par
    \vspace{2pt}
    \centering
    \includegraphics[width=1\linewidth]{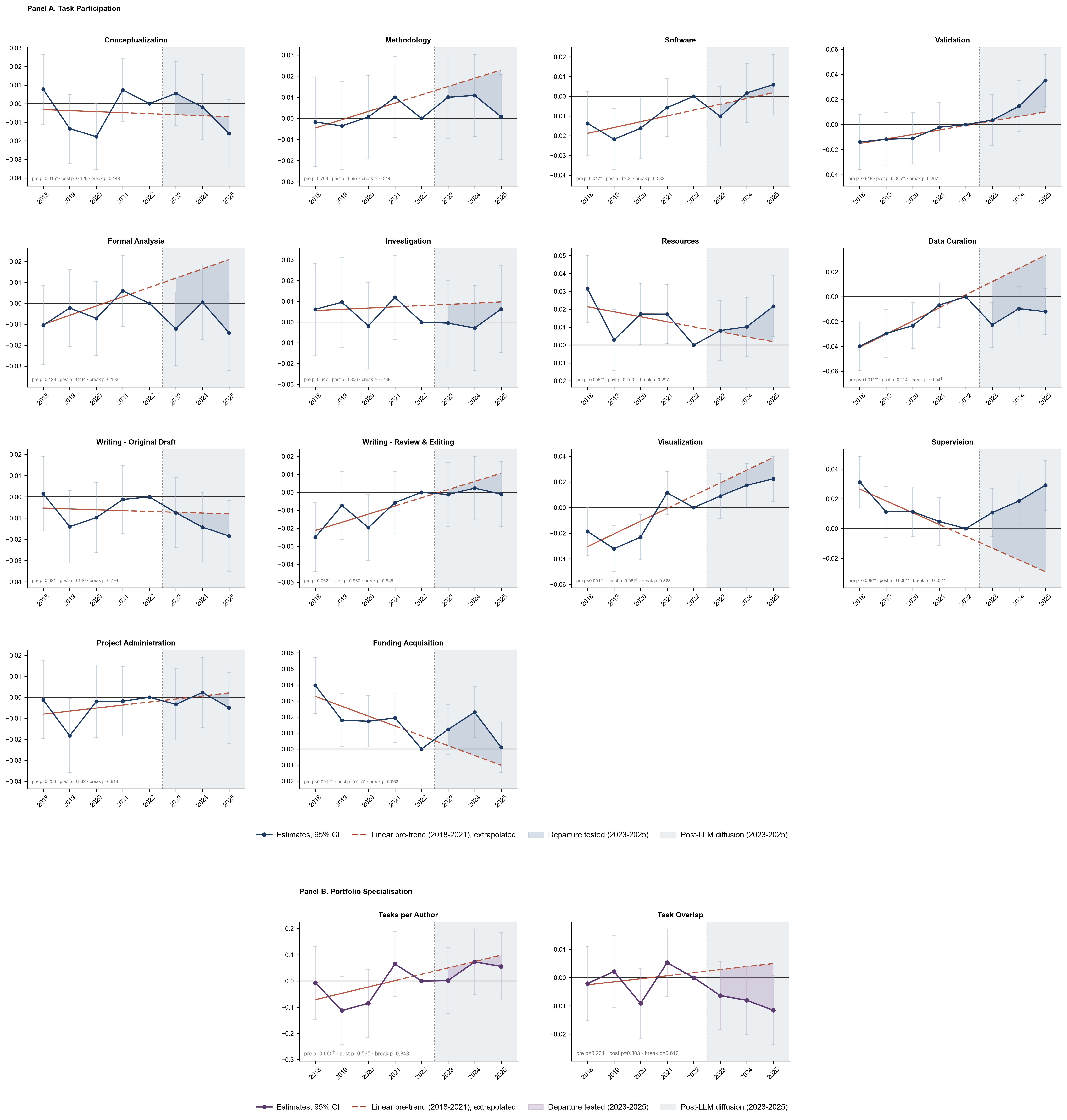}
    \par\vspace{1em}

   \caption{\textbf {Event-study estimates of task participation and portfolio specialization across the 14 CRediT roles, established teams.} Estimated changes relative to 2022, estimated using Equation~\eqref{eq:event_study}
   with field fixed effects and team-level controls, for established teams whose exact, complete co-author configuration had appeared previously ($n=14{,}983$). Panel A reports the share of a paper's authors credited with each of the 14 CRediT contributor roles, analyzed separately rather than
   grouped into the six research stages used in the main analysis. Panel B reports the two portfolio measures computed over the same 14 roles: the average number of roles per author, and the average pairwise overlap between co-authors' sets of roles. Error bars show 95\% confidence intervals based on robust standard errors. The gray band marks the period following the launch  of ChatGPT in late 2022. The red dashed lines extrapolate the linear trends
   estimated over 2018--2021; the shaded areas between these trends and the post-2022 estimates illustrate the deviations assessed by the trend-break tests. Each panel reports joint Wald tests assessing whether the pre-period
   (2018--2021) and post-period (2023--2025) coefficients are jointly equal to zero relative to 2022, along with the trend-break test. Significance level: $^{\dagger}p<0.10$, $^{*}p<0.05$, $^{**}p<0.01$, $^{***}p<0.001$.}
    \label{fig:existing14_panels}
\end{figure}

\begin{figure}[H]
    \centering
    \includegraphics[width=1\linewidth]{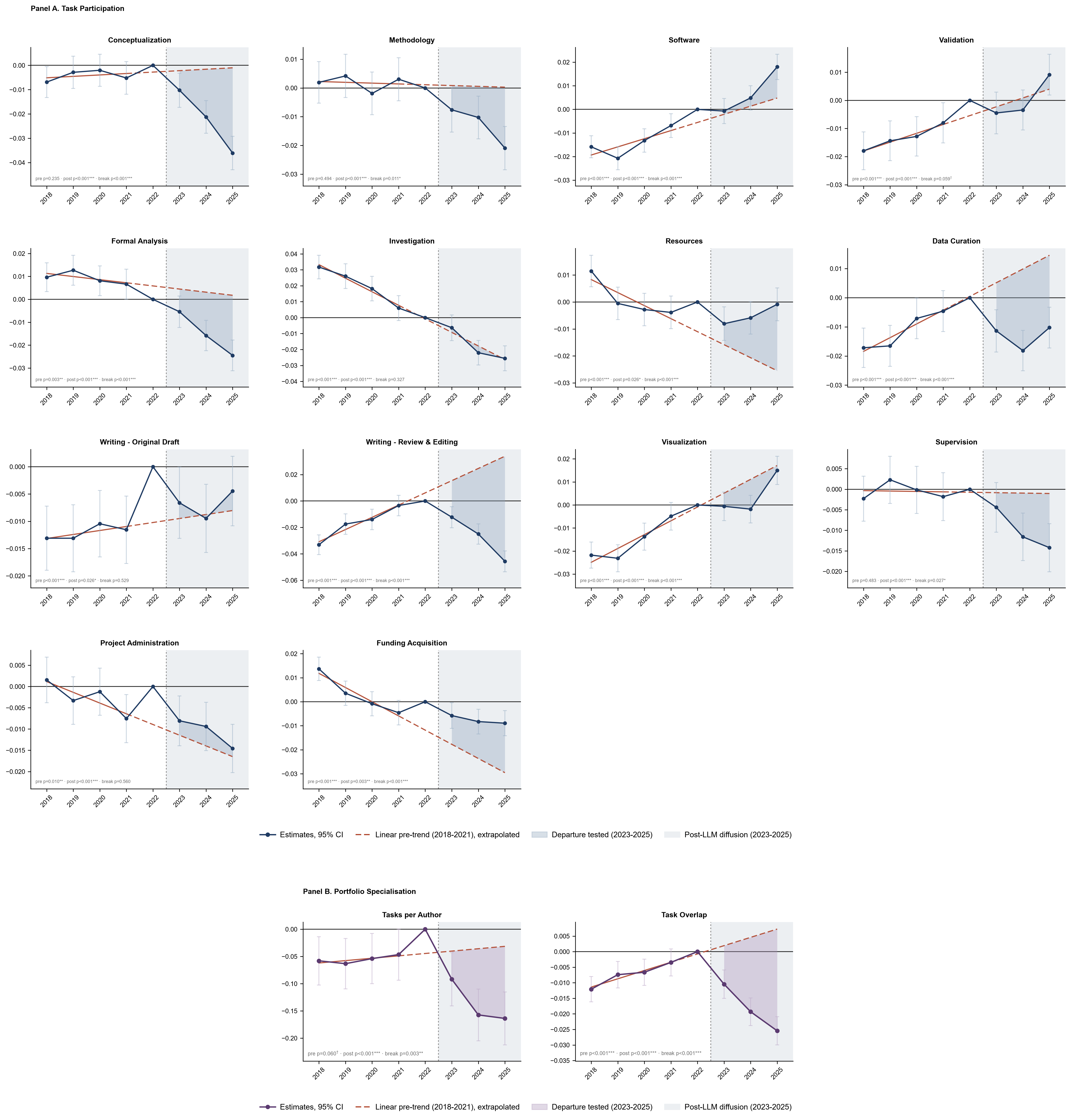}
    \par\vspace{1em}
    \caption{\textbf{Event-study estimates of task participation and portfolio specialization across the 14 CRediT roles, newly formed teams.} Estimated changes relative to 2022, estimated using Equation~\eqref{eq:event_study} with field fixed effects and team-level controls, for newly formed configurations, including those containing previously collaborating subsets ($n=102{,}228$). Panel A reports the share of a paper's authors credited with each of the 14 CRediT contributor roles, analyzed separately rather than grouped into the six research stages used in the main analysis. Panel B reports the two portfolio measures computed over the same 14 roles: the average number of roles per author, and the average pairwise overlap between co-authors' sets of roles. Error bars show 95\% confidence intervals based on robust standard errors. The gray band marks the period following the launch of ChatGPT in late 2022. The red dashed lines extrapolate the linear trends estimated over 2018--2021; the shaded areas between these trends and the post-2022 estimates illustrate the deviations assessed by the trend-break tests. Each panel reports joint Wald tests assessing whether the pre-period (2018--2021) and post-period (2023--2025) coefficients are jointly equal to zero relative to 2022, along with the trend-break test. Significance level:  $^{\dagger}p<0.10$, $^{*}p<0.05$, $^{**}p<0.01$, $^{***}p<0.001$.}
    \label{fig:new14_panels}
\end{figure}

\newpage
\section{Decomposing newly formed teams}
\label{sec: decomposing_new_teams}

As set out in Subsection~\ref{subsec: teamclassification}, our main classification of team type is binary, distinguishing between established teams, whose exact co-author configuration has appeared previously, and newly formed teams, which bring together a configuration not observed before. Within the latter category, however, collaboration histories may differ. Some teams consist entirely of researchers who have never worked together before, whereas others incorporate subsets of members with prior collaborative experience.

To explore this heterogeneity, we decompose newly formed teams into three mutually exclusive categories based on the share of members linked by prior collaboration: below 25\%, between 25\% and 50\%, and between 50\% and 75\%. Purely new teams, whose members share no prior collaboration ties, fall into the lowest band. Established teams and newly formed teams with a prior collaboration share of 75\% or higher are excluded from this decomposition. This exercise allows us to examine whether the organizational adjustments documented in our analysis extend to newly configured teams that nevertheless contain previously collaborating subsets.

\begin{figure}[t!]
    \centering
    \includegraphics[width=\textwidth]{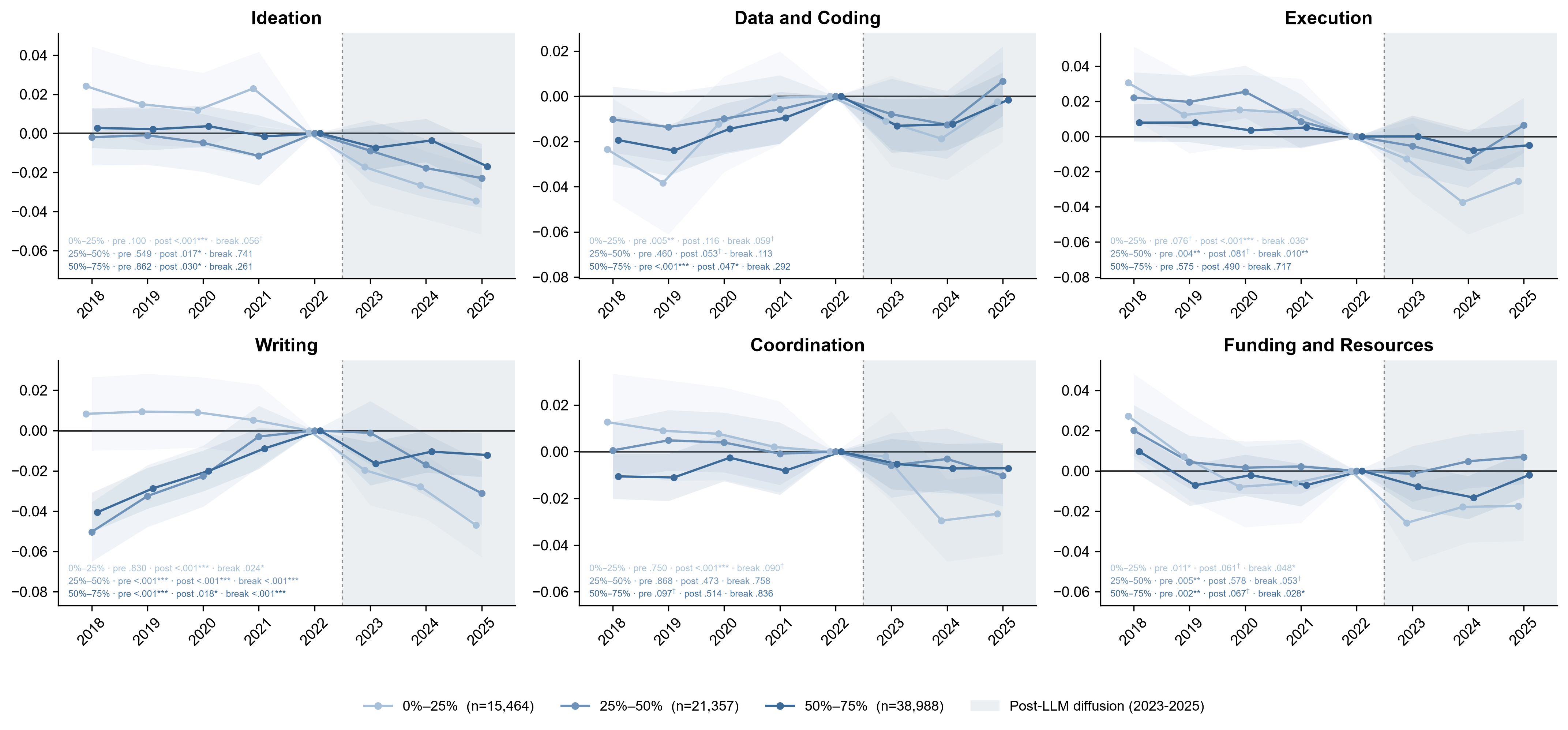}
    \caption{\textbf {Task participation in newly formed teams by prior-collaboration band.}
    Dynamic estimates of the author share credited with each CRediT stage, relative to 2022, from Equation~\eqref{eq:event_study}, with field fixed effects and team-level controls. Specifications are estimated separately for three mutually exclusive groups: partially overlapping teams with collaboration shares below 25\%, from 25\% to below 50\%, and from 50\% to below 75\%. Purely new teams are included in the first partial-overlap band.
    Established teams and partially overlapping teams with shares of
    75\% or more are excluded. Sample sizes are reported in the legend.
    Markers are horizontally offset for readability but refer to the year shown on the axis. Ribbons denote 95\% confidence intervals based on heteroskedasticity-robust standard errors. For each band, panels report joint Wald tests that the 2018--2021 and 2023--2025 coefficients equal zero relative to 2022, and a joint test of
    departures in 2023--2025 from the linear trend fitted over 2018--2021. Extrapolated trends are omitted for readability. Significance level: $^{\dagger}p<0.10$, $^{*}p<0.05$, $^{**}p<0.01$, $^{***}p<0.001$.}
\label{fig:sweep_share_new}
\end{figure}

\begin{figure}[htbp]
    \centering
    \includegraphics[width=\textwidth]{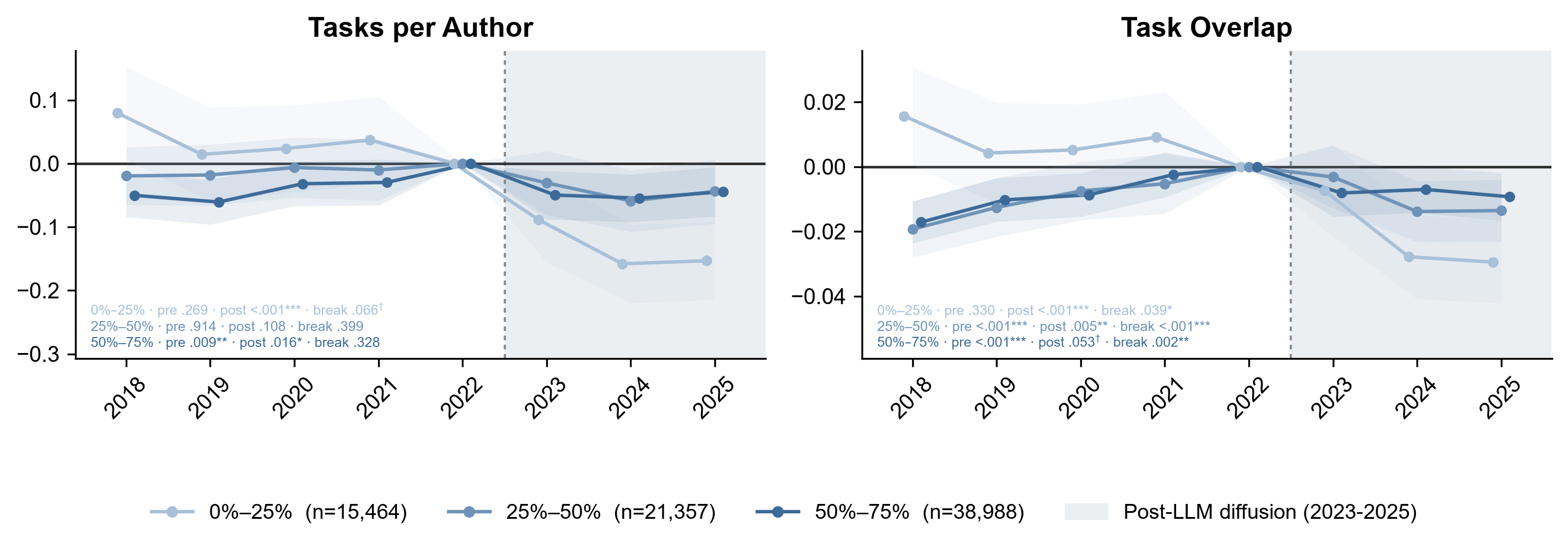}
    \caption{\textbf {Task specialization in newly formed teams by prior-collaboration band.} Dynamic estimates relative to 2022 for average task overlap across co-authors (left) and average tasks per author (right), from Equation~\eqref{eq:event_study}, with field fixed effects and team-level controls. Specifications are estimated separately for the three mutually exclusive groups defined in Figure~\ref{fig:sweep_share_new}; sample sizes are reported in the legend. Markers are horizontally offset for readability. Ribbons denote 95\% confidence intervals based on heteroskedasticity-robust standard errors. For each band, panels report joint Wald tests that the 2018--2021 and 2023--2025 coefficients equal zero relative to 2022, and a joint test of departures in 2023--2025 from the linear trend fitted over 2018--2021. Extrapolated trends are omitted for readability. Significance level: $^{\dagger}p<0.10$, $^{*}p<0.05$, $^{**}p<0.01$, $^{***}p<0.001$.}
    \label{fig:sweep_a2_new}
\end{figure}

\newpage
Figures~\ref{fig:sweep_share_new} and~\ref{fig:sweep_a2_new} reveal that the changes observed among newly formed teams are not confined solely to configurations with little or no prior collaboration. For instance, participation in \textit{Writing} and task overlap exhibit significant downward departures from their pre-LLM trends in all three collaboration bands ($p = 0.024$, $p < 0.001$ and $p < 0.001$ for \textit{Writing}; $p = 0.039$, $p < 0.001$ and $p = 0.002$ for task overlap). Participation in \textit{Execution} also exhibits a significant downward departure in the two lower bands ($p = 0.036$ and $p = 0.010$), though not in the highest band ($p = 0.717$). The evidence is somewhat less consistent for other outcomes. Although participation in \textit{Ideation} and average tasks per author generally decline relative to 2022, these changes are not accompanied by significant departures from pre-existing trends at the 5\% level in any of the three bands.

Overall, these findings provide two additional insights. First, some of the organizational adjustments documented in the main analysis extend to newly formed teams that include members with a history of collaboration. Second, the decomposition does not reveal a smooth gradient in organizational adjustment scaling neatly with the extent of prior collaboration. These findings remain descriptive, and separate significance tests within each band do not establish differences between bands. Nevertheless, they are consistent with the possibility that the organizational response to LLM diffusion depends less on the extent of prior collaboration per se than on whether teams reproduce existing working arrangements or must adapt them to a new configuration.

\end{document}